\documentclass[graphics,floatfix, footinbib,tightenlines,nobibnotes, aps, prb, twocolumn]{revtex4-1}
\usepackage{amsmath,amssymb,wasysym}
\usepackage{graphicx}
\usepackage{dcolumn}
\usepackage{bm}
\usepackage{braket}
\usepackage{subcaption}
\usepackage{verbatim}
\usepackage{float}
\usepackage{bbold}
\usepackage{color}
\usepackage{xcolor}
\usepackage{relsize}
\usepackage{amsthm}
\usepackage{enumerate}
\usepackage{soul,xcolor}
\usepackage[T1]{fontenc}
\usepackage[colorlinks=true ,urlcolor=blue,urlbordercolor={0 1 1}]{hyperref}

\newcommand{\beq}{\begin{eqnarray}}
\newcommand{\eeq}{\end{eqnarray}}

\newcommand{\bsp}{\begin{split}}
\newcommand{\esp}{\end{split}}

\newcommand{\be}{\begin{equation}}
\newcommand{\ee}{\end{equation}}

\begin{document}

\setstcolor{red}

\title{Doping a Mott insulator with excitons in moir\'e bilayer: fractional superfluid, neutral Fermi surface and Mott transition}
\author{ Ya-Hui Zhang}
\affiliation{Department of Physics and Astronomy, Johns Hopkins University, Baltimore, Maryland 21218, USA
}

\date{\today}

\begin{abstract}
In this paper we explore possible phases arising from doping neutral excitons into a Mott insulator in the context of moir\'e bilayers.  We consider two moir\'e layers coupled together only through inter-layer repulsion and there is a U(2)$\times$ U(2) symmetry. The densities of the two layers can be tuned to be $n_t=x,n_b=1-x$ with $n_t+n_b=1$. $x=0$ limit is a layer polarized Mott insulator and small $x$ regime can be reached by doping inter-layer excitons at density $x$. Charge gap can remain finite at small $x$, as is demonstrated experimentally in the WSe$_2$-hBN-WSe$_2$/WS$_2$ system. To capture the intertwinement of the spin and exciton degree of freedom, we propose a four-flavor spin model. In addition to the obvious possibility of single exciton condensation phase, we also identified more exotic physics with fractionalization: (I) We define spin gap $\Delta_t, \Delta_b$ for the two layers respectively.  As long as the spin gap at either layer is finite, single exciton condensation is impossible and we can only have paired exciton condensation. If both spin gaps are finite, it can be a fractional exciton superfluid with paired exciton condensation coexisting with $Z_2$ spin liquid. Numerical evidences for such a phase will be provided. (II) If the layer polarized Mott insulator at $x=0$ is in a U(1) spin liquid with spinon Fermi surface, the natural phase at $x>0$ hosts neutral Fermi surface formed by fermionic excitons.  There are metallic counter-flow transport and also Friedel oscillations in layer polarization in this exotic phase. Our numerical simulation in one dimension  observes an analog of this neutral Fermi surface phase. (III) There could be a metal-insulator transition (MIT) by tuning $x$ in the universality class of a bandwidth tuned MIT. We provide one theory for such a continuous Mott transition and predicted a universal drag resistivity.

\end{abstract}

\pacs{Valid PACS appear here}
\maketitle

\section{Introduction}

\begin{figure}[ht]
\centering
\includegraphics[width=0.5\textwidth]{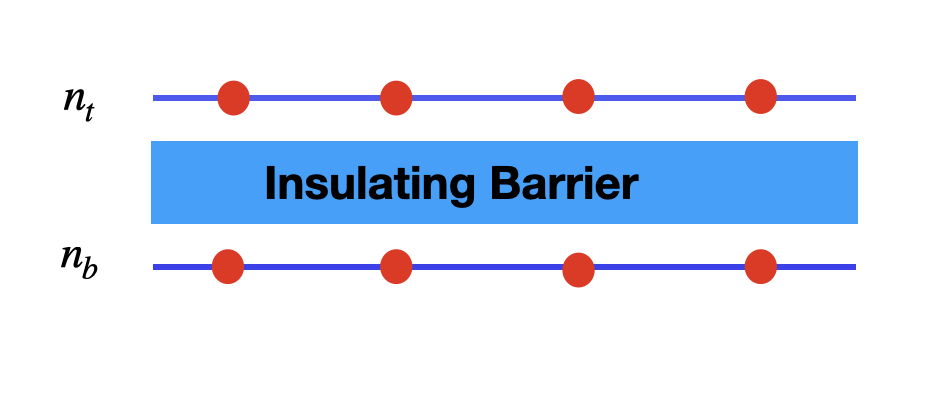}
\caption{Illustration of a moir\'e bilayer. Two layers are separated by an insulating barrier and each layer has the same superlattice with lattice constant $a_M$.}
\label{fig:moire_bilayer_illustration}
\end{figure}

Recently the idea of moir\'e bilayer\cite{zhang_2020,zhang20214} was introduced by us to simulate ``spin'' physics and realize quantum spin liquid (QSL)\cite{savary2016quantum,zhou2017quantum,knolle2019field} phases using the layer pseudospin. Similar to the quantum Hall bilayer\cite{eisenstein2014exciton,li2017excitonic,liu2017quantum}, we consider a bilayer system where two layers are coupled through inter-layer repulsion without inter-layer tunneling (see Fig.~\ref{fig:moire_bilayer_illustration}). Thus the total charge of each layer is separately conserved.  If we label the charge of the two layers as $Q_t$ and $Q_b$ respectively,  we can define  two conserved quantum numbers: $Q=Q_t+Q_b$ and $P_z=\frac{1}{2}(Q_t-Q_b)$. We can view $Q$ as the total charge, then $P_z$ is a pseudospin and its transport can be conveniently probed by electrically counter-flow measurement\cite{zhang_2020}. We also need the inter-layer distance $d$ to be much smaller than the intra-layer lattice constant $a_M$ so that the inter-layer repulsion is comparable to the intra-layer repulsion. To achieve that, it is natural to  build up a bilayer using  moire superlattices based on graphene\cite{cao2018correlated,Wang2019Signatures,Wang2019Evidence,yankowitz2019tuning,Wang2019Signatures,chen2019tunable,lu2019superconductors,Cao2019Electric,Liu2019Spin,Shen2019observation,polshyn2020nonvolatile,chen2020electrically,sharpe2019emergent,serlin2020intrinsic} or transition metal dichalcogenides (TMD)\cite{tang2020simulation,regan2019optical,wang2019magic,gu2021dipolar,zhang2021correlated,li2021continuous,ghiotto2021quantum}. Here we will focus on TMD moir\'e superlattice in which Mott physics and anti-ferromagnetic spin coupling have been well studied experimentally in single moir\'e layer\cite{tang2020simulation,regan2019optical,wang2019magic,li2021continuous,ghiotto2021quantum}.  Previous theoretical studies have also shown that an extended spin 1/2 Hubbard model is a good description of the single moir\'e layer\cite{wu2018hubbard,wu2019topological,pan2020band,pan2020quantum,pan2021interaction,pan2022interaction,zhang2020moire}. Especially for TMD hetero-bilayer there is a good SU(2) spin rotation symmetry because valley contrasting flux is negligible\cite{wu2018hubbard}.

Unlike the quantum Hall bilayer, the spin within each moir\'e layer is still active and we have both the layer pseudospin and the real spin. In our previous work\cite{zhang_2020,zhang20214}, we consider the ideal limit where the real spin and the layer pseudospin together form a SU(4) spin. Although various interesting phases including a chiral spin liquid was predicted theoretically in the SU(4) symmetric point, we note that this ideal limit requires the interlayer distance to be almost zero, which is in tension with the requirement of zero inter-layer tunneling. Hence the realization of the ideal SU(4) model in realistic systems is a challenge. Twisted AB stacked TMD homobilayer has been experimentally realized\cite{xu2022tunable}, where inter-layer tunneling is suppressed by spin conservation instead of hBN barrier. However, it is not clear now whether the inter-layer tunneling in that system is completely zero, which is required for counterflow transport.  Experimentally it is much easier to completely suppress the inter-layer tunneling by putting a thick hexagon boron nitride (hBN) between the two layers\cite{shimazaki2020strongly,gu2021dipolar,zhang2021correlated}. In Ref.~\onlinecite{gu2021dipolar,zhang2021correlated}, a TMD moir\'e layer based on WSe$_2$/WS$_2$ on the bottom is separated from a moir\'e-less TMD monolayer WSe$_2$ on the top by a hBN barrier.  Let us label the density of the two layers as $n_t,n_b$ per moir\'e unit cell. A correlated insulator with charge gap is found for the filling $n_t=x, n_b=1-x$ in the range $x\in (0,x_c)$ with the critical exciton density $x_c$ as large as $0.5$\cite{zhang2021correlated}. The SU(4) symmetry is strongly broken in this case and we have only U(2)$\times$ U(2) symmetry corresponding to the separate conservation of charge and spin in the two layers. This moir\'e+monolayer setup is easier to fabricate than the moir\'e+moir\'e system which we previously proposed\cite{zhang_2020,zhang20214}. However, it is not clear whether interesting spin liquid phase can still be realized in this anisotropic case with lower symmetry. The main purpose of this work is to propose a theoretical framework to analyze the spin and exciton dynamics and demonstrate the existence of exotic phases with fractionalization even in this simpler system.

We first want to highlight the essential difference between the system considered here and the traditional excitonic insulator\cite{jerome1967excitonic,zittartz1967theory,halperin1968excitonic,comte1982exciton}. In the conventional case, a bosonic exciton operator is well defined on top of a band insulator.  In contrast, the vacuum of the inter-layer excitons in our moir\'e bilayer is a layer polarized Mott insulator at the $n_t=0,n_b=1$ limit. Unlike a band insulator, the spin configuration of a Mott insulator is not uniquely frozen and the spin remains as an active degree of freedom at low energy.  In the usual excitonic insulator, one can make a particle hole transformation relative to the band insulator, so a hole operator is well defined. Under such a particle hole transformation, the band insulator becomes the vacuum of the exciton. Then the active degree of freedoms are just small densities of electron and holes on top of the band insulator, which then form excitons.   In contrast, the spin configuration of a Mott insulator is not frozen and Mott insulator can not be viewed as a vacuum. Excitons are not the only active degree of freedom at finite $x$ and they need to interact with the spin 1/2 moments.  It is clear that we need a more sophisticated model to capture the exciton-spin dynamics than a simple boson model usually assumed for excitonic insulator. If one applies a strong magnetic field to polarize the spin, then the Mott insulator becomes a band insulator and we are in the conventional excitonic insulator. In this paper we will be mainly focused on the zero magnetic field case where spin is not dead and more interesting physics than single exciton condensation may arise.

We propose a four-flavor spin model to capture the low energy spin-exciton dynamics.   This four dimensional Hilbert space can be constructed by the tensor product of two spin $1/2$ corresponding to $\vec P$ and $\vec S$. Here $\vec S$ represents the real spin and $\vec P$ represents the layer pseudospin. $P_z=\frac{1}{2} (n_t-n_b)$ is the layer charge. The physical operators can be constructed as $P_\mu S_\nu$ with $\mu,\nu=0,1,2,3$. $P_0 S_0$ is the identity operator and there are $15$ non-trivial operators representing the neutral spin-exciton degree of freedom. They correspond to the $15$ generators of SU(4) group. Actually at leading order the low energy model is just a four-flavor Heisenberg spin model similar to the SU(4) spin model studied in Ref.~\onlinecite{zhang20214}. Now we need to include anisotropy terms coming from finite inter-layer distance $d>0$ and imbalanced filling $n_t<<n_b$, but the structure of the spin model remains similar to the SU(4) model, though the symmetry is now reduced to $(SU(2)_t\times SU(2)_b \times U(1)_l)/Z_4$\footnote{$Z_4$ here corresponds to $c_{i;a \sigma}\rightarrow e^{i\frac{2\pi n}{4}} c_{i;a\sigma}, n=0,1,2,3$, which is shared by the global charge transformation and the spin transformation such as generated by $e^{i \frac{1}{2} P_z 2\pi n}e^{i S_{t;z} 2\pi n}$.}. Here $SU(2)_t$ is generated by $\vec S_t=(P_z+\frac{1}{2}) \vec S$, $SU(2)_b$ is generated by $\vec S_b=(\frac{1}{2}-P_z) \vec S$ and $U(1)_l$ is generated by $P_z$. Physically $\vec S_t, \vec S_b$ are the spin operators of the two layers respectively and $U(1)_l$ corresponds to the layer polarization or a dipole charge.

This spin model can be naturally derived from a four-flavor Hubbard model suitable for moir\'e+moir\'e case.  In the strong interaction limit, the system is in a Mott insulator at total filling $\nu_T=n_t+n_b=1$. But even for the moir\'e+monolayer case, we argue that the same model works for the small x limit because the top layer can inherit the same moir\'e superlattice from the bottom layer through inter-layer repulsion.  Actually  at leading order, the effective exciton-spin model  below the charge gap is constrained to be in the form we propose by the $U(2)\times U(2)$ symmetry of the system.  We note a simple Hubbard model is not sufficient\cite{pan2022interaction} to capture the long range Coulom interaction. However, at the filling $\nu_T=1$ below the charge gap, the effect of longer range interaction is to renormalize the values of the parameters in our four-flavor spin model. In this paper we will treat these couplings as phenomenological parameters and explore the phase diagram.  Derivation of these parameters for a specific realistic material will be left to future work. We also emphasize that our model only works at energy scale well below the charge gap.  When the charge gap is small, additional ring exchange terms beyond bilinear terms are needed. When the charge gap approaches zero, a bosonic model is not sufficient anymore and one needs a fermionic model to also capture the possibility of electron-hole gas. We also do not include disorder in our study, which may lead to Anderson localization at small $x$\cite{ahn2022disorder}. Future experiments are needed to address the question whether the insulator seen in Ref.~\onlinecite{gu2021dipolar,zhang2021correlated} is actually an Anderson insulator of electrons or excitons.


We study the four-flavor spin model on triangular lattice using either Schwinger boson mean field or Abrikosov fermion mean field theory. Deep inside the Mott insulator, Schwinger boson theory is convenient to capture the $120^\circ$ magnetic order of $\vec S_b$ in the layer polarized limit with $x=0$. If the $120^\circ$ order at the bottom layer is strong, the physics is reduced to a spin $1/2$ boson model and we obtain either an exciton condensation phase with $\vec S_t$ ordered or a paired exciton condensation phase with $\vec S_t$ gapped. In paired exciton condensation phase, only a Cooper pair of excitons (bi-exciton) is condensed and the single exciton is gapped. In a more interesting case, we find the $120^\circ$ order in the bottom layer can also be depleted by reducing the density $n_b$, after which we get a paired exciton condensation phase coexisting with $Z_2$ spin liquid. We will provide numerical evidences for spin gaps through density matrix renormalization group (DMRG) simulation\cite{white1992density,white1993density,mcculloch2008infinite}.  We actually find that there are two different $Z_2$ spin liquids  with $0$ or $\pi$ flux for the Schwinger bosons. Both $0$ and $\pi$ flux $Z_2$ spin liquids have been proposed before for a single layer spin $1/2$ model on triangular lattice\cite{wang2006spin}. But they are usually unstable to Schwinger boson condensation because the boson density is too large. Here because we can reduce the density of the boson in the bottom layer to $n_b=1-x$, it is possible to stabilize the $Z_2$ topological order. The $Z_2$ fractionalization coexists with the paired exciton condensation, and thus the phases can be called as fractional superfluids. The possibility of $Z_2$ spin liquids in this system opens the exciting future direction to search for superconductivity upon doping the bilayer Mott insulator, following Anderson's resonating-valence-bond (RVB) theory\cite{anderson1987resonating}.

For the weak Mott regime, magnetic order may not exist even in the layer polarized limit\cite{li2021continuous}. In the single moir\'e layer based on MoTe$_2$/WSe$_2$, the weak Mott insulator was shown to be more consistent with a U(1) spin liquid phase with spinon Fermi surface\cite{li2021continuous} given the large spin susceptibility at lowest temperature. Smoking gun evidence for such a neutral fermi surface has remained elusive because of the lack of the probe of neutral excitations. We propose to add a monolayer TMD separated from MoTe$_2$/WSe$_2$ by a hBN similar to the setup of Ref.~\onlinecite{gu2021dipolar,zhang2021correlated}. If the layer polarized limit is indeed a spinon Fermi surface state, then we should use the four-flavor Abrikosov fermion $f_{i;a\sigma}$ to attack the four-flavor spin model. In the $x=0$ limit, the band for $f_{i;t\sigma}$ is empty while $f_{i;b\sigma}$ forms the spinon Fermi surface. Upon increasing $x$, there is a one component to two components Lifshitz transition and we reach a U(1) spin liquid phase where both layers host a spinon Fermi surface. Physically $f_{t;\sigma}$ can be interpreted as a fermionic exciton formed by binding electron $c_{i;t\sigma}^\dagger$ with a holon in the bottom layer. This phase will support metallic counter-flow transport $\rho_{\text{counterflow}}(T) \sim T^{\alpha}$ with some non-Fermi-Liquid (NFL) exponent $\alpha$ due to gauge fluctuation. The exact value of $\alpha$ is still not well established theoretically\cite{lee1992gauge,maslov2011resistivity,hartnoll2014transport,lee2018recent,lee2021low} and we hope the future experiment in moir\'e bilayer could provide experimental constraint on the theory of spinon Fermi surface phase in 2+1d. The neutral Fermi surface could also be revealed by Friedel oscillation in terms of the layer polarization, which may be measured electrically.

In the experiments of Ref.~\onlinecite{gu2021dipolar,zhang2021correlated}, the charge gap closes beyond a critical doping $x_c$ for the exciton density( see Fig.~\ref{fig:moire_bilayer_illustration}(b)). It is thus interesting to study the metal insulator transition (MIT) tuned by $x$. We will argue that such a transition is in the class of bandwidth tuned MIT. The easiest way to understand the transition is from tuning the inter-layer distance $d$. When $d$ is large, the two layers decouple and must be in metallic phases with Fermi surfaces. At fixed x, when we decrease $d$ to be smaller than a critical distance $d_c$, the system becomes insulating because of the inter-layer repulsion $U' n_t n_b$, which is analogous to the Hubbard U term. Therefore the distance tuned MIT is similar to the MIT tuned by $U/t$ in the simple spinful Hubbard model. It is easy to imagine that $d_c$ decreases with $x$ and then one can drives the same MIT by tuning $x$ at fixed $d$. We also provide a theory of this MIT by generalizing the theory of Ref.~\onlinecite{senthil2008theory}. We assume the Mott insulator side hosts spinon Fermi surfaces and the MIT is driven by the condensation of the slave boson. The theory is essentially the same as in the single layer case, but in the moir\'e bilayer case one can drive currents in the two layers separately. Hence we can get a $2\times 2$ tensor for the resistivity $\rho_{xx}$. In the slave boson theory of MIT, there is a universal drag resistivity $\rho_{xx;12}$ at order $h/e^2$, which may be easier to measure than the jump of the residual resistivity proposed in the single layer case\cite{senthil2008theory}. Drag resistivity essentially measures the pseudospin-charge separation and thus may also be very useful to characterize the metallic state upon doping the bilayer Mott insulator by tuning $n_t+n_b$ away from $1$.

 The rest of the paper is structured in the following way. In Sec.~\ref{sec:model} we introduce the four-flavor spin model and its derivation. We clarify the relation ship between exciton and layer pseudospin language in Sec.~\ref{sec:dictionary}.  In Sec.~\ref{sec:boson_fermion_parton_construction} we introduce the framework to deal with the model using either schwinger boson or Abrikosov fermion parton construction.  In Sec.~\ref{sec:bosonic_exciton_superfluids} we discuss various possible exciton superfluid phases coexisting with magnetic order or Z$_2$ spin liquids. This includes a spin polarized BEC of excitons, a paired superfluid of excitons with a spin gap in one of the two layers.  Most interestingly we show that it is possible to deplete the magnetic order of the Mott insulator by doping excitons, leading to a Z$_2$ spin liquid coexisting with exciton superfluid.  A phase diagram in $x-J'_b$ space will be provided, with $x$ as the exciton density and $J'_b$ as the next nearest neighbor coupling in the Mott layer. The phase diagram hosts  two different Z$_2$ spin liquids corresponding to zero and $\pi$ flux ansatz of Schwinger boson, which connect to the $120^\circ$ ordered phase and stripe ordered antiferromagnetic phase respectively. In Sec.\ref{sec:neutral_fermi_surface} we discuss the possibility of fermionic exciton and neutral Fermi surface and in Sec.~\ref{sec:MIT} we show a critical theory of continuous mott insulator transition tuned by the exciton density $x$.

\section{Anisotropic SU(4) spin model\label{sec:model}} 

We consider a moir\'e bilayer with two moir\'e superlattice layers separated by an insulating barrier, as shown in Fig.~\ref{fig:moire_bilayer_illustration}. The two moir\'e superlattices are assumed to be aligned with each other and share the same triangular lattice.  The inter-layer tunneling is forbidden by insulating barrier. In the following we are going to derive a four-flavor exciton-spin model for filling $\nu_T=1$ at the energy scale well below the charge gap. We will justify this model using a simple four-flavor Hubbard model. But we note at leading order the form of this four-flavor spin model is constrained by the U(2)$\times $ U(2) symmetry of the system and we expect the same model even for more complicated models, for example with longer range Coulomb interaction.

We consider the following four-flavor Hubbard model\cite{zhang20214}

\begin{align}
H&=- \sum_{a=t,b} t_a \sum_{ij}\sum_{\sigma=\uparrow,\downarrow} c^\dagger_{i;a \sigma}c_{j;a \sigma}-\frac{D}{2}\sum_i (n_{i;t}-n_{i;b})\notag\\
&+\frac{1}{2} \sum_i \sum_a U_a n_{i;a}(n_{i;a}-1)+U' \sum_i n_{i;t}n_{i;b} \notag\\
&+\sum_a V_a \sum_{\langle ij \rangle} n_{i;a}n_{j;a}+V' \sum_{\langle ij \rangle}(n_{i;t}n_{j;b}+n_{i;b}n_{j;t})
\label{eq:Hubbard_model_moire_bilayer}
\end{align}
where $a=t,b$ labels the layer index and $\sigma=\uparrow,\downarrow$ labels the spin. In TMD, spin is locked to valley due to a strong ising spin-orbit coupling. Here we focus on two narrow moir\'e bands from the two valleys on top of the original valence band. Within this two flavor space, there is a good SU(2) spin rotation symmetry\cite{wu2018hubbard}. . $n_{i;a}=\sum_{\sigma}c^\dagger_{i;a\sigma}c_{i;a\sigma}$ is the density at layer $a$ on site $i$. $U_a$ and $U'$ are intra-layer and inter-layer on-site Hubbard repulsion. $V_{a}$ and $V'$ are intra-layer and inter-layer nearest neighbor repulsion. $D$ is a displacement field. Note here we assume that the two layers can have different hopping and interaction, so one layer can be made less correlated, for example with $U_b>>U_t$. The model has exact U(4) symmetry if $t_t=t_b$, $U_t=U_b=U'$, $V_t=V_b=V'$ and $D=0$.  We will work in the case with significant anisotropic terms breaking the U(4) symmetry. Even in the anisotropic case, we still have a large symmetry $U(2)\times U(2)$, where each U(2) corresponds to the charge and spin conservation at one of the two layers. Note the spin  $\vec{S}_{i;a}=\frac{1}{2} c^\dagger_{i;a\sigma} \vec{\sigma}_{\sigma \sigma'}c_{i;a\sigma'}$ at each layer $a=t,b$ is separately conserved, the same as the charge $n_{i;a}=\sum_{\sigma}c^\dagger_{i;a\sigma}c_{i;a\sigma}$. In the above we only included on-site and nearest neighbor repulsion. For filling $\nu_T=1$, long ranger Coulomb interaction should only renormalizes the values of the parameters in the low energy spin model we are going to study.

\begin{figure}[ht]
\centering
\includegraphics[width=0.5\textwidth]{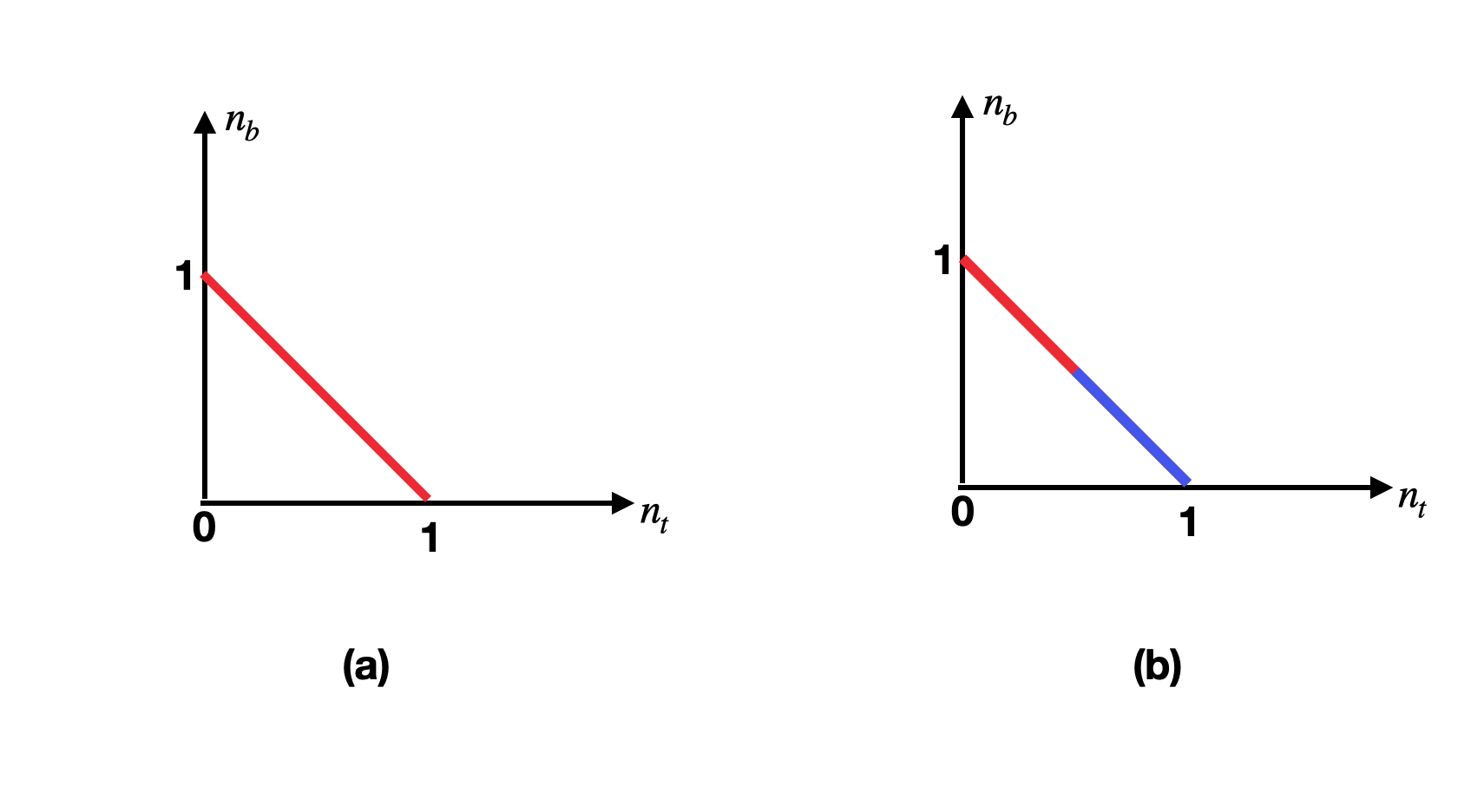}
\caption{Illustration of the bilayer Mott insulator at total filling $n_t+n_b=1$. The red color labels the region with a Mott charge gap. The blue color labels the region without charge gap. (a) If $U_t,U_b,U'>>t_t,t_b$, the whole line of $n_t+n_b=1$ is in a Mott insulator. This applies to moir\'e bilayer formed in WSe$_2$-WS$_2$-WSe$_2$ or twisted AB stacked WSe$_2$ bilayer where each layer feels a strong moir\'e superlattice potential.  (b) If the top layer is less correlated with a large $t_t$, then the charge gap is finite only in the region with $n_t$ smaller than a critical value $x_c$. This is true in moir\'e+monolayer setup such as in the WSe$_2$-hBN-WSe$_2$/WS$_2$ system\cite{gu2021dipolar,zhang2021correlated}.}
\label{fig:moire_Mott_insulator_illustration}
\end{figure}

 If we assume $U_t,U_b,U'$ are all much larger than the hopping $t_t,t_b$, we can have a Mott insulator at total filling $n_t+n_b=1$, as illustrated in Fig.~\ref{fig:moire_Mott_insulator_illustration}(a).  This is true in WSe$_2$-WS$_2$-WSe$_2$ system or twisted AB stacked WSe$_2$ bilayer\cite{zhang20214}. However, we will also consider the case that the top layer is less correlated and has a larger $t_t$. This is true in the moir\'e+monolayer system such as in the recent experiments\cite{gu2021dipolar,zhang2021correlated}. In this case we expect that the Mott gap survives only when the density of the top layer is smaller than a critical value, as illustrated in Fig.~\ref{fig:moire_Mott_insulator_illustration}. In this paper we mainly focus on the case with $n_t=x, n_b=1-x$ with small $x$, so there is a charge gap and the difference between the moir\'e+moir\'e and moir\'e+monolayer system does not matter too much.

 Within the region with a finite charge gap, the low energy physics is governed by an anisotropic SU(4) model obtained from the standard $t/U$ expansion:

\onecolumngrid

\begin{align}
	H_S&=J_t \sum_{\langle ij \rangle}\vec{S}_t(i) \cdot \vec{S}_t(j)+J_b \sum_{\langle ij \rangle}\vec{S}_b(i) \cdot \vec{S}_b(j)
    +J'_b \sum_{\langle \langle ij \rangle \rangle}\vec{S}_b(i) \cdot \vec{S}_b(j)
	+\frac{1}{2}J_{pz}\sum_{\langle ij \rangle} P_z(i)P_z(j)\notag \\ 
    &+\frac{1}{2}J_p \sum_{\langle ij \rangle}\big(P_x(i)P_x(j)+P_y(i)P_y(j)\big)\left(4\vec{S}(i) \cdot \vec{S}(j)+S_0(i)S_0(j)\right)
	\label{eq:spin_layer_model}
\end{align}
\twocolumngrid

where $J_t=\frac{4 t_t^2}{U_t}$, $J_b=\frac{4 t_b^2}{U_b}$, $J_p=\frac{4 t_t t_b}{U'}$ and $J_{pz}=2\delta V-\frac{1}{2}(J_t+J_b)+(\frac{4 t_t^2}{U'}+\frac{4 t_b^2}{U'})$, where $\delta V=V_t+V_b-2 V'$.  In the above we ignored the constant term and the term linear to $P_z(i)$, which is just a chemical potential.  We also include a term $J'_b$ coming from the next nearest neighbor hopping in the bottom layer. Such a term will be included in the discussion of Sec.~\ref{subsection:fractional_SF}, but can be ignored for now.

In the $SU(4)$ symmetric limit with $J_t=J_b=J_p=J_{pz}=J$, we recover the SU(4) spin model

\onecolumngrid

\begin{align}
	H_S=\frac{J}{8}\left(4 \vec P(i)\cdot \vec P(j)+P_0(i)P_0(j)\right)\left(4 \vec S(i)\cdot \vec S(j)+S_0(i)S_0(j)\right)
\end{align}

\twocolumngrid

In the above we keep up to the order of $t^2/U$, but higher order terms can be generated in the standard way.  In the case where the top layer is not correlated, we have large value of $\frac{t_t}{U_t}$ and $\frac{t_t}{U'}$, so $J_t$ can be significantly larger than $J_b$ and $J_p$.  We may also worry that the $t/U$ expansion fails because $\frac{t_t}{U_t}$  or $\frac{t_t}{U'}$ are too large.  However, the term $\frac{t_{t}^2}{U_t}$ exists only when there are two nearby particles in the top layer and is associated with a factor $n_t(i) n_t(j) \sim x^2$. Similarly the terms $\frac{t_t t_b}{U'}$ and $\frac{t_t^2}{U'}$ are associated with a factor $n_t(i) n_b(j)\sim x$ in the $x<<1$ limit. Therefore the terms including $t_t$ are suppressed at the dilute exciton limit and the $t/U$ expansion is valid in the $x<<1$ limit. When we increase the exciton density, the expansion will fail eventually, which indicates a Mott transition with the charge gap closes.  This Mott transition will be discussed in Sec.~\ref{sec:MIT}.

\subsection{Moir\'e+monolayer case}

In the $x<<1$ limit, we can have a Mott insulator even if the top layer is not strongly correlated. Actually the top layer even does not need to have a moir\'e superlattice. The point is that the doped electrons in the top layer inherit the moir\'e potential of the bottom layer and are trapped in the triangular moir\'e superlattice sites due to the attraction from the hole in the bottom layer.  The only assumption we need to make is that the exciton is tightly bound with the 2D position of the electron-hole pair vertically aligned. At low energy the electron in the top layer can not hop away from the hole in the bottom layer, which will break the exciton and costs energy. With this assumption, we only need to keep bosonic degree of freedom in the low energy. As we will argue below, the low energy physics is captured by a similar spin-layer four-flavor model as in Eq.~\ref{eq:spin_layer_model}.

When the top layer is moir\'e-less, in principle we need to keep $O\big((\frac{a_M}{a})^2\big)$ states for the top layer within each moir\'e unit cell of the bottom layer. But because of the inter-layer coulomb interaction, it is reasonable to imagine that there are only finite number of states needed to be kept. For example, if there is a hole at the moir\'e site $i$ in the bottom layer, then the relevant state for the top layer is state $c^\dagger_{i;t\sigma} \ket{0}= \int d^2 \mathbf x \Phi(\mathbf x-\mathbf{R_i})f^\dagger_\sigma(\mathbf x) \ket{0}$. Here $f^\dagger_\sigma(\mathbf x)$ is the microscopic electron creation operator on the top layer, which lives in the original lattice with small lattice constant $a$.  Note we still have two flavors $\sigma=\uparrow,\downarrow$ coming from the spin-valley locking.  $\Phi(\mathbf x-\mathbf{R_i})$ is the wavefunction for the lowest exciton state when the hole is localized at moir\'e site $i$. $c^\dagger_{i;t\sigma}$ can be viewed as a coarse graining operator for the top layer living on the moir\'e site $i$. At low energy, it is reasonable to assume that we only need to keep the exciton created by $c^\dagger_{i;t\sigma}c_{i;b\sigma'}$. Note here $c_{i;b\sigma'}$ is the annihilation operator of electron at the moir\'e site $i$ of the bottom layer. In low energy we can safely ignored other states in the top layer and keep only four states at each site: $\ket{1}_i=c^\dagger_{i;t\uparrow}\ket{0}$, $\ket{2}=c^\dagger_{i;t\downarrow}\ket{0}$, $\ket{3}_i=c^\dagger_{i;b\uparrow}\ket{0}$, $\ket{4}=c^\dagger_{i;b\downarrow}\ket{0}$.  This is exactly the same Hilbert space  defined in Eq.~\ref{eq:spin_layer_model}. The symmetry of this system is still the same as the moir\'e bilayer case, with spin of the two layers separately conserved. The pseudospin $P_z$ is still conserved.  Within this restricted Hilbert space, the low energy effective Hamiltonian is constrained by symmetry to be in the same form as in Eq.~\ref{eq:spin_layer_model}. Thus we conclude that Eq.~\ref{eq:spin_layer_model} is also the correct effective model for the  moir\'e+monolayer case.

The parameters can of course be different from the case with both layers to be moir\'e layers. We will treat the couplings $J_t, J_b, J_p, J_{pz}$ as phenomenological parameters.  $J_b$ is the spin-spin coupling of the bottom layer and is fixed by the physics of the moir\'e layer itself and we know $J_b>0$. $J_t$ is the spin-spin coupling of the to layer. $J_p$ is basically the hopping of the exciton.  $J_{pz}$ is the nearest neighbor repulsion between two excitons.

\begin{figure}
\centering
\includegraphics[width=0.5\textwidth]{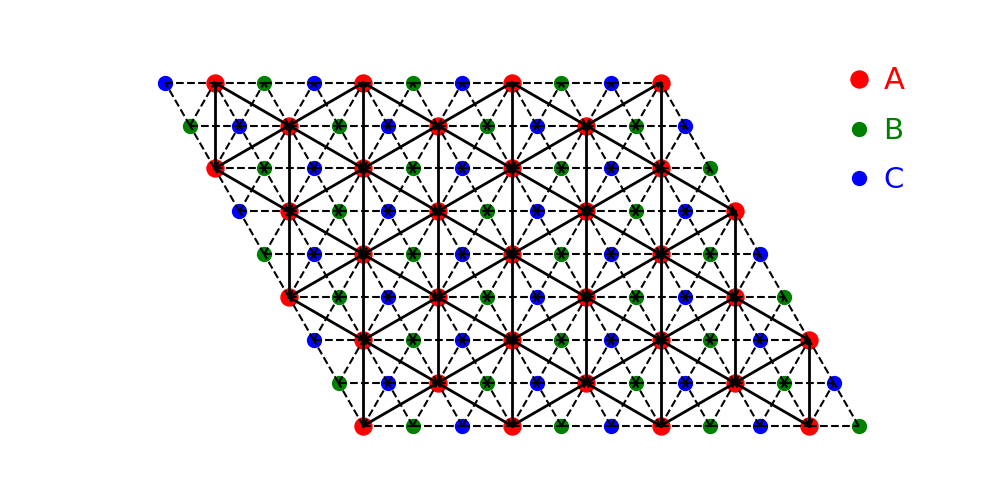}
\caption{Illustration of a toy model for the moir\'e-monolayer case. The bottom layer has a moir\'e superlattice labeled by the solid line. The top layer is just a monolayer TMD without moir\'e superlattice. However, it feels interaction from the bottom moir\'e superlattice. We assume that only three orbitals in the top layer need to be kept  per moir\'e unit cell. They live on the A,B,C sublattice, which together form a triangular lattice with lattice constant $\frac{a_M}{\sqrt{3}}$ denoted by the dashed line. Only the A sublattice is on top of the moir\'e site in the bottom layer. If we fix $n_b=1$ and increase $n_t$, the doped electron in the top layer will be repelled to B,C sublattice and form a honeycomb lattice. However, if we increase $n_t$ while keeping $n_b+n_t=1$, the doped electron in the top layer will be bound to the hole in the bottom layer at the moir\'e site $A$. In this case, states at $B$ and $C$ sublattice are penalized, but they can assist super-exchange through virtual hopping.}
\label{fig:lattice_moireless}
\end{figure}

Next we offer a simple toy model to show that we still expect $J_t>0$ and $J_p>0$. As argued before, we only need to keep one coarse graining orbital on top of the moir\'e site for the top layer. This is the A sublattice in Fig.~\ref{fig:lattice_moireless}. But there are some other orbitals which may be important in the virtual process to generate the $J_p$ and $J_t$ terms. For example, it is also important to keep the B,C sublattice in Fig.~\ref{fig:lattice_moireless}.  With these three orbitals, we can write down an effective model for the moir\'e-monolayer system:

\begin{align}
H&=-t_b \sum_{ij}c^\dagger_{i;b\sigma} c_{j;b\sigma}+\frac{U_b}{2} \sum_i n_{i;b}^2\notag\\
&-t_t \sum_{\tilde i \tilde j} c^\dagger_{\tilde i;t\sigma}c_{\tilde j;t\sigma}+\frac{U_t}{2}\sum_{\tilde i} n_{\tilde i;t}^2+V_t \sum_{\langle \tilde i \tilde j \rangle}n_{\tilde i;t}n_{\tilde j;t}\notag\\
&+U' \sum_{i\in A} n_{i;t}n_{i;b}+V'\sum_{\langle i \tilde j \rangle}n_{i;b}n_{\tilde j;t}
\label{eq:lattice_model_moireless}
\end{align}
where $i,j \in A$ live on the bottom layer in the big lattice formed by A sites. $\tilde i, \tilde j$ live on the small lattice formed by A,B,C sites together for the top layer.  Note $V_t$ is between the nearest neighbor of the smaller triangular lattice denoted by the dashed line. $V'$ is the inter-layer  repulsion between the moir\'e site $i$ in the bottom layer with the nearest neighbor B or C site in the top layer. $U'$ is the on-site inter-layer repulsion. Generically we expect $U'>V'$, $V_t>V'$. In the above we ignored the intra-layer and inter-layer interaction between A sites.

Let us fix our density to be $n_t=x,n_b=1-x$. In the low energy, we only need to keep the states with one electron in either layer at site $A$.  We can not occupy both layers of the same A site, which is forbidden by $U'$. The state at B and C is also penalized. For example, if we move one electron in the top layer from site A to site B, there will be energy cost of $2V'$, assuming nearest neighbor A sites of B are occupied in the bottom layer (otherwise the energy cost is $2V_t$ or $V_t+V'$, which is even larger).  With this constraint, the low energy model must be a spin-layer four-flavor model only on A site. 

Next we try to derive the parameters in Eq.~\ref{eq:spin_layer_model}. First, the $J_b$ term is not changed and should still be $J_b=\frac{4 t_b^2}{U_b}$.  For $J_t$, we now need to consider a four-step process through the intermediate B or C sites.  This gives a term $\frac{4 t_t^4}{U_t (V_t+V')^2}(4 \vec{S}_t(i) \cdot \vec{S}_t(j)-n_t(i)n_t(j))+\frac{2 t_t^4}{ V' (V_t+V')^2} (4 \vec{S}_t(i)\cdot \vec{S}_t(j)+n_t(i)n_t(j))$.  In total we get: $J_t=\frac{4 t_t^4}{U_t (V_t+V')^2}+\frac{2 t_t^4}{ V' (V_t+V')^2}$.  The $n_t(i)n_t(j)$ term will modify the $J_{pz}$ term. The most important term is the hopping term $J_p$.  This is generated by a three-step process:  electron in the top layer hops from site i to site j through an intermediate B or C site;  the electron in the bottom layer hops from site j to site i. We find that $J_p=\frac{4 t_t^2 t_b}{U' V'}+\frac{t_t^2 t_b}{V'^2}$.  As for the repulsion term $J_{pz}$, there will be contributions from the super-exchange. But the dominant term is still from the anisotropy part of the nearest neighbor repulsion $\delta V$, which is ignored in Eq.~\ref{eq:lattice_model_moireless}. Thus we can just use $J_{pz}=2\delta V$.  For the moir\'e-monolayer system, the asymmetry of the two layers are larger. Hence we expect a larger $J_{pz}$ term compared to the moir\'e bilayer case. However, when the exciton density is low, the repulsion term should not be too important.

Based on the toy model, we can show that the low energy Hamiltonian is still in the form of Eq.~\ref{eq:spin_layer_model} with $J_p>0, J_t>0, J_{pz}>0$. In this simple model we only keep three states for the top layer. Strictly speaking, there are $O((\frac{a_M}{a})^2)\sim O(1000)$ orbitals within each moir\'e unit cell. Therefore we should keep more orbitals at the sublattice A,B,C, but they will have higher on-site energy $\Delta$ compared to those we are keeping now.  We can also get super-exchange coupling by hopping to these high energy orbitals in the virtual process, but this does not change the physics qualitatively. 

One additional implication of the toy model in Eq.~\ref{eq:lattice_model_moireless} is the following: if we fix $n_b=1$ at each site and only dope the top layer, then the electron in the top layer will hop in the honeycomb lattice formed by B and C sublattice, while the state at site A is pushed to higher energy by the $U'$ term. This may be one interesting way to realize a honeycomb lattice Hubbard model, which we leave to future analysis.

\section{Dictionary between exciton and layer-pseudospin language\label{sec:dictionary}}

Conventionally the physics in coulomb coupled double layer is thought to be associated with the inter-layer excitons. One may wonder why we use the notation ``layer pseudo-spin'' instead of exciton in this paper.     In this section we want to show that the layer pseudo-spin is mathematically equivalent to inter-layer exciton in the spinless case and is more convenient in the spinful case.

\subsection{Spinless case}

We consider the simple case with spinless electron first. With a strong Zeeman field, the lattice Hubbard model in Eq.~\ref{eq:Hubbard_model_moire_bilayer} reduces to

\begin{align}
H&=- \sum_{a=t,b} t_a \sum_{ij} c^\dagger_{i;a}c_{j;a}-D\sum_i P_z(i) +U' \sum_i n_{i;t}n_{i;b}\notag\\
&+V \sum_{\langle ij \rangle} n_i n_j +\delta V \sum_{\langle ij \rangle}P_z(i) P_z(j) +\frac{V_t-V_b}{2} \sum_{ ij } n_i P_z(j)
\end{align}
where $V=\frac{V_t+V_b}{2}+V'$ and $\delta V=V_t+V_b-2V'$. $P_z(i)=\frac{1}{2}(n_t(i)-n_b(i))$.  In the above we absorbed term linear to $P_z$ into the D term.

If we set $t_a=t_b=t$, $V_t=V_b=V'$, we can see that the above model just reduces to the standard spin 1/2 Hubbard model if we interpret $a=t,b$ as the spin index. The $D$ term is basically a Zeeman field in spin language.  If we fix our density to be $n_t=x,n_b=1-x$, then the system is in an insulating state when $U'/t>>1$. The insulator can be called inter-layer excitonic insulator because electron and hole in the two layers are bound. Alternatively in the Hubbard model language the insulator is clearly a Mott insulator, with a possible partial spin polarization $P_x=x-\frac{1}{2}$. If we decrease $U'/t$, there can be a metal-insulator transition. In exciton language, the metal side can be viewed as an electron-hole liquid and the Mott transition is associated with the exciton dissociation.  The situation is a little different if $t_t>t_b$. In this case the Mott transition may happen if increasing $x$. Nevertheless, the insulating side is still captured by a spin $1/2$ model, though with anisotropy term breaking the SU(2) spin rotation down to $U(1)$.

Deep inside the insulating phase, the spin model Eq.~\ref{eq:spin_layer_model} reduces to

\begin{equation}
	H_S=J_p \sum_{\langle ij \rangle}\big(P_x(i)P_x(j)+P_y(i)P_y(j)\big)+J_z\sum_{\langle ij \rangle} P_z(i)P_z(j)
	\label{eq:spin_less_model}
\end{equation}
where $J_z=\frac{1}{2}J_{pz}+\frac{1}{4}(J_t+J_b)$. Note in the above we rewrite every operator in the form $P_\mu S_\nu$ and then replace $S_z=\frac{1}{2}, S_0=1, S_{x,y}=0$. We have also ignored the constant term and the term linear to $P_z$.

Eq.~\ref{eq:spin_less_model} is just a  spin $1/2$ model with only XY symmetry if we view $\vec P$ as a spin $1/2$ degree of freedom. It is well-known that such a model can be mapped to a hard-core bosonic model:

\begin{equation}
	H_b=t_b \sum_{\langle ij \rangle}(b^\dagger_i b_j+h.c.)+V\sum_{\langle ij \rangle}n^b_i n^b_j
	\label{eq:boson_spinless}
\end{equation}
where $t_b=\frac{1}{2}J_p$ and $V=J_z$. Again we ignored the constant term and the term linear to $n=P_z+\frac{1}{2}$.  We have used the mapping $P^\dagger_i\rightarrow b_i^\dagger, P^-_i \rightarrow b_i, \frac{1}{2}+P_z \rightarrow n^b_i$.  In the hard-core boson language, we label $\ket{P_z=-\frac{1}{2}}$ as $\ket{0}$ with $n^b=0$. We label $\ket{P_z=\frac{1}{2}}$ as $\ket{1}$ with $n^b=1$.  We list a dictionary between the layer pseudo-spin and the exciton language in Table.~\ref{table:dictionary_spin_exciton}. We also list several possible spin phases as examples. In the spinless case one can use the exciton language and write down the low energy Hamiltonian as a boson model, which is mathematically equivalent to certain spin $1/2$ model.  On triangular lattice, the ground state is known to be in a $120^\circ$ order, or an exciton supersolid in the exciton language.

\begin{table}[ht]
\centering
\begin{tabular}{|c|c|c|}
\hline
 & pseudo-spin language & exciton language\\ \hline
 $c^\dagger_t\ket{0}$ & $\ket{P_z=\frac{1}{2}}$ & $b^\dagger\ket{0}$ \\ \hline
 $c^\dagger_b\ket{0}$ & $\ket{P_z=-\frac{1}{2}}$ & $\ket{0}$ \\ \hline
 $\frac{1}{2}(n_t-n_b)$ & $P_z$ & $n_b-\frac{1}{2}$ \\ \hline
 $c^\dagger_b c_t$ & $P^-$ & $b$ \\ \hline
 $c^\dagger_t c_b$ & $P^\dagger$ & $b^\dagger$ \\ \hline
 Phase I &  XY FM & exciton condensation \\ \hline
 Phase II &  120$^\circ$ Neel order & exciton supersolid \\ \hline
 Phase III & valence bond solid & crystallized insulator \\ \hline
 Phase IV & chiral spin liquid & exciton FQHE \\ \hline
 Phase V & spinon Fermi surface & exciton metal \\ \hline
 phase VI & Fermi liquid & electron hole liquid \\ \hline
 Counter-flow & spin conductivity & exciton conductivity \\ \hline
 \end{tabular}
 \caption{A dictionary between the layer pseudo-spin language and the exciton language in the spinless case. $b^\dagger$ creates exciton starting from the bottom layer polarized limit. XY FM denotes the XY ferromagnetism with $\langle S_x \rangle \neq 0$. FQHE denotes fractional quantum Hall state. The 120$^\circ$ order (supersolid phase) is known to be the ground state of Eq.~\ref{eq:spin_less_model} (Eq.~\ref{eq:boson_spinless}). But we also list some other phases which may be stabilized by additional ring exchange terms. Counter-flow measures the pseudo-spin transport or equivalently the exciton transport, which can distinguish these different phases unambiguously.}
 \label{table:dictionary_spin_exciton}
\end{table}

\subsection{Spinful case}
Next we discuss the case without any magnetic field and the real spin $\vec S$ must be included in the low energy dynamics.  In this case, one can not capture the full dynamics with a simple boson Hubbard model. We are forced to keep the spin operator $\vec S$ in the Hamiltonian. We can still use the mapping $P^\dagger \rightarrow b^\dagger$ to write Eq.~\ref{eq:spin_layer_model} as a boson-spin model:

\begin{align}
	H_S&=J_t \sum_{\langle ij \rangle}\vec{S}_t(i) \cdot \vec{S}_t(j)+J_b \sum_{\langle ij \rangle}\vec{S}_b(i) \cdot \vec{S}_b(j)\notag \\ 
    &~~~+\frac{1}{2}J_{pz}\sum_{\langle ij \rangle} n^b_in^b_j \notag\\
	&~~~+\frac{1}{4} J_p \sum_{\langle ij \rangle}\big(b^\dagger_i b_j+b^\dagger_j b_i\big)\left(4\vec{S}(i) \cdot \vec{S}(j)+S_0(i)S_0(j)\right)
	\label{eq:spin_boson_model}
\end{align}

The above Hamiltonian is equivalent to Eq.~\ref{eq:spin_layer_model}. However, the connection to the SU(4) symmetric limit is not explicit. Also the model looks complicated and is not a simple boson Hubbard model anymore. Besides, now the inter-layer exciton creation operator $c^\dagger_{b \sigma}c_{t \sigma'}$ must carry spin index $\sigma,\sigma'=\uparrow,\downarrow$. One can see that there are several different flavors of exciton. Both $b^\dagger$ and $b^\dagger \vec S$ are physical exciton creation operators. Therefore, exciton is not a very convenient representation in the spinful case. In the following we will use the $\vec P$ instead and take the anisotropic SU(4) model in Eq.~\ref{eq:spin_layer_model}. The ``exciton-spin'' physics is fully captured in this model.

\section{Exciton fractionalization: bosonic and fermionic parton theory \label{sec:boson_fermion_parton_construction}}

\begin{figure}
\centering
\includegraphics[width=0.5\textwidth]{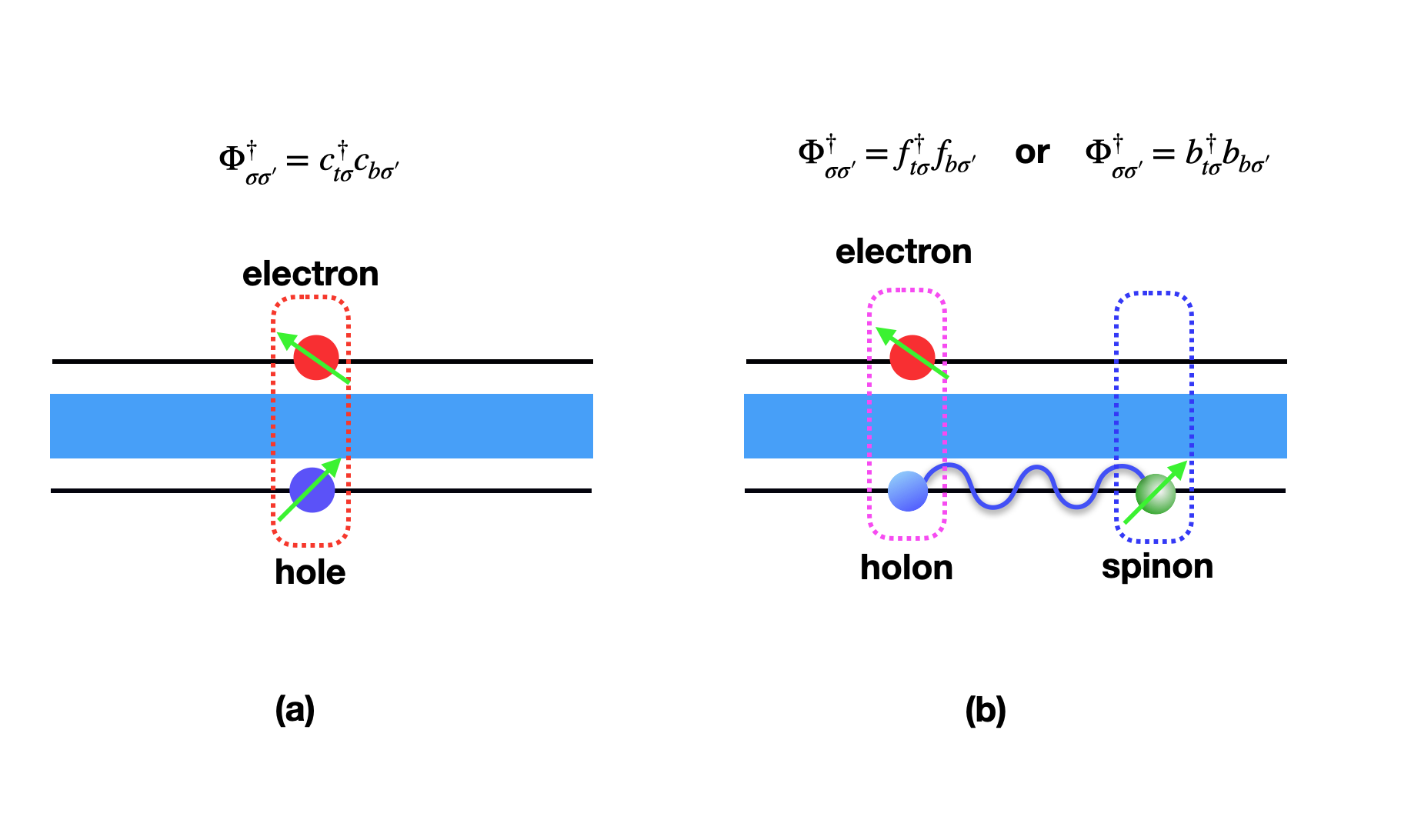}
\caption{(a)A conventional exciton formed by a $S_t=\frac{1}{2}$ electron in the top layer and a  $S_b=\frac{1}{2}$ hole in the bottom layer. The exciton carries the dipole charge $P_z=1$ and $S_t=S_b=\frac{1}{2}$. It carries two spin indexes: $\Phi^\dagger_{\sigma \sigma'}=c^\dagger_{t\sigma}c_{b\sigma'}$. In the model of Eq.~\ref{eq:spin_layer_model},the exciton creation operator $\Phi^\dagger_{\sigma \sigma'}$ can be identified as $P^\dagger$ and $P^\dagger \vec{S}$. (b) Fractionalization of exciton. The hole operator in the bottom layer is divided into a holon and a spinon, which interact with each other through a gauge field (blue wave line). Then an electron in the top layer and a holon bind into a fractionalized exciton (surrounded by the pink dashed square), which carries $P_z=\frac{1}{2}$, $S_t=\frac{1}{2}$ and $S_b=0$. Meanwhile the spinon (surrounded by the dashed blue square) carries $P_z=-\frac{1}{2}$, $S_t=0$ and $S_b=\frac{1}{2}$. The dashed pink square ($f^\dagger_{t\sigma}$ or $b^\dagger_{t\sigma}$) and the dashed blue square ($f_{b\sigma}$ or $b_{b\sigma}$) can be viewed as one half of the conventional exciton and they can be either fermionic or bosonic. Both the conventional exciton condensation phase and more exotic phase with neutral Fermi surface can be conveniently described with these fractionalized degree of freedom.}
\label{fig:exciton_fractionalization}
\end{figure}

Usually we think the inter-layer exciton is bosonic because it is formed by electron hole pair. In our case, the exciton creation operator $\Phi^\dagger_{\sigma \sigma'}=c^\dagger_{t\sigma}c_{b\sigma'}$ carries two spin indexes and there are four of them forming a vector representation of $SO(4) \cong (SU(2)\times SU(2))/Z_2$. They can be identified as the $P^\dagger$ and $P^\dagger \vec{S}$ operators in Eq.~\ref{eq:spin_layer_model}. A phase with single exciton condensation corresponds to $\langle \Phi^\dagger_{\sigma \sigma'} \rangle \neq 0$, which needs to also break the spin rotation symmetry of both layers. This means that single exciton condensation needs to be accompanied by magnetic orders in both layers.  Description of the various phases directly using these exciton operators $\Phi^\dagger_{\sigma \sigma'}$ is too complicated.  It is more convenient to use fractionalized operators which correspond to one half of the conventional exciton operator and carry a spin $1/2$ in one of the two layers.

 The fractionalization of the exciton can be understood in the following way. When we dope holes into the Mott insulator of the bottom layer, it is known that a useful picture is through spin-charge separation\cite{lee2006doping}: holon carries charge and neutral spinon carries a spin $1/2$.  Then we can imagine an ``exciton'' formed by the electron in the top layer with the holon in the bottom layer. This is the object inside the pink square in Fig.~\ref{fig:exciton_fractionalization}(b). The holon is a fractionalized degree of freedom, and its statistics can be either bosonic or fermionic depending on the exact spin state of the Mott insulator. As a result, this ``exciton'' formed by electron-holon pair can be either bosonic or fermionic.  As shown in Fig.~\ref{fig:exciton_fractionalization}(b), we can think that the conventional exciton in Fig.~\ref{fig:exciton_fractionalization}(a) is divided into two parts: the pink square represents the exciton-holon pair and the blue square represents the spinon in the bottom layer. There are two possibilities: (I) If the spinon in the bottom layer is fermionic and labeled as $f_{i;b\sigma}$, then the hole operator is $c_{i;b\sigma}=\varphi_i f_{i;b\sigma}$ with $\varphi_i$ as a bosonic holon. In this case, the pink square is labeled as $f^\dagger_{i;t\sigma}=c^\dagger_{i;t\sigma} \varphi_i$.  The conventional exciton operator can be written as $\Phi^\dagger_{i;\sigma \sigma'}=f^\dagger_{i;t\sigma}f_{i;b\sigma'}$. $f^\dagger_{i;\sigma}=c^\dagger_{i;t\sigma}\varphi_i$ and $f_{i;b\sigma'}$ correspond to the pink and blue square in Fig.~\ref{fig:exciton_fractionalization}(b).  (II) The bottom layer hole operator is written as $c_{i;b\sigma}=h_i b_{i;\sigma}$. $h_i$ is a fermionic holon and $b_{i;\sigma}$ is a bosonic spinon. In this case we have $\Phi^\dagger_{i;\sigma \sigma'}=b^\dagger_{i;t\sigma}b_{i;b\sigma'}$. $b^\dagger_{i;\sigma}=c^\dagger_{i;t\sigma}h_i$ and $b_{i;b\sigma'}$ correspond to the pink and blue square in Fig.~\ref{fig:exciton_fractionalization}(b).

The parton operators $f_{i;a\sigma}$ are just a four-flavor generalization of the familiar Abrikosov fermion\cite{abrikosov1965electron} parton theory of spin $1/2$ system. Similarly $b_{i;a\sigma}$ is the Schwinger boson parton\cite{arovas1988functional,read1991large}.  We can just view the system in a Mott insulator of a four-flavor Hubbard model ad defined in Eq.~\ref{eq:Hubbard_model_moire_bilayer}. Inside the Mott insulator, we can just restrict ourselves to the four-flavor spin model defined in Eq.~\ref{eq:spin_layer_model}.  We can then simply use either a four-flavor Schwinger boson  or four-flavor Abrikosov fermion to represent the anisotropic SU(4) spin. 

In the Schwinger boson theory, the neutral particle-hole operators are represented as:

\begin{align}
\vec{S}_{i;a}&=\frac{1}{2} \sum_{\sigma,\sigma'}b^\dagger_{i;a \sigma} \vec{\sigma}_{\sigma \sigma'}b_{i;a\sigma'}\notag\\
P^\dagger_i&= \sum_\sigma b^\dagger_{i;t\sigma}b_{i;b\sigma} \notag\\
P^{-}_i&=\sum_\sigma b^\dagger_{i;b\sigma}b_{i;t\sigma}\notag\\
P_{i;z}&=\frac{1}{2}\sum_{\sigma}(b^\dagger_{i;t\sigma}b_{i;t\sigma}-b^\dagger_{i;b\sigma}b_{i;b\sigma})\notag\\
P^\dagger_i \vec{S}_i&= \frac{1}{2}\sum_{\sigma \sigma'}b^\dagger_{i;t\sigma}\vec{\sigma}_{\sigma \sigma'}b_{i;b\sigma'}\notag\\ 
P^{-}_i \vec{S}_i&= \frac{1}{2}\sum_{\sigma \sigma'}b^\dagger_{i;b\sigma}\vec{\sigma}_{\sigma \sigma'}b_{i;t\sigma'}
\label{eq:Schwinger_boson}
\end{align}
with the constraint $\sum_{a=t,b}\sum_{\sigma=\uparrow,\downarrow} b^\dagger_{i;a\sigma}b_{i;a\sigma}=1$.  $b_{i;a\sigma}$ is a four-flavor boson labeled by layer index $a=t,b$ and spin index $\sigma=\uparrow, \downarrow$.

In the  Abrikosov fermion theory, the particle-hole operators are written as:

\begin{align}
\vec{S}_{i;a}&=\frac{1}{2} \sum_{\sigma,\sigma'}f^\dagger_{i;a \sigma} \vec{\sigma}_{\sigma \sigma'}f_{i;a\sigma'}\notag\\
P^\dagger_i&= \sum_\sigma f^\dagger_{i;t\sigma}f_{i;b\sigma} \notag\\
P^{-}_i&=\sum_\sigma f^\dagger_{i;b\sigma}f_{i;t\sigma}\notag\\
P_{i;z}&=\frac{1}{2}\sum_{\sigma}(f^\dagger_{i;t\sigma}f_{i;t\sigma}-f^\dagger_{i;b\sigma}f_{i;b\sigma})\notag \\ 
P^\dagger_i \vec{S}_i&= \frac{1}{2}\sum_{\sigma \sigma'}f^\dagger_{i;t\sigma}\vec{\sigma}_{\sigma \sigma'}f_{i;b\sigma'}\notag\\ 
P^{-}_i \vec{S}_i&= \frac{1}{2}\sum_{\sigma \sigma'}f^\dagger_{i;b\sigma}\vec{\sigma}_{\sigma \sigma'}f_{i;t\sigma'}
\label{eq:Schwinger_fermion}
\end{align}
with the constraint $\sum_{a=t,b}\sum_{\sigma=\uparrow,\downarrow} f^\dagger_{i;a\sigma}f_{i;a\sigma}=1$.  $f_{i;a\sigma}$ is a four-flavor boson labeled by layer index $a=t,b$ and spin index $\sigma=\uparrow, \downarrow$.

In the above we only focus on the neutral spin and layer pseudospin sector. If we are also interested in the dynamics of the charge, we can simply extend the construction to a parton construction of the four-flavor electron: (I) $c_{i;a\sigma}=h_i b_{i;a\sigma}$ or (II) $c_{i;a\sigma}=\varphi_i f_{i;a\sigma}$. The statistics and physical quantum numbers of the parton operators are listed in Table.~\ref{table:spinon_quantum number}. In either construction, there is an emergent $U(1)$ gauge field $\mathbf a$ corresponding to gauge transformation:  (I)$h_i \rightarrow h_i e^{-i \alpha_i}, b_{i;a\sigma}\rightarrow b_{i;a\sigma}e^{i\alpha_I}$ or (II)$\varphi_i \rightarrow \varphi_i e^{-i \alpha_i}, f_{i;a\sigma}\rightarrow f_{i;a\sigma}e^{i\alpha_I}$.

\begin{table}[ht]
\centering
\begin{tabular}{|c|c|c|c|c|c|c|}
\hline
Operator &  Statistics & $Q$ & $ P_z $ & $\vec{S}_t$ & $\vec{S}_b$ &$Q_g$ \\ \hline
\hline
$c^\dagger_{t\sigma}$ &fermion  & $1$ & $\frac{1}{2}$ & $\frac{1}{2}$ & $0$  &$0$\\  \hline
$c^\dagger_{b\sigma}$ &fermion  & $1$ & $-\frac{1}{2}$ & $0$ & $\frac{1}{2}$ &$0$ \\  \hline 
\hline
$h$ &fermion & $-1$ & $0$ &$0$ &$0$ &$1$ \\ \hline
$b^\dagger_{b\sigma}$  &boson&   $0$ & $-\frac{1}{2}$ & $0$ & $\frac{1}{2}$ &$1$ \\ \hline
$b^\dagger_{t\sigma}$ &boson&  $0$ & $\frac{1}{2}$ & $\frac{1}{2}$ & $0$ &$1$ \\  \hline
$c^\dagger_{t\sigma} h$ &boson&  $0$ & $\frac{1}{2}$ & $\frac{1}{2}$ & $0$ &$1$ \\  \hline
\hline
$\varphi$ &boson & $-1$ & $0$ &$0$ &$0$ &$1$ \\ \hline
$f^\dagger_{b\sigma}$  &fermion&   $0$ & $-\frac{1}{2}$ & $0$ & $\frac{1}{2}$ &$1$ \\ \hline
$f^\dagger_{t\sigma}$ &fermion&  $0$ & $\frac{1}{2}$ & $\frac{1}{2}$ & $0$ &$1$ \\  \hline
$c^\dagger_{t\sigma} \varphi$ &fermion&  $0$ & $\frac{1}{2}$ & $\frac{1}{2}$ & $0$ &$1$ \\  
\hline
\end{tabular}
\caption{Table: quantum numbers of various operators. $c^\dagger_{a\sigma}$ is electron operator. There are two different parton constructions: (I) $c_{i;a\sigma}=\varphi_i f_{i;a\sigma}$; (II) $c_{i;a\sigma}=h_i b_{i;a\sigma}$.  In both parton constructions, there is an emergent U(1) gauge field.  $Q$ is the physical charge. $P_z$ is the z component of the layer pseudospin. $\vec{S}_t$ and $\vec{S}_b$ are spin of the top and bottom layers separately. $Q_g$ is the gauge charge. $Q_g=0$ is the condition for gauge invariant operators. We note that the spinon $b^\dagger_{t\sigma}$ or $f^\dagger_{t\sigma}$ can be identified as electron-holon pair $c^\dagger_{t\sigma} h$ or $c^\dagger_{t\sigma} \varphi$. We note that the assignment of the dipole charge $P_z$ can be adjusted, as long as the total $P_z$ quantum number for $f^\dagger_{t;\sigma}$ and $f_{b;\sigma}$ is $1$. Here we choose an assignment of $1/2$+$1/2$, but one can also use $1$+$0$, which simply corresponds to a redefinition of the internal gauge field $a_\mu$.}
 \label{table:spinon_quantum number}
\end{table}

\textbf{Ioffe-Larkin rule for the counterflow resistivity} Before a detailed application of either bosonic or fermionic parton theory, we first provide a formula to calculate the counterflow resistivity. Because the two layers have separate conserved charges, there are two U(1) probing gauge fields $A^t_\mu$ and $A^b_\mu$, which couple to the currents $J^t_\mu$ and $J^b_\mu$ in the two layers separately in the form $-J^a_\mu A^a_\mu$ with $a=t,b$. We can do a linear combination $A^c_\mu=\frac{1}{2}(A^t_\mu+A^b_\mu)$ and $A^s_\mu=A^t_\mu-A^b_\mu$. Correspondingly we can define the charge current $J^c_\mu=J^t_\mu+J^b_\mu$ and $J^s_\mu=\frac{1}{2}(J^t_\mu-J^b_\mu)$ so that the minimal coupling is in the form $-J^c_\mu A^c_\mu-J^s_\mu A^s_\mu$. Physically it is easier to see that the current $J^c_\mu$ is the current associated with the total charge $Q=n^t+n^b$, while $J^s_\mu$ is the current associated with the layer polarization $P_z=\frac{1}{2}(n_t-n_b)$. In our model, we assume that there is a charge gap. Hence we always have $\vec J^c=0$. The only possible non-zero current below the Mott gap is $\vec J^s$, which corresponds to the configuration $\vec J^t=-\vec J^b$. Because the currents in the two layers are opposite, this current is called counter-flow current\cite{liu2017quantum}.

Let us use the fermionic parton as an example and derive a Ioffe-Larkin rule for the counterflow resistivity defined as $\rho_{s;xx}=\sigma_{s;xx}^{-1}$, with $\sigma_{s;xx}=\frac{J^s_x}{E^s_x}$. $\vec E^s=-\vec \partial A^s_0+\frac{\partial \vec A^s}{\partial t}$ is the counter-flow electric field. Physically $\vec E^s=\vec E^t-\vec E^b$, where $\vec E^a$ is the electric field in the layer $a$.  Suppose that we have fermionic partons $f_{t;\sigma}$ and $f_{b;\sigma}$ in both layers. There is a U(1) gauge field $\vec a$. From Table.~\ref{table:spinon_quantum number} we know $f^\dagger_{t;\sigma}$ couples to $a_\mu+\frac{1}{2} A^s_\mu$, while $f^\dagger_{b;\sigma}$ couples to $a_\mu-\frac{1}{2}A^s_\mu$. Then it is easy to write down a linear response formula for each layer:

\begin{align}
J^t_x&=\sigma_{t;xx} (\frac{1}{2} E^s_x+e_x) \notag\\
J^b_x&=\sigma_{b;xx}(-\frac{1}{2} E^s_x+e_x)
\end{align}
where $\vec e$ is the electric field associated with the internal gauge field $a_\mu$.  $\sigma_{a;xx}$ is the conductivity of the fermion $f_{a}$ in each layer separately.

Because the charge current $\vec J^c=\vec J^t+\vec J^b=0$ below the charge gap, we have $J^t_x=J^s_x$ and $J^b_x=-J^s_x$. Then we get

\begin{align}
J^s_x&=\sigma_{t;xx} (\frac{1}{2} E^s_x+e_x) \notag\\
-J^s_x&=\sigma_{b;xx}(-\frac{1}{2} E^s_x+e_x)
\end{align}

From the above two equations, it is easy to obtain the Ioffe-Larkin rule:

\begin{equation}
    \rho_{s;xx}=\rho_{t;xx}+\rho_{b;xx}
    \label{eq:ioffe_larkin_rule_resistivity}
\end{equation}
where $\rho_{a;xx}=\sigma_{a;xx}^{-1}$ is the resistivity of the parton at the layer $a$.  The above formula basically tells us that in a counter-flow transport, the partons from the two layers form a series circuit.

The above formula applies to both fermionic and bosonic partons. One immediate consequence of the formula is that the counter-flow transport is insulating as long as parton in one of the layers is insulating. In the $x=0$ limit, we have $\rho_t=\infty$ because $n_t=0$. In this limit there can not be any counter-flow current. Once $x>0$, both layers may have finite or even zero $\rho_a$, then the counter-flow resistivity can be finite or zero.

Counter-flow transport measures the response to $\vec A^s$. A similar formula also exists for the compressibility $\kappa_s=\frac{J^s_0}{A^s_0}$, which is essentially the inter-layer polarizability $\frac{\partial P_z}{\partial D}$. Following the same argument as above, the formula for $\kappa_s$ is:

\begin{equation}
    \kappa_s^{-1}=\kappa_t^{-1}+\kappa_b^{-1}
    \label{eq:ioffe_larkin_rule_compressibility}
\end{equation}
where $\kappa_a$ is the compressibility of the parton at layer $a$.  To have a finite $\kappa_s$, the partons in both layers must be in a compressible phase.

In the following we will apply the Schwinger boson theory or the Abrikosov fermion theory to analyze the case where the layer polarized Mott insulator is magnetically ordered or in a spin liquid phase with spinon Fermi surface respectively. We need to emphasize that the usage of these fractionalized parton operators does not necessarily leads to an exotic phase. Actually even for a conventional symmetry breaking phase with magnetic order and simple exciton condensation, it is convenient to start from the Schwinger boson parton theory, as will be demonstrated in Sec.~\ref{sec:bosonic_exciton_superfluids}.

\section{Inter-layer superfluids with bosonic excitons\label{sec:bosonic_exciton_superfluids}}

Deep inside the Mott insulator with a large charge gap, we can restrict ourselves to the spin-exciton model in Eq.~\ref{eq:spin_layer_model} without ring-exchange terms.  We can also add next-nearest neighbor spin-spin coupling $J'_b$ to tune the spin phase in the layer polarized limit. In the limit $J'_b/J_b<<1$, we know the spin phase of the bottom layer is in the $120^\circ$ ordered phase in the $x=0$ limit.  We will see that doping excitons can lead to various  different inter-layer superfluid phases with either magnetic order or spin gap. 

To capture the magnetic order, it is convenient to use the four-flavor Schwinger boson representation in Eq.~\ref{eq:Schwinger_boson}. The spin model (Eq.~\ref{eq:spin_layer_model}) can be rewritten as:

\begin{align}
H&=\frac{1}{4}J_t \sum_{\langle ij \rangle} (b^\dagger_{i;t \sigma_1} \vec{\sigma}_{\sigma_1 \sigma'_1} b_{i;t \sigma'_1})\cdot (b^\dagger_{j;t \sigma_2} \vec{\sigma}_{\sigma_2 \sigma'_2} b_{j;t \sigma'_2})\notag\\
&+\frac{1}{4}J_b \sum_{\langle ij \rangle} (b^\dagger_{i;b \sigma_1} \vec{\sigma}_{\sigma_1 \sigma'_1} b_{i;b \sigma'_1})\cdot (b^\dagger_{j;b \sigma_2} \vec{\sigma}_{\sigma_2 \sigma'_2} b_{j;b \sigma'_2}) \notag\\
&+\frac{1}{2} J_p \sum_{\langle ij \rangle} (b^\dagger_{i;t \sigma} b_{i;b \sigma'} b^\dagger_{j;b \sigma'} b_{j;t \sigma}+ b^\dagger_{i;b \sigma} b_{i;t\sigma'} b^\dagger_{j;t\sigma'} b_{j;b \sigma})\notag\\
&-\frac{1}{4} J_{pz} \sum_{\langle ij \rangle} (n_{i;t} n_{j;b}+n_{i;b}n_{j;t})\notag\\
&-D \sum_i (n_{i;t}-n_{i;b})
\label{eq:boson_model}
\end{align}
where in the last term we added a constant term $-\frac{1}{8} J_{pz} \sum_{\langle ij \rangle}  (n_{i;t}+n_{i;b})(n_{j;t}+n_{j;b})$ by using the constraint $n_{i;t}+n_{i;b}=1$.  The $D$ term is added if we use grand canonical ensemble. Alternatively we can also fix the density to be $n_t=x, n_b=1-x$ and then the $D$ term is not needed.

\subsection{Classification of possible superfluid phases}

\onecolumngrid

\begin{table}[H]
\centering
\begin{tabular}{|c|c|c|}
\hline
 &  $\langle b_{b;\sigma} \rangle\neq 0$ & $\langle b_{b;\sigma}\rangle=0$, $\langle \epsilon_{\sigma \sigma'}b_{b;\sigma}b_{b;\sigma'} \rangle\neq 0$ \\ \hline
$\langle b_{t;\sigma} \rangle\neq 0$ & I: $\langle P^\dagger \rangle \neq 0$, $\Delta_t=0$, $\Delta_b=0$ & III: $\langle P^\dagger\rangle=0$, $\langle P^\dagger P^\dagger \rangle \neq 0$, $\Delta_t=0$, $\Delta_b>0$ \\ \hline
$\langle b_{t;\sigma}\rangle=0$, $\langle \epsilon_{\sigma \sigma'}b_{t;\sigma}b_{t;\sigma'} \rangle\neq 0$  & II: $\langle P^\dagger\rangle=0$, $\langle P^\dagger P^\dagger \rangle \neq 0$, $\Delta_t>0$, $\Delta_b=0$ & IV: $\langle P^\dagger\rangle=0$, $\langle P^\dagger P^\dagger \rangle \neq 0$, $\Delta_t>0$, $\Delta_b>0$ \\ \hline
\hline
\end{tabular}
\caption{Four different types of phases described by the Schwinger boson mean field theories. The Schwinger boson at each layer can be either single condensed or pair condensed. If it is pair condensed, then there is a spin gap $\Delta_a$ for the corresponding layer $a=t,b$. As long as one of $\Delta_t,\Delta_b$ is finite, the exciton condensation carries charge $2$ under $P_z$ with only $\langle P^\dagger P^\dagger \rangle \neq 0$. In this case the condensed object is formed by two electrons and two holes. When both $\Delta_t>0$ and $\Delta_b>0$, the paired exciton condensation coexists with a $Z_2$ topological order. }
 \label{table:schwinger_boson_phases}
\end{table}

\twocolumngrid

We will show that the model in Eq.~\ref{eq:boson_model} hosts a rich phase diagram depending on the parameters.  Before going into details, we first list possible phases and their physical properties.  We have three quantum numbers: $P_z$, $S_t$, $S_b$.  From Eq.~\ref{eq:boson_model}, we can obtain a mean field Hamiltonian in the form $H_M=H_t+H_b$, where $H_t,H_b$ are quadratic Hamiltonians for $b_{t;\sigma}$ and $b_{b;\sigma}$ separately. Because of the conservation of $P_z$, $\vec{S}_t$ and $\vec{S}_b$, terms like $b^\dagger_{t;\sigma}b_{b;\sigma}$ and $b_{t;\sigma}b_{b;\sigma'}$ are forbidden. Therefore $b_{t;\sigma}$ and $b_{b;\sigma}$ remain decoupled in the mean field level.

\begin{figure}
\centering
\includegraphics[width=0.5\textwidth]{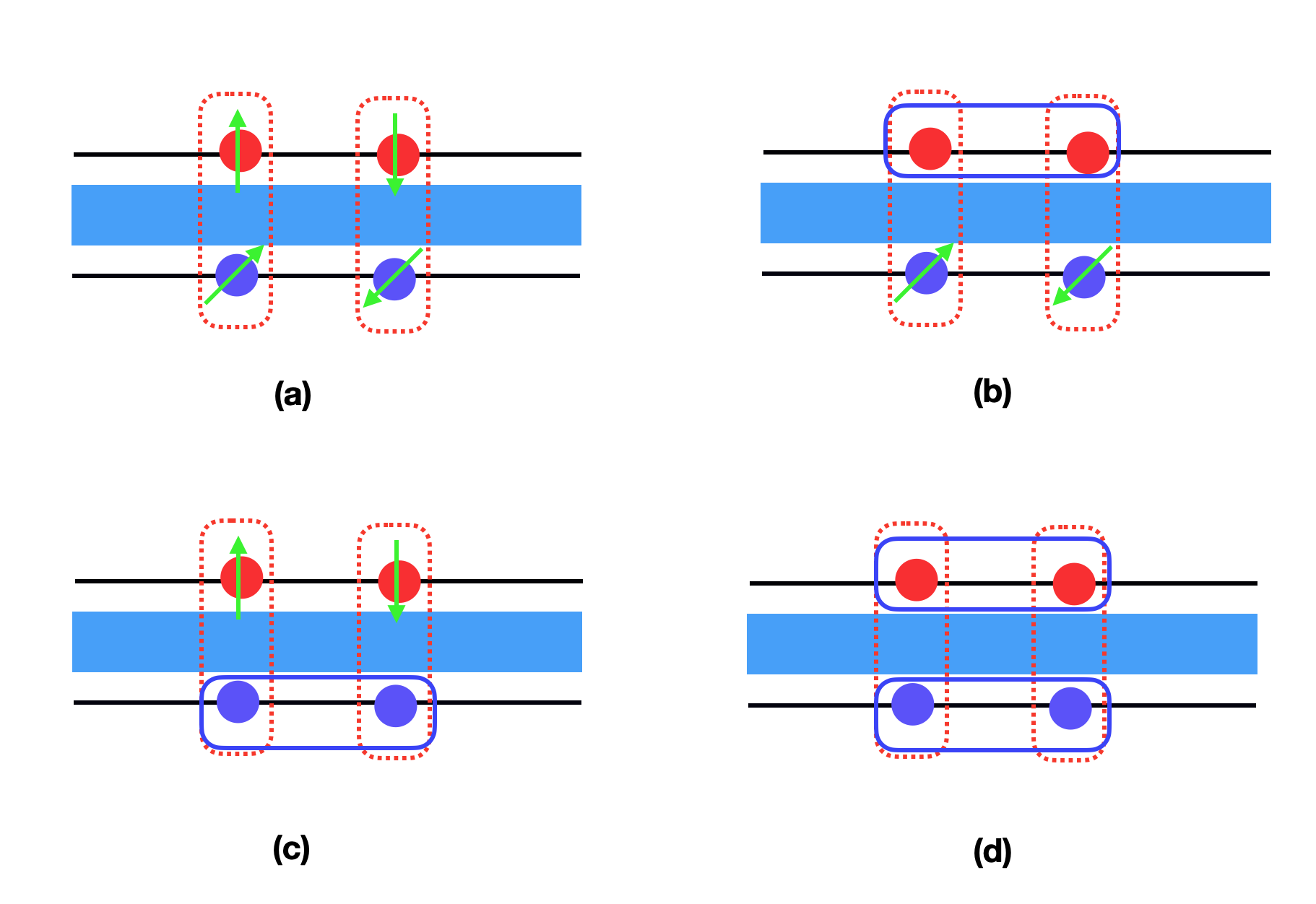}
\caption{Four possible exciton superfluid phases described by the Schwinger boson mean field theories. Here the vertical red dashed square surrounds an exciton. The horizontal blue solid square denotes a spin-singlet pairing term within one layer, which causes a finite spin gap $\Delta_a$ in this layer and also a gap for single exciton. (a) Single exciton condensation phase coexisting with magnetic orders in both layers; (b) There is magnetic order in the bottom layer, but the spin in the top layer is gapped. Only a pair of exciton condenses. (c) Magnetic order in the top layer and spin gap in the bottom layer. Only a pair of exciton condenses. (d) Both layers have a spin gap. Paired exciton condensation coexists with $Z_2$ spin liquid. These four phases in (a)(b)(c)(d) correspond to phases I, II, III, IV in Table.~\ref{table:schwinger_boson_phases} respectively.}
\label{fig:exciton_superfluid_phases}
\end{figure}

The bosons at the two layers have densities $n_t=x$ and $n_b=1-x$. With finite density, the bosons are known to form a condensation phase. But for each layer $a$, there are two different possibilities: (1) Single boson $b_{a;\sigma}$ is condensed, leading to magnetic ordering of the spin $\vec{S}_a$. (2) Only a pair of boson is condensed: $\langle \epsilon_{\sigma \sigma'}b_{i;a\sigma}b_{j;a\sigma'}\rangle \neq 0$. In this case, the spin $\vec{S}_a$ is gapped due to nearest neighbor pairing of the spinons.  Combining the two layers, we have four types of superfluid phases as listed in Table.~\ref{table:schwinger_boson_phases} and illustrated in Fig.~\ref{fig:exciton_superfluid_phases}. Note that each type may contain subcategories specified by detailed symmetry realization such as the momentum associated with the magnetic ordering.  As long as spin in one of the layer is gapped, the superfluid has a condensation of charge $2e$ under $P_z$, which supports paired exciton condensation with $\langle P^\dagger P^\dagger\rangle \neq 0$ order instead of the simple exciton condensation with $\langle P^\dagger \rangle \neq 0$.

One particular case is the phase IV with both $\Delta_t>0$ and $\Delta_b>0$. In the Schwinger boson theory, the spin gap is caused by the paired condensation of the Schwinger boson.  In this class, the emergent U(1) gauge field is only higgsed down to $Z_2$. This means that there is still a deconfined $Z_2$ gauge field coexisting with the inter-layer superfluid order. The paired condensation of the Schwinger boson is the standard description of the $Z_2$ spin liquid in the single layer model. Our bilayer model provides an interesting realization of it.   In contrast, if $b_{i;t}$ or $b_{i;b}$ is condensed, the $Z_2$ gauge field would be completely higgsed and we are left only with a conventional symmetry breaking order without fractionalization.

In our case because $n_b=1-x$ is large, it is easy to have a magnetic order ($\langle b_{b;\sigma}\rangle \neq 0$) in the bottom layer. Then we can have either the phase I or the phase II, which will be discussed in Sec.~\ref{subsection:with_magnetic_order}. We will find that the phase I is favored if $\frac{J_t}{J_p}$ is small, with the spin in the top layer in a ferromagnetic phase due to kinetic energy. But with a reasonably large value of $\frac{J_t}{J_b}$, the spin gap in the top layer is finite and we get the phase II.  In Sec.\ref{subsection:fractional_SF} we will show that increasing $x$ or adding a $J^\prime_b$ term can further destroy the magnetic order in the bottom layer and favors the phase IV, which is a fractional superfluid with deconfined Z$_2$ topological order.

\subsection{With magnetic order in the bottom layer: exciton condensation and paired exciton condensation\label{subsection:with_magnetic_order}}

We deal with the case with a robust $120^\circ$ order in the bottom layer first. The magnetic order can be suppressed by $J'_b/J_b$, which will be discussed in the next subsection.

We can simplify our analysis by condensing the boson $b_b$ at momentum $\mathbf Q=K,K'$ to form the $120^\circ$ order. We represent the Schwinger boson at the bottom layer as $\chi_i=\begin{pmatrix} b_{i;b\uparrow} \\ b_{i;b \downarrow} \end{pmatrix}$. After the condensation of $\chi_i$ at momentum $\mathbf Q=\mathbf{K},\mathbf{K'}$, we get a three-sublattice magnetic order. We can always rotate the spin on B and C sublattice so that the ordered moment at each site is pointing along the $z$ direction.  More specifically, we can write the Schwinger boson as

  \begin{align}
 \chi_{A;i}&=\tilde \chi_i \notag\\
\chi_{B;i}&=e^{-i \frac{2\pi}{3} \sigma_y} \tilde \chi_i\notag\\
\chi_{C;i}&=e^{i \frac{2\pi}{3} \sigma_y} \tilde \chi_i
\label{eq:schwinger_boson_transformation_120order}
  \end{align} 

with
\begin{equation}
	\tilde \chi_i=\left(
	\begin{array}{c}
	\sqrt{n_b-a^\dagger_i a_i}\\
	a_i
	\end{array}
	\right)
	\label{eq:schwinger_boson_HP_120order}
\end{equation}
where $a_i$ is the Holstein-Primakoff boson representing the spin wave on top of the magnetic order.   $n_b$ is the density of Schwinger boson at the bottom layer.

We can do a large $n_b$ expansion to capture the dynamics of the goldstone mode $a_i$. Substituting Eq.~\ref{eq:schwinger_boson_HP_120order} into Eq.~\ref{eq:boson_model}, we get

\begin{align}
H_M&=H_b+H_t+H_{int}
\end{align}
with

\begin{align}
H_t&=-\frac{1}{4} J_p M \sum_{ij} b_{i;t\sigma}^\dagger b_{j;t \sigma}+J_t \sum_{\langle ij \rangle}\vec{S}_{i;t}\cdot \vec{S}_{j;t}+\frac{1}{2}J_{pz}\sum_{\langle ij \rangle}n_{i;t}n_{j;t}
\end{align}

\begin{align}
H_b&=+\frac{1}{8}J_b M \sum_{\langle ij \rangle}(a_i a_j+a_i^\dagger a_j^\dagger)-\frac{3}{8} J_b M \sum_{\langle ij \rangle} (a_i^\dagger a_j+a_j^\dagger a_i)\notag\\
&+\frac{3}{2} J_b M \sum_i a_i^\dagger a_i +...\notag\\
\end{align}

and

\begin{align}
H_{int}&=-\frac{1}{4} J_p \sum_{ij} a_j^\dagger a_i b^\dagger_{i;t\sigma} b_{j;t\sigma}\notag\\
&+\frac{\sqrt{3}}{4} J_p \sqrt{M} \sum_{\langle ij \rangle} \big((a_i -a_j^\dagger) b_{i;t\sigma}^\dagger b_{j;t\sigma}+h.c.\big)
\end{align}
where the summation of $\langle ij \rangle$ is on $(i,j)=(A,B), (B,C), (C,A)$. Note the order of $i,j$ matters for the second term of $H_{int}$.  In $H_t$ we have rewritten the $J_{pz}$ term using the identity $n_{i;b}=1-n_{i;t}$ and ignore the constant term and terms linear to $n_{i;t}$, which can be absorbed into chemical potential.  We have defined $M=n_b-\langle a_i^\dagger a_i\rangle$, which is the magnetization of the magnetic order in the bottom layer.  For $H_b$ we have ignored higher order terms such as $a_i a^\dagger_j a_j$.

In the following we will ignore the dynamics of the spin wave $a_i$ to simplify our analysis, which represents the gapless spin waves on top of the $120^\circ$ order in the bottom layer.  The the main dynamical degree of freedom is $b_{t\sigma}$. Because of the condensation of $b_{i;b\sigma'}$, now $b_{i;t\sigma}$ is a physical operator. Physically $b^\dagger_{i;t\sigma} \sim b^\dagger_{i;t\sigma} \langle b_{i;b \sigma'}\rangle \sim c^\dagger_{i;t\sigma}c_{i;b\sigma'}$ now creates an exciton with $S_t=\frac{1}{2}$ and $P_z=1$.  Note that the exciton carries spin 1/2 under the SU(2) spin rotation in the top layer. The SU(2) spin rotation in the bottom layer is already broken and $S_b$ is no longer well defined.

We can then just relabel $b_{i;\sigma}=b_{i;t\sigma}$, $n_i=n_{i;t}$ and $\vec{S}_{i;t}=\vec{S}_i$. Then we basically have a gas of spin $1/2$ bosons at density $n=x$, described by the Hamiltonian:

\begin{align}
H_{exciton}&=-t \sum_{ ij }P b^\dagger_{i;\sigma}b_{j;\sigma} P-\mu\sum_i b^\dagger_{i;\sigma}b_{i;\sigma}\notag\\
&+J \sum_{\langle ij \rangle} \vec{S}_i \cdot \vec{S}_j+V \sum_{\langle ij \rangle}n_i n_j
\label{eq:boson_t_J}
\end{align}
where $t=\frac{1}{4}(M_0-x)J_p$, $V=\frac{1}{2} J_{pz}$ and $J=J_t$.  The boson is at density $x$ per site. We have $\vec{S}_i=\frac{1}{2} \sum_{\sigma \sigma'} b^\dagger_{i;\sigma}\vec{\sigma}_{\sigma \sigma'}b_{i;\sigma'}$ and $n_i=\sum_{\sigma}b^\dagger_{i;\sigma}b_{i;\sigma}$. $M_0$ is the magnetization of the $120^\circ$ order at the layer polarized limit with $x=0$. With a finite x, the magnetization is further depleted and we have $M=M_0-x$. For the model with $J_b'=0$, we know $M_0\approx 1/2$. We also introduce the projection operator $P$ to implement the constraint that $n_i=0,1$, inherited from the Schwinger boson formulation.

The above effective Hamiltonian works at the small x regime. We can see that the dynamics of doped excitons on the background of $120^\circ$ magnetic order is basically captured by a spin 1/2 hard core boson gas problem, an observation already made in our previous paper\cite{zhang20214}. Note that Eq.~\ref{eq:boson_t_J} can be viewed as a bosonic t-J model.  Next we analyze this effective model. We will find a spin polarized BEC at small $J/t$ and a spin gapped paired superfluid at large $J/t$. In the intermediate regime there is a spiral phase where the boson $b$ condenses at non-zero momentum.

\subsubsection{Spin polarized BEC at small J/t\label{subsubsection:spin_polarized_BEC}}

If we set $J=0$, then it is natural to expect that the ground state of this model is a spin polarized BEC, similar to what is usually found for BEC with integer spin\cite{kawaguchi2012spinor}. For the spin polarized BEC, we take an ansatz with $\langle b_{i;\uparrow}\rangle=\sqrt{n_0}$ and $\langle b_{i;\downarrow} \rangle=0$. Next we analyze the goldstone modes above this condensation using the Bogoliubov mean field theory and show that it is unstable if $\frac{J}{t}$ is large enough.

In momentum space we assume $b_{\uparrow}(\mathbf k=0)=\sqrt{N_s} \sqrt{n_0}$, where $N_s$ is the number of sites. At mean field level we have energy:

\begin{equation}
	E_M=-(6t+\mu)n_0N_s+3(V+\frac{1}{4}J)n_0^2N_s
\end{equation}

Variation with respects to $n_0$ leads to:
\begin{equation}
	\mu=6(V+\frac{1}{4}J)n_0-6t
\end{equation}

Next we consider excitations by also including $b_\sigma(\mathbf q\neq 0)$.

 the mean field Hamiltonian is

\onecolumngrid

\begin{align}
H_M&=-t \sum_{ij}b_{i;\sigma}^\dagger b_{j;\sigma}-\mu \sum_{i}b^\dagger_{i;\sigma}b_{i;\sigma}-(6t+\mu) n_0 N +3(V+\frac{1}{4}J) n_0^2 N +6 (V+\frac{1}{4}J) n_0 \sum_{\mathbf q \neq 0}b^\dagger_{\uparrow}(\mathbf q) b_{\uparrow}(\mathbf q)\notag\\
&+\sum_{\mathbf q\neq 0}  (6V-\frac{3}{2}J+\frac{J}{2} F(\mathbf q)) n_0 b^\dagger_{\downarrow}(\mathbf q) b_{\downarrow}(\mathbf q)
+(V +\frac{1}{4}J)n_0 \sum_{\mathbf q\neq 0}b^\dagger_{\uparrow}(\mathbf q) b_{\uparrow}(\mathbf q) F(\mathbf q) \notag\\
&+\frac{Vn_0+\frac{1}{4}J n_0}{2}\sum_{\mathbf q\neq 0} \big(F(\mathbf q) b_{\uparrow}(\mathbf q) b_{\uparrow}(-\mathbf q)+F(\mathbf q)b^\dagger_{\uparrow}(\mathbf q) b^\dagger_{\uparrow}(-\mathbf q)\big)
\end{align}

\twocolumngrid

where $F(\mathbf q)=\sum_{j} \cos(\mathbf q \cdot \mathbf{r_j})$ and $j=1,2,...,6$ lists the six nearest neighbors of site $i=(0,0)$.

It is easy to see that the $b_{\uparrow}$ and $b_{\downarrow}$ part decouple.  We can diagonalize the Hamiltonian using the Bogoliubov transformation and express the Hamiltonian with new bosonic operator $\gamma(\mathbf q)$.

In the end we get:

\begin{equation}
	H_M=E_0+\sum_{\mathbf q}\omega_{\uparrow}(\mathbf q)\gamma^\dagger(\mathbf q)\gamma(\mathbf q)+\sum_{\mathbf q}\omega_{\downarrow}(\mathbf q)b^\dagger_{\downarrow}(\mathbf q)b_{\downarrow}(\mathbf q)
\end{equation}
where,
\begin{align}
	&\omega_{\uparrow}(\mathbf q)=\notag \\ 
    &\sqrt{\big(t(6-F(q))+(V+\frac{1}{4}J)n_0F(\mathbf q)\big)^2-\big((V+\frac{1}{4}J)n_0F(\mathbf q)\big)^2}
\end{align}
and
\begin{equation}
	\omega_{\downarrow}(\mathbf q)=(t-\frac{J}{2} n_0)(6-F(\mathbf q))
\end{equation}

Using $F(\mathbf q)=\sum_{j} \cos(\mathbf q \cdot \mathbf{r_j})\approx 6-\frac{3}{2}q^2$ at small $|\mathbf q|$, we find $\omega_{\uparrow}(\mathbf q)\approx 3\sqrt{2t(V+\frac{1}{4}J)} |\mathbf q| $ and $\omega_{\downarrow}(\mathbf q)=\frac{3}{2}(t-\frac{1}{2} J n_0)|\mathbf q|^2$ at small $|q|$.  Physically the linear mode from $b_{\uparrow}$ corresponds to the Goldstone mode breaking the U(1) symmetry generated by $P_z$. The quadratic model from $b_{\downarrow}$ is the spin wave from breaking the $SU(2)$ spin rotation symmetry of the top layer.

The stabilization of this spin polarized BEC requires that $t-\frac{1}{2} J n_0>0$.  We expect the condensation $n_0\approx x$, and we have $t=\frac{1}{4}(M_0-x)J_p$ with $M_0\approx \frac{1}{2}$ and $J=J_t$.  Then the stabilization condition becomes

\begin{equation}
	\frac{J_t}{J_p}<\frac{M_0}{2x}-1
\end{equation}

When $\frac{J_t}{J_p}>\frac{M_0}{2x}-1$, the spin polarized BEC is unstable because of the antiferromagnetic spin-spin coupling. We will discuss the possibility in the remaining of this section.

\subsubsection{Paired superfluid at large J/t\label{subsubsection:paired_sf}}

When $J<<t$, we believe the ground state of Eq.~\ref{eq:boson_t_J} is a spin polarized BEC. But when $J/t$ is large, it is obvious that the spin polarization is disfavored. In this regime, a promising candidate of Eq.~\ref{eq:boson_t_J} is a paired superfluid phase with $\langle \epsilon_{\sigma \sigma'} b_{i;\sigma}b_{j;\sigma'}\rangle \neq 0$, but $\langle b_{i;\sigma}\rangle=0$. In the following we will describe this paired condensation phase of the excitons and discuss its instability when decreasing $J/t$.

We define operators $\hat T_{ij}= \sum_{\sigma} b^\dagger_{i; \sigma} b_{j; \sigma}$ and $\hat \Delta_{ij}=\sum_{\sigma \sigma'}\epsilon_{\sigma \sigma'} b_{i; \sigma}b_{j; \sigma'}$. A general symmetric mean field ansatz for the exciton from decoupling Eq.~\ref{eq:boson_t_J} is in the form

\begin{align}
H_M&= \sum_{ij} T_{ij} \hat T_{ij } -\sum_{\langle ij \rangle}(\Delta_{ij} \hat \Delta_{ij}+h.c.)-\mu \sum_i n_{i}
\label{eq:mean_field_boson}
\end{align}
where $T_{ij}=T_{ji}^*$ and $\Delta_{ij}=-\Delta_{ji}$. 

We have self consistent equations:

\begin{align}
T_{ij }&=\frac{3}{8} J \langle \hat T_{ji;a} \rangle -t\notag\\
\Delta_{ij}&=\frac{3}{8} J \langle \hat \Delta^\dagger_{ij} \rangle \notag\\
\langle n_{i} \rangle &= x \notag\\
\label{eq:paired_exciton_self_consistent_equations}
\end{align}
where $J=J_t$ and $t=\frac{1}{4}J_p(M_0-x)$. $M_0$ is the magnetization of the $120^\circ$ order in the bottom layer. $x$ is the density of the excitons.

We will only consider uniform ansatz with $T_{ij}=T$. The pairing term needs to be in the odd angular momentum channel. We consider ansatz with $\Delta_{i,i\pm \hat x}=\pm \Delta$, while the other bonds can be generated by $C_3$ rotation. This ansatz is in the $f$ wave channel and it still preserves the time reversal symmetry. It is symmetric under $C_6$ rotation if we apply a gauge transformation afterwards. The symmetry property of the ansatz is the same as the zero-flux ansatz in the projection symmetry group (PSG) of $Z_2$ spin liquids on triangular lattice\cite{wang2006spin}. This is expected as the sShwinger boson $b_i=b_{i;t}$ shares the same PSG property as the Schwinger boson $b_{i;b}$ in the bottom layer, which is in the $120^\circ$ ordered phase and is known to be proximate to the zero-flux ansatz\cite{wang2006spin}.  The mean field Hamiltonian can be rewritten in momentum space as:

\begin{align}
	H_M&=\sum_{\mathbf k}(\xi(\mathbf k)-\mu)b^\dagger_{\sigma}(\mathbf k)b_{\sigma}(\mathbf k)\notag\\
	&+\Delta(\mathbf k)(b_{\uparrow}(\mathbf k) b_{\downarrow}(-\mathbf k)-b_{\downarrow}(\mathbf k)b_{\uparrow}(-\mathbf k)+h.c.)
\end{align}

with 
\begin{align}
\xi(\mathbf k)&=2T (\cos k_x+2 \cos \frac{1}{2} k_x \cos \frac{\sqrt{3}}{2} k_y)\notag\\
\Delta(\mathbf k)&=2i \Delta (\sin k_x-2 \sin \frac{1}{2} k_x \cos \frac{\sqrt{3}}{2}k_y)
\end{align}

Using $\psi_a(\mathbf k)=\begin{pmatrix} b_{\uparrow}(\mathbf k)\\ b^\dagger_{\downarrow}(-\mathbf k) \end{pmatrix}$, the Hamiltonian can be rewritten as:

\begin{equation}
	H_a=\sum_{\mathbf k}\psi^\dagger_a(\mathbf k) \begin{pmatrix} \xi_a(\mathbf k) -\mu_a  & \Delta^*(\mathbf k) \\ \Delta(\mathbf k)& \xi_a(-\mathbf k)-\mu_a \end{pmatrix} \psi_a(\mathbf k)+ \sum_{\mathbf k} \mu
\end{equation}

\begin{figure}[ht]
\centering
\includegraphics[width=0.5\textwidth]{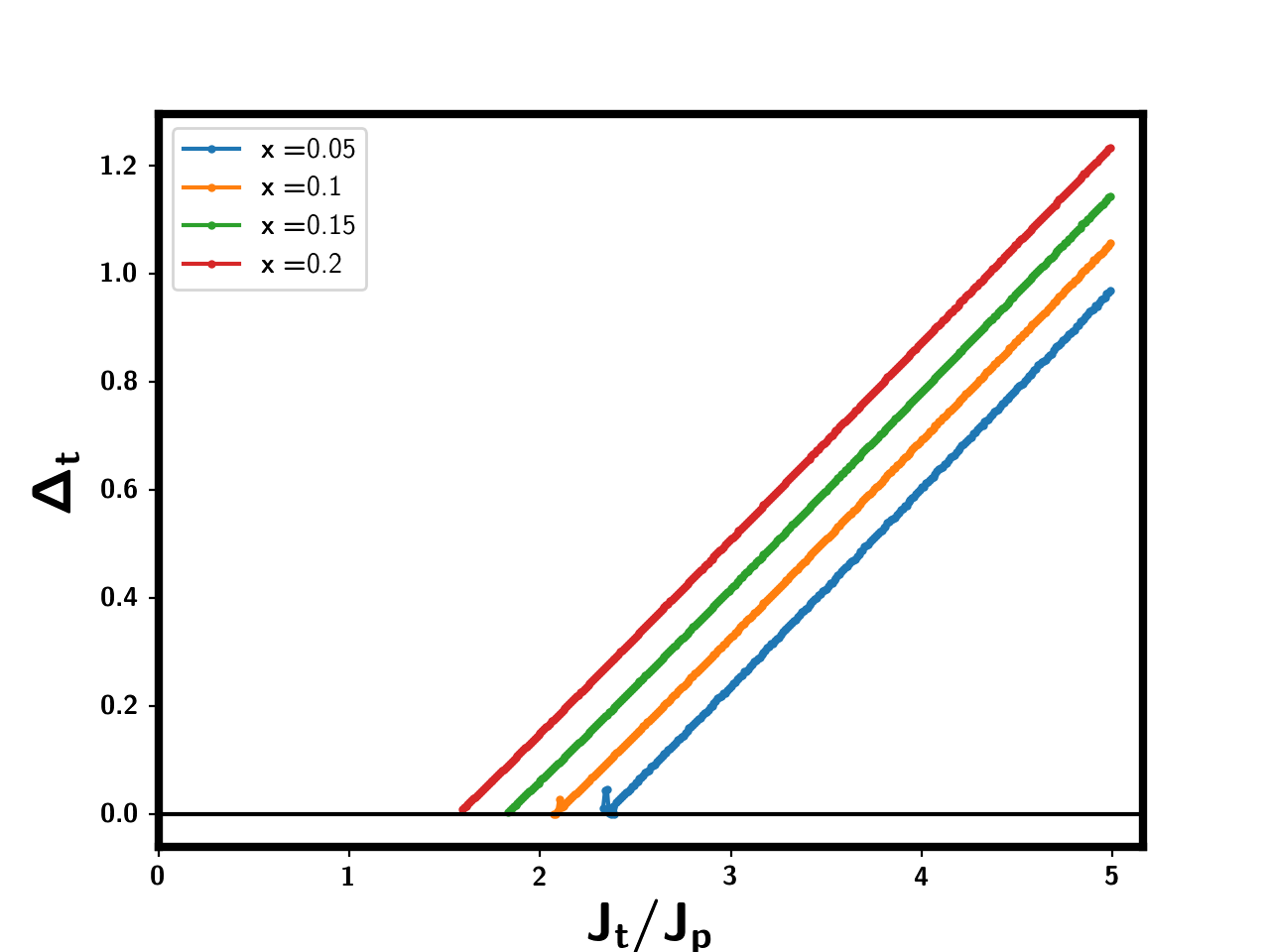}
\caption{The spin gap $\Delta_t$ for the paired superfluid phase from mean field calculation using Eq.~\ref{eq:calculate_T_Delta} and Eq.~\ref{eq:paired_exciton_self_consistent_equations}. }
\label{fig:paired_sf_mean_field}
\end{figure}

Using the Bogoliubov transformation, we can get:

\begin{equation}
	H_a=\sum_{\mathbf k} \omega_a(\mathbf k) \sum_{\sigma}\gamma^\dagger_{\sigma}(\mathbf k) \gamma_{\sigma}(\mathbf k)+\sum_{\mathbf k}(\omega_a(\mathbf k)+\mu)
\end{equation}
where $\gamma_\sigma(\mathbf k)$ is bosonic operator which diagonalizes the Hamiltonian. 

We have

\begin{equation}
	\omega(\mathbf k)=\sqrt{(T A_a(\mathbf k)-\mu)^2- \Delta^2 B(\mathbf k)^2}
\end{equation}
where $A(\mathbf k)=2(\cos k_x+2 \cos \frac{1}{2} k_x \cos \frac{\sqrt{3}}{2} k_y)$ and $B(\mathbf k)=2(\sin k_x-2 \sin \frac{1}{2} k_x \cos \frac{\sqrt{3}}{2}k_y)$.

$\langle \hat T_{ij}\rangle$ can be calculated as $\sum_{ ij }\langle \hat T_{ij} \rangle=\frac{\partial \langle H_M \rangle}{\partial T}$. Similarly $\sum_{\langle ij \rangle}\langle \hat \Delta_{ij} +\hat \Delta^\dagger_{ij}\rangle=-\frac{\partial \langle H_M \rangle}{\partial \Delta}$ and $\sum_i n_{i}=-\frac{\partial \langle H_M \rangle}{\partial_{\mu}}$. Eventually we get

\begin{align}
\langle \hat T_{ij} \rangle&=\frac{1}{6N_s} \sum_{\mathbf k} \frac{ A(\mathbf k)( T A(\mathbf k)-\mu)}{\omega(\mathbf k)}\notag\\
\langle \hat \Delta_{i,i+\hat{x}} \rangle&=\frac{1}{6N_s} \sum_{\mathbf k} \frac{\Delta B(\mathbf k)^2}{\omega(\mathbf k)}\notag\\
\langle n \rangle&=\frac{1}{N_s} \sum_{\mathbf k} \frac{T A(\mathbf k)-\mu}{\omega(\mathbf k)}-1
\label{eq:calculate_T_Delta}
\end{align}
where $N_s$ is the number of sites in the mesh grid of the momentum space.

Combination of Eq.~\ref{eq:calculate_T_Delta} and Eq.~\ref{eq:paired_exciton_self_consistent_equations} gives a set of self consistent equations to obtain the mean field parameters $T,\Delta,\mu$ for given values of $t,J,x$ in the Hamiltonian in Eq.~\ref{eq:boson_t_J}. We show the obtained spin gap $\Delta_t$ in Fig.~\ref{fig:paired_sf_mean_field}. For each $x$, we find the spin gap closes when $\frac{J_t}{J_p}<\frac{J_{t;c}}{J_p}$, indicating the instability of the paired superfluid phase for smaller value of $J_t$. Intuitively the single exciton condensation without spin gap is favored when $\frac{J_t}{J_p}$ is small because the kinetic energy dominates.

\subsubsection{Intermediate spiral phase\label{subsubsection:intermediate_spiral}}

In the previous two parts, we show that the exciton is in a spin polarized BEC phase when $\frac{J_t}{J_p}<\frac{M_0}{2x}-1$ and a paired superfluid phase when $\frac{J_t}{J_p}>\frac{J_{t;c}}{J_p}$.  Generically, we expect $\frac{M_0}{2x}-1<\frac{J_{t;c}}{J_p}$. For example, when $x=0.1$, the spin polarized BEC is in the regime $\frac{J_t}{J_p}<1.5$ and the paired superfluid phase is in the regime $\frac{J_t}{J_p}>2$.  In this case, we are left with an intermediate regime. Note that the spin gap $\Delta_t$ must be zero when $\frac{J_t}{J_p}<\frac{J_{t;c}}{J_p}$.  So the single exciton must be condensed with $\langle b_{i;t\sigma} \rangle \neq 0$. But it can not be a uniform spin polarized BEC because the full spin polarization is unstable due to the analysis in Sec.~\ref{subsubsection:spin_polarized_BEC}. A natural possibility is then a BEC phase at a non-zero momentum $\mathbf Q$, so the spin correlation has anti-ferromagnetic component.  The momentum $\mathbf Q$ may be generically incommensurate and we can dub this phase as a spiral BEC phase.  Note that both the spin polarized BEC and the spiral BEC are labeled as phase I in Table.~\ref{table:schwinger_boson_phases}, as we do not distinguish the momentum of the exciton condensation and magnetic order there.  We will show numerical evidence for such spiral phase in the next subsection.

\subsubsection{Numerical results}

From the theoretical analysis, we know that the ground state goes through spin polarized exciton condensation, spiral exciton condensation and paired exciton condensation when we increasing $\frac{J_t}{J_p}$ while assuming the bottom layer is in the $120^\circ$ ordered phase. Here we offer numerical evidences for this picture based on infinite density matrix renormalization group (DMRG) simulation\cite{mcculloch2008infinite,white1992density,white1993density}. The infinite DMRG is performed using the TeNPy Library (version 0.4.0)\cite{tenpy}. We define the triangular lattice unit vectors as: $\mathbf{a}_1=(1,0)$ and $\mathbf{a}_2=(-\frac{1}{2},\frac{\sqrt{3}}{2})$. We will use a system with periodic boundary along $\mathbf{a}_2$ direction: $O(\mathbf r+L_y \mathbf a_2)=O(\mathbf r)$, where $O$ is an arbitrary operator and $L_y$ is the system size along the $\mathbf{a}_2$ direction.  We will use $L_y=6$.  Along $\mathbf{a}_1$ the system size is infinite. However we need to choose a unit cell with size $L_x\times L_y$. We will use $L_x=6$ so that we can fix exciton density $x$ to be a small value such as $\frac{1}{18}$.  We use bond dimension up to $m=4000$. Typical truncation error is at order $10^{-4}$.

 We fix $J_b=J_p=J_{pz}=1$ and then vary $J_t$ in our DMRG simulation of Eq.~\ref{eq:spin_layer_model}. The details of the implementation are shown in the Append.~\ref{append:dmrg}. We label the four states as $\ket{1}=\ket{t\uparrow},\ket{2}=\ket{t\downarrow},\ket{3}=\ket{b\uparrow},\ket{4}=\ket{b\downarrow}$. The operators can be labeled as $S_{ab}=\ket{a}\bra{b}$ with $a,b=1,2,3,4$. We used three U(1) quantum numbers $Q_1=S_{22}-S_{11},Q_2=S_{33}-S_{11},Q_3=S_{44}-S_{11}$.  We can extract correlation lengths within a sector fixed by quantum number using the transfer matrix technique.  We label $\xi_{S_t}$ for the largest correlation length in the sector $(Q_1,Q_2,Q_3)=(2,1,1)$, where the typical operator is $S_t^{-}$. Similarly we label $\xi_{S_b}$ for the correlation length in the sector $(Q_1,Q_2,Q_3)=(0,-1,1)$, where the typical operator is $S_b^{-}$. We label $\xi_{P}$ for the largest correlation length in the sector $(Q_1,Q_2,Q_3)=(1,2,1)$, where the typical operator is $S_{31}=P^-(I+S_z)$, which carries charge $-1$ under $P_z$. We label $\xi_{PP}$ for the largest correlation length in the sector $(Q_1,Q_2,Q_3)=(0,2,2)$, where the typical operator is $S_{31}S_{42}$, which carries charge $-2$ under $P_z$. 

\begin{figure*}[tbp]
\centering
\includegraphics[width=0.8\linewidth]{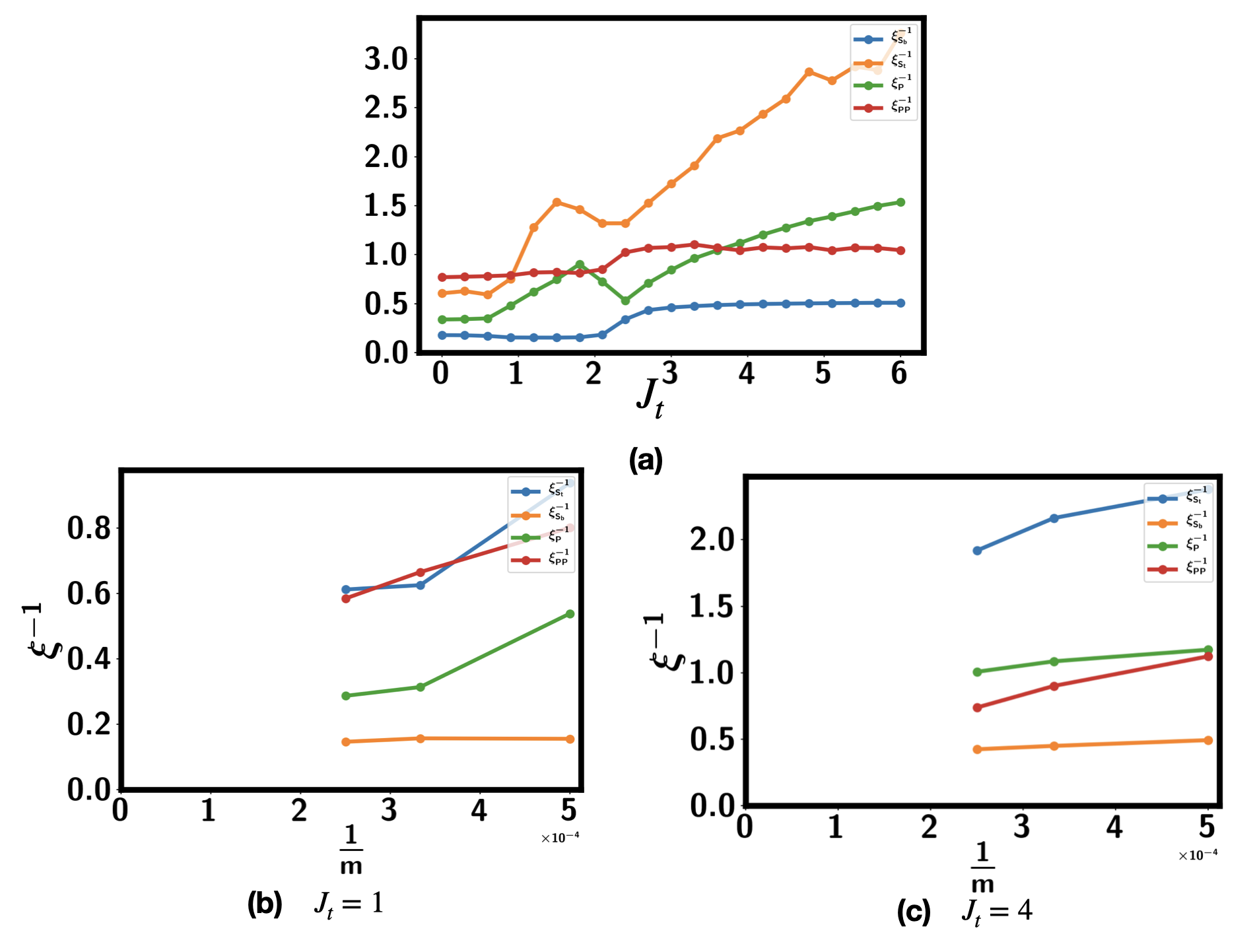}
\caption{Correlation length from infinite DMRG. We fix $J_b=J_p=J_{pz}=1$. We use system size $L_y=6$ and a unit cell size $L_x=6$. The exciton density is fixed at $x=\frac{1}{18}$. (a) Inverse correlation lengths with $J_t$ using bond dimension $m=2000$; (b) Results at $J_t=1$. We used bond dimension $m=2000,3000,4000$. (c) Results at $J_t=4$. We used bond dimension $m=2000,3000,4000$.}
\label{fig:corr_len}
\end{figure*}

We plot the inverse of these various correlation lengths in Fig.~\ref{fig:corr_len} with exciton density $x=\frac{1}{18}$ for a $L_y=6$ cylinder. $\xi^{-1}$ is never zero at a finite bond dimension $m$ even if it is zero when $m=\infty$. The extrapolation is hard given that we only have results for $m\leq 4000$ due to computation cost.  Therefore it is tricky to determine whether there is a spin gap $\Delta_t$, $\Delta_b$ for the spin in the two layers. However, it is useful to compare $\xi^{-1}_P$ and $\xi^{-1}_{PP}$. If the phase is in an exciton condensation phase, single exciton is more elementary and we expect $\xi^{-1}_{PP}=2\xi^{-1}_P$ because $\xi^{-1}_{PP}$ is roughly the gap of a pair of exciton when the effective system size is finite (controlled by the bond dimension $m$). This is exactly what we found for $J_t=1$ as shown in Fig.~\ref{fig:corr_len}(b). Thus we believe that the phase is in the single exciton condensation phase for $J_t=1$. This would imply that $\Delta_t=\Delta_b=0$ because single exciton $c^\dagger_{t;\sigma}c_{b;\sigma'}$ carries spin indexes of both layers and its condensation must lead to gapless spin waves for both layers. In contrast, when $J_t=4$, we can clearly see that $\xi^{-1}_{PP}<\xi^{-1}_P$ and $\xi^{-1}_{PP}$ decreases significantly faster than $\xi^{-1}_P$ when we increase the bond dimension $m$. This strongly suggests that the phase is in a paired exciton condensation phase with a gap for single exciton. This would imply that at lease one of $\Delta_t,\Delta_b$ is finite. We note that $\xi^{-1}_{S_b}<<\xi^{-1}_{S_t}$ for the entire range of $J_t$, so we conclude that $\Delta_t>0$, while $\Delta_b$ may still be zero, though it really needs data at $m \rightarrow \infty$ to prove $\xi^{-1}_{S_b}=0$. A finite $\Delta_t$ is further supported by the fact that $\xi^{-1}_{S_t}$ grows almost linearly with $J_t$ after $J_t\geq 2.2$ (see Fig.~\ref{fig:corr_len}(a)), which resembles the behavior of $\Delta_t$ from mean field calculation shown in Fig.~\ref{fig:paired_sf_mean_field}. Thus we conclude that the ground state is in the phase II (see Table.~\ref{table:schwinger_boson_phases}) when $J_t\geq 2.2$ with $\Delta_t>0$. In the Schwinger boson theory, this corresponds to the phase with $\langle b_{b;\sigma}\rangle \neq 0$ and $\langle b_{t;\sigma}\rangle=0$. 

\begin{figure*}[tbp]
\centering
\includegraphics[width=0.8\linewidth]{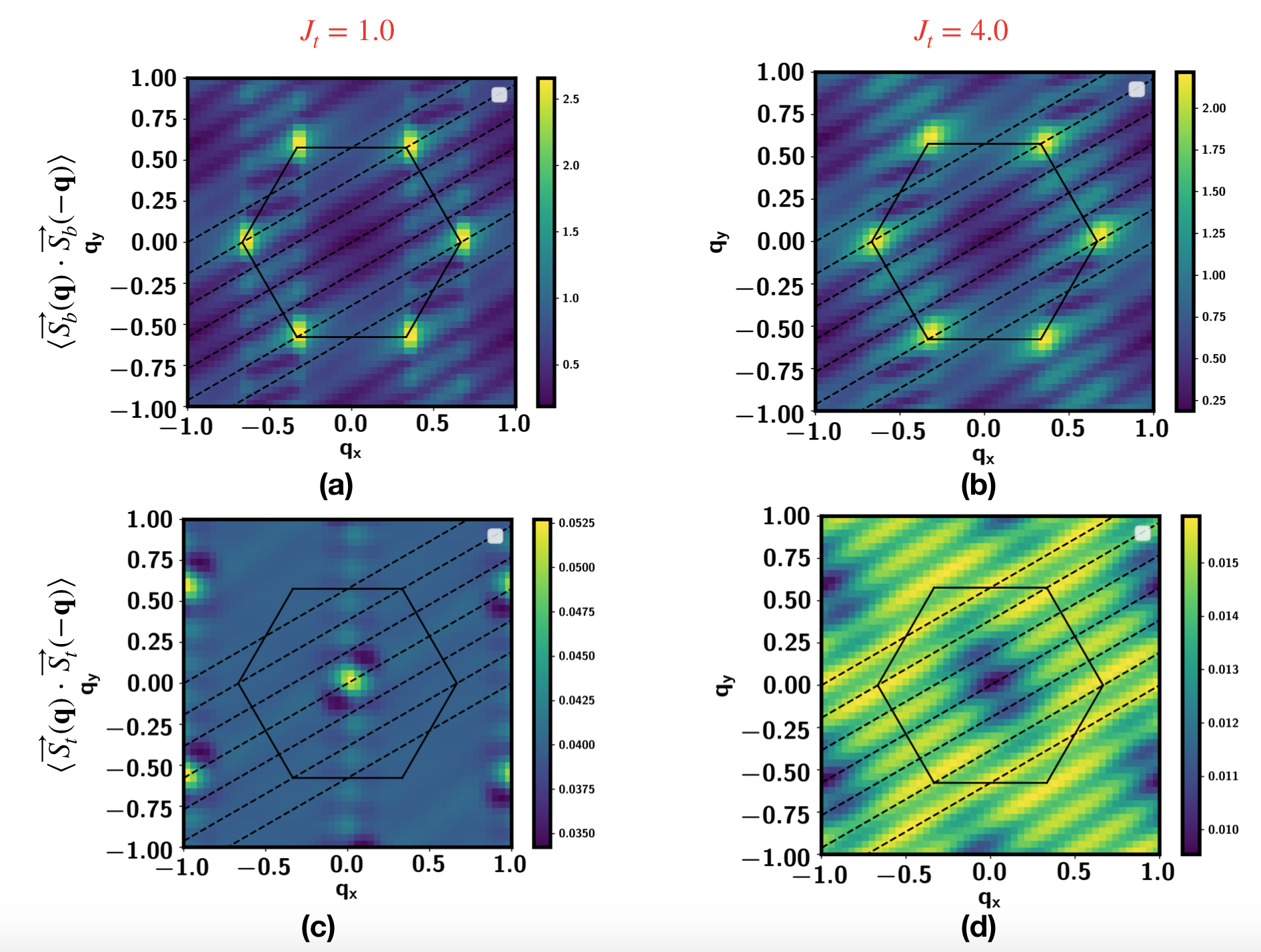}
\caption{Spin-spin structure factors in the two layers. We fix $J_b=J_p=J_{pz}=1$. We use system size $L_y=6$ and a unit cell size $L_x=6$. Bond dimension $m=4000$ is used. The exciton density is fixed at $x=\frac{1}{18}$. (a)(b)$\langle \vec{S_b}(\mathbf q)\cdot \vec{S_b}(-\mathbf q)\rangle$ in the bottom layer for $J_t=1,4$.  (c)(d)$\langle \vec{S_t}(\mathbf q)\cdot \vec{S_t}(-\mathbf q)\rangle$ in the top layer for $J_t=1,4$. }
\label{fig:SS}
\end{figure*}

To further support this picture, we show the spin-spin structure factors $\vec{S_a}(\mathbf q)\cdot \vec{S_a}(-\mathbf q)$ for the two layers $a=t,b$ in Fig.~\ref{fig:SS}. In the bottom layer, there are peaks at momentum $K,K'$ for both $J_t=1,4$, consistent with the $120^\circ$ order. In the top layer, the spin structure factor has a peak at $\mathbf q=0$ for $J_t=1$, consistent with the spin polarized state discussed in Sec.~\ref{subsubsection:spin_polarized_BEC}. When $J_t=4$, there is no peak in the spin structure factor in the top layer, consistent with a spin gap $\Delta_t>0$. This agrees with the paired superfluid phase discussed in Sec.~\ref{subsubsection:paired_sf}.

\begin{figure*}[ht]
\centering
\includegraphics[width=0.8\linewidth]{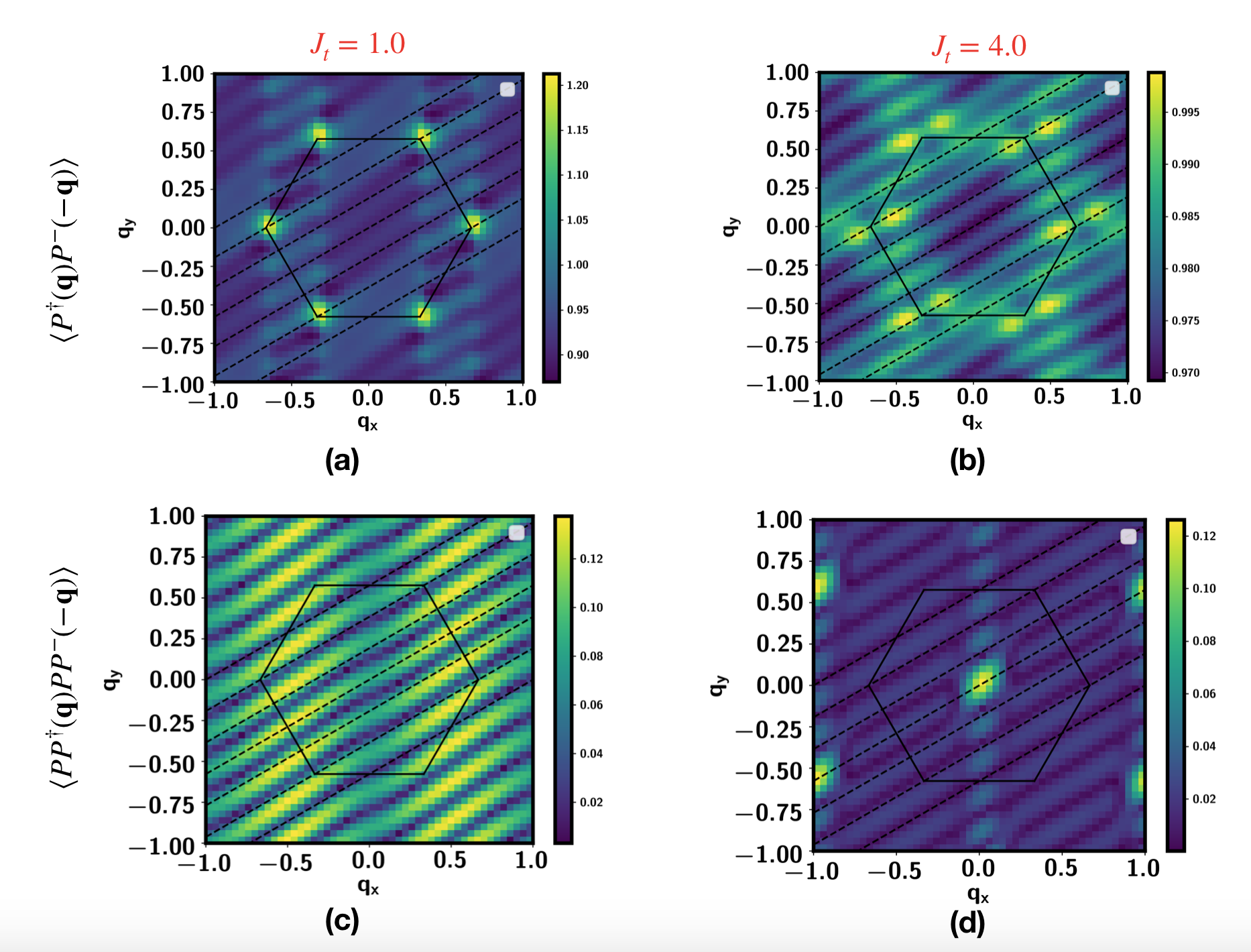}
\caption{Exciton and paired exciton structure factors in the two layers. We fix $J_b=J_p=J_{pz}=1$. We use system size $L_y=6$ and a unit cell size $L_x=6$.  Bond dimension $m=4000$ is used. The exciton density is fixed at $x=\frac{1}{18}$. (a)(b)$\langle P^\dagger(\mathbf q) P^-(-\mathbf q)\rangle$ in the bottom layer for $J_t=1,4$.  (c)(d)$\langle PP^{\dagger}(\mathbf q)\cdot PP^{-}(-\mathbf q)\rangle$ in the top layer for $J_t=1,4$. }
\label{fig:PP}
\end{figure*}

We also show the single exciton correlation and paired exciton correlation in the momentum space in Fig.~\ref{fig:PP}. $\langle P^\dagger(\mathbf q) P^{-}(-\mathbf q) \rangle$ is the structure factor for single exciton operator $P^\dagger \otimes I$, where $I$ is the identity in the spin sector. $\langle PP^\dagger(\mathbf q)PP^{-}(-\mathbf q)\rangle$ is the structure for the paired exciton operator $PP^\dagger_i=(\epsilon_{\sigma \sigma'}c^\dagger_{i;t\sigma}c^\dagger_{i+\hat{x};t\sigma'})(\epsilon_{\alpha \beta}c_{i;b\alpha}c_{i+\hat{x};b\beta})$(see Sec.~\ref{append:dmrg}). For $J_t=1$, we can see that $\langle P^\dagger(\mathbf q) P^{-}(-\mathbf q)\rangle$ has a peak at $\mathbf q=K,K'$. Because $P^\dagger \sim b^\dagger_{t;\sigma}b_{b;\sigma'}$ and we have $b_t$ condenses at momentum $\mathbf q=0$ and $b_{b}$ condenses at momentum $K,K'$ to get the $120^\circ$ order, $P^\dagger$ needs to carry a momentum $K,K'$. On the other hand, $\langle PP^\dagger(\mathbf q) PP^{-}(-\mathbf q)$ is featureless.  In contrast, for $J_t=4$, we find that $\langle P^\dagger(\mathbf q) P^{-}(-\mathbf q)\rangle$ only has broad peaks at momentum close to $K,K'$ and $\langle PP^\dagger(\mathbf q) PP^{-}(-\mathbf q)\rangle$ has a peak at momentum $\mathbf q=0$. Combined with the correlation length $\xi^{-1}_P,\xi^{-1}_{PP}$ in Fig.~\ref{fig:corr_len}, we conclude that the phase is in the paired exciton condensation phase at momentum $\mathbf q=0$, while the single exciton is gapped.

\begin{figure*}[ht]
\centering
\includegraphics[width=0.8\linewidth]{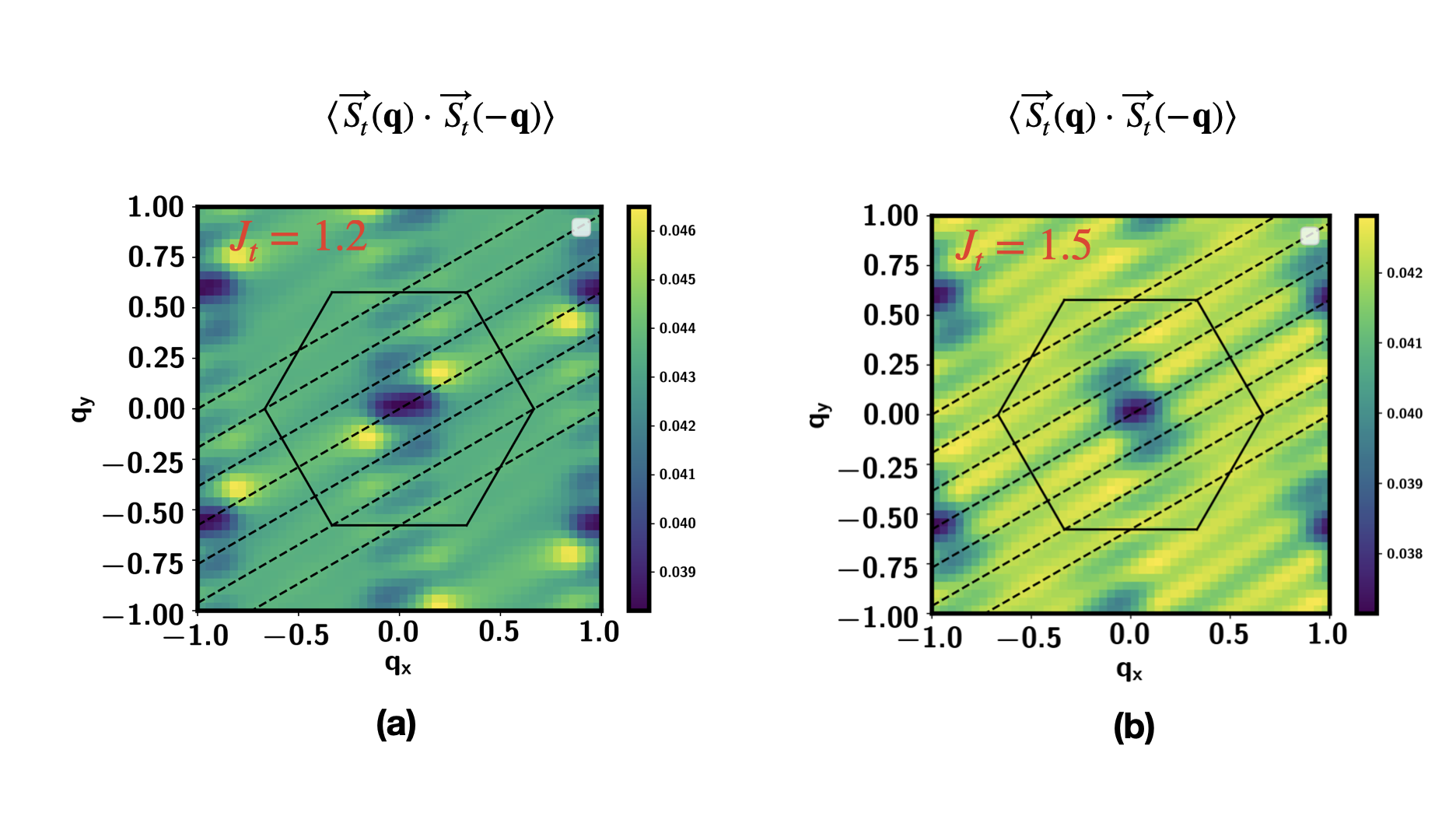}
\caption{Spin-spin structure factors in the top layers. We fix $J_b=J_p=J_{pz}=1$. We use system size $L_y=6$ and a unit cell size $L_x=6$.  (a)$J_t=1.2$ with bond dimension $m=2000$. One can see peaks at momentum away from $\mathbf q=0$, consistent with the spiral phase discussed in Sec.~\ref{subsubsection:intermediate_spiral} for intermediate value of $\frac{J_t}{J_p}$. (b)$J_t=1.5$ with bond dimension $m=4000$. No clear peak is seen, suggesting that the spin in the top layer may already be gapped at $\frac{J_t}{J_p}=1.5$. Then it enters the paired exciton condensation phase. }
\label{fig:spiral}
\end{figure*}

When $J_t$ is in the intermediate regime ($J_t=1.2$), we notice that the spin structure factor in the top layer has peaks at momentum away from $\mathbf q=0$ as shown in Fig.~\ref{fig:spiral}(a), consistent with the spiral phase discussed in Fig.~\ref{subsubsection:intermediate_spiral} with $b_t$ condensed at a non-zero momentum. Because $P^\dagger \sim b^\dagger_t b_b$, $\langle P^\dagger(\mathbf q) P^{-}(-\mathbf q) \rangle$ will also have peaks at momentum away from $K,K'$. But this spiral phase seems to be gone already when $J_t=1.5$ (see Fig.~\ref{fig:spiral}(b)) and then it is replaced by the paired exciton condensation phase.

In the above we have assumed that the spin $\vec{S_b}$ in the bottom layer is always in the $120^\circ$ ordered phase. This assumption should be valid in the $x \rightarrow 0$ limit if there is a $120^\circ$ order with ordering moment $M_0$ at $x=0$. However, a finite $x$ will deplete the ordering moment $M\sim M_0-x$ to be zero when $x$ is large enough. We will discuss this interesting case in the following part of this section.

\subsection{Paired superfluid coexisting with $Z_2$ spin liquid\label{subsection:fractional_SF}}

Previously we discussed the case with a magnetic order in the bottom layer. In this limit the Schwinger boson $b_{b\sigma}$ in the bottom layer is condensed at momentum $K,K'$ with a condensation density $M$, then the low energy dynamics is captured by a spin 1/2 boson gas represented by the Schwinger boson $b_t$ in the top layer.  We can get phase I and phase II in Table.~\ref{table:schwinger_boson_phases} depending on the competition between the $J_t$ and the $J_p$ term.  Especially we find that there is a finite spin gap at the top layer when $J_t$ is large enough. The next natural question is whether we can also get rid of the single Schwinger boson condensation in the bottom layer and get a spin gap also for $\vec{S}_b$. 

We expect the condensation density for the Schwinger boson $b_{b\sigma}$ at the bottom layer is $M=M_0-x$, where $M_0$ is the ordering moment at $x=0$. Then we can imagine that $M$ is going to vanish when increasing $x$ to be larger than a critical value $x_c=M_0$, especially in the case when $M_0$ is small already in the layer polarized limit.  This leads to the possibility of the phase IV, which has spin gap for both layers and a $Z_2$ topological order.

We can obtain a mean field ansatz for such a phase in the Schwinger boson mean field theory. We define operators $\hat T_{ij;a}= \sum_{\sigma} b^\dagger_{i;a \sigma} b_{j;a \sigma}$ and $\hat \Delta_{ij;a}=\sum_{\sigma \sigma'}\epsilon_{\sigma \sigma'} b_{i;a \sigma}b_{j;a \sigma'}$. A general spin rotation symmetric mean field ansatz  is in the form

\begin{align}
H_M&=\sum_{a=t,b} \sum_{ij} T_{ij;a} \hat T_{ij;a } -\sum_{\langle ij \rangle}(\Delta_{ij;a} \hat \Delta_{ij;a}+h.c.)\notag\\
&-(\mu+D) \sum_i n_{i;t}-(\mu-D) \sum_i n_{i;b}
\label{eq:mean_field_boson}
\end{align}
where $T_{ij}=T_{ji}^*$ and $\Delta_{ij}=-\Delta_{ji}$.  $D$ is the displacement field between the two layers. $\mu \pm D$ basically gives the chemical potential for the two layers separately. There is no hybridization between the two layers such as  the term $b^\dagger_{a\sigma}b_{b\sigma'}$ or $b_{t;\sigma}b_{b;\sigma'}$, which are forbidden by the $SU(2)\times SU(2)$ spin rotation symmetry.

We have self consistent equations:

\begin{align}
T_{ij;a }&=\frac{3}{8} J_a \langle \hat T_{ji;a} \rangle +\frac{1}{2} J_p \langle \hat T_{ji;\bar a } \rangle \notag\\
\Delta_{ij;a}&=\frac{3}{8} J_a  \langle \hat \Delta^\dagger_{ij;a} \rangle \notag\\
\langle n_{i;t} \rangle &=  x \notag\\
\langle n_{i;b} \rangle &= (1-x)
\label{eq:boson_self_consistent_equations}
\end{align}
where $\bar \sigma$ labels the opposite spin of $\sigma$.  $\bar a$ is the opposite layer index of $a$. The last equations are used to solve the chemical potential $\mu$ and the displacement field $D$ at fixed exciton density $x$.

It is known that there are two different possible symmetric ansatz for the Schwinger boson in the spin $1/2$ model: the zero flux ansatz and the $\pi$ flux ansatz\cite{wang2006spin}. They are distinguished by the projective symmetry group (PSG). In our case the Schwinger bosons in the two layers share the same gauge transformation and it is easy to show that the PSG classification is the same as the spin $1/2$ case and we still only have two possible ansatz. The zero-flux ansatz is just a uniform ansatz for $b_{a\sigma}$. The $\pi$ flux ansatz has a $2\times 1$ unit cell and a projective translation symmetry $T_x T_y=-T_y T_x$. If the density of the Schwinger boson $n_a$ is too large, it is known that the ansatz is unstable to single boson condensation $\langle b_{a\sigma} \rangle \neq 0$, resulting magnetic order in the layer $a$.  Because $n_t=x<<1$, we do not need to worry about the top layer, which should be in a paired condensation phase provided $J_t/J_p$ is reasonably large according to the discussion in the previous subsection. Then the main question is whether $b_{b\sigma}$ condenses or not.

 In the $x=0$ limit, it is known that the single Schwinger boson condensation leads to the $120^\circ$ ordered phase with $\mathbf Q=K$ and the stripe ordered phase with $\mathbf Q=M$ respectively from the zero flux and the $\pi$ flux ansatz\cite{wang2006spin}. The zero flux ansatz and the $\pi$ flux ansatz are separated by a first order transition by tuning the next nearest neighbor coupling $\frac{J^\prime_b}{J_b}$ in the $J_b-J^\prime_b$ model\cite{wang2006spin}. Indeed previous numerical studies find the $120^\circ$ and the stripe ordered phases in the small and large $J^\prime_b$ regimes. Our goal is to deplete the Schwinger boson condensation $\langle b_{b\sigma}\rangle$ from $M_0$ to $M=M_0-x$ by reducing the density $n_b$ from $n_b=1$ to $n_b=1-x$. When $x>x_c$ with $x_c=M_0$, we believe the symmetric ansatz such as in Eq.~\ref{eq:mean_field_boson} is stable. In principle we can get $x_c$ by solving self consistent equations in Eq.~\ref{eq:boson_self_consistent_equations}. However, the Schwinger boson self consistent equations do not give quantitatively precise results due to the lack of the quantum fluctuations.  Therefore we mainly use the Schwinger boson mean field ansatz as a guidance and relies on DMRG simulation to decide the phase diagram.

\begin{figure}[ht]
\centering
\includegraphics[width=0.95\linewidth]{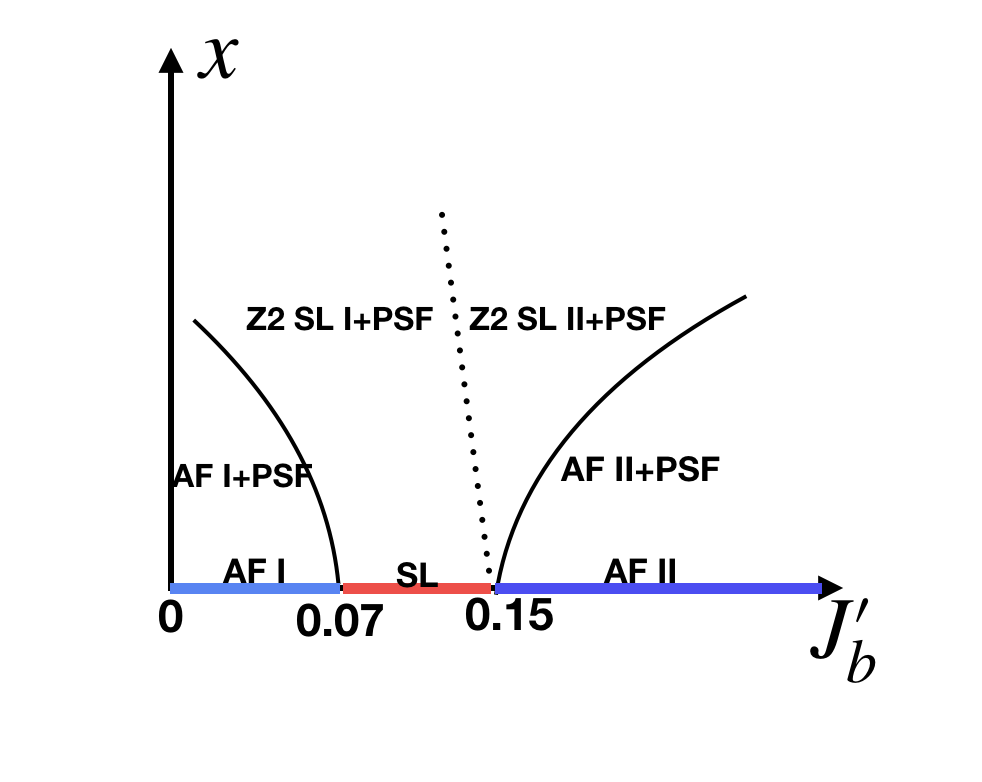}
\caption{Illustrated phase diagram with next nearest neighbor coupling $J'_b$ and the exciton density $x$.  We assume $J_b=1$. In the $x=0$ line, AF I and AF II stand for the 120$^\circ$ ordered phase and the stripe phase respectively\cite{zhu2015spin,hu2015competing}. SL stands for the spin liquid in the spin $1/2$ model studied previously\cite{zhu2015spin,hu2015competing,saadatmand2016symmetry,iqbal2016spin}. We assume that $J_t$ is large enough so that there is always a spin gap in the top layer and we can only obtain a paired superfluid (PSF) when doping excitons. After the ordering moment in the bottom layer is depleted, we naturally obtain two different Z$_2$ spin liquid phases coexisting with paired exciton superfluid (PSF). Z$_2$ SL I and Z$_2$ SL II are described by the zero flux and $\pi$ flux ansatz of Schwinger bosons respectively and the condensation of $b_{b\sigma}$ leads to the AF I and AF II phases nearby. The dashed line is a first order transition separating the two different Z$_2$ spin liquids+ PSF phases. }
\label{fig:phase_diagram_Z2}
\end{figure}

\begin{figure}[ht]
\centering
\includegraphics[width=0.95\linewidth]{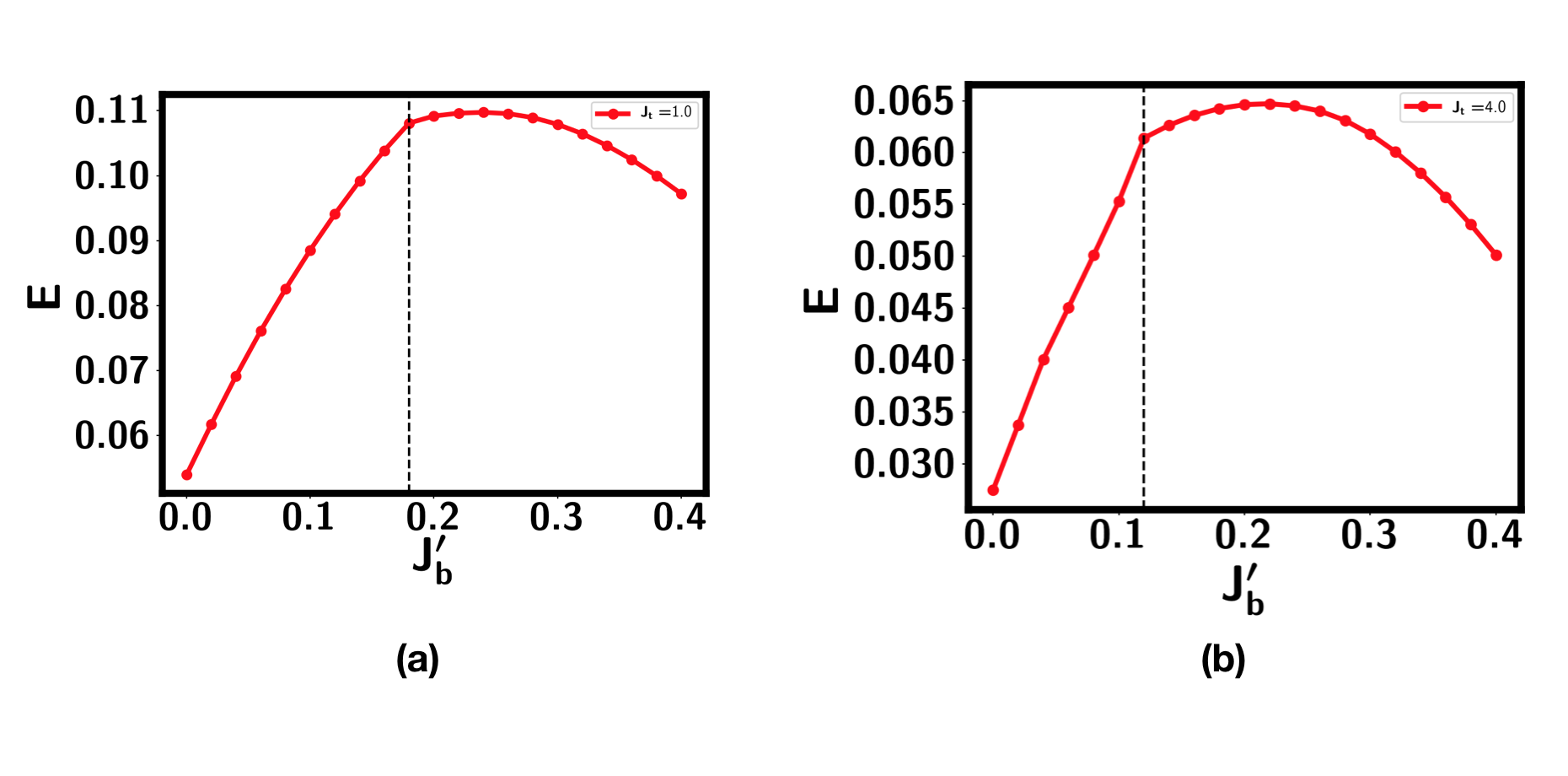}
\caption{Energy $E$ vs $J^\prime_b$ at fixed $J_t$. We use $J_b=J_p=J_{pz}=1$ and $x=\frac{1}{18}$.  The results are from infinite DMRG with bond dimension $m=2000$ for system size $L_y=6$. (a) $J_t=1$. The dashed line is at $J^\prime_b=0.18$. (b) $J_t=4$. The dashed line is at $J^\prime_b=0.12$. }
\label{fig:energy_Z2}
\end{figure}

We assume that there is also a next nearest neighbor hopping $t'$ in the bottom layer, which results in a next nearest neighbor spin spin coupling $J'_b$ entering as $J'_b \sum_{\langle \langle ij \rangle \rangle}\vec{S}_{i;b}\cdot \vec{S}_{j;b}$ as included in Eq.~\ref{eq:spin_layer_model}. In the $x=0$ limit, we can focus just on the $J_b-J'_b$ model for the spin 1/2 moment in the bottom layer. Such a model has been studied before\cite{kaneko2014gapless,li2015quasiclassical,zhu2015spin,hu2015competing,saadatmand2016symmetry,iqbal2016spin,wietek2017chiral,misumi2017mott,gong2017global,jiang2022nature}. It was found that the ground state is in the $120^\circ$ order ($\mathbf Q=K$) in $J'_b/J_b\in [0,0.07]$, a stripy ordered phase with momentum $\mathbf Q=M$ in $J'_b/J_b\in [0.15,\infty)$ and in a spin liquid phase when $J'_b/J_b\in [0.07,0.15]$. When approaching the intermediate spin liquid phase from either magnetic ordered phase, we expect that the ordered moment $M_0$ decreases and vanishes.    Then around the intermediate $J'_b/J_b$, the ordered moment $M_0$ can be easily depleted by a finite $x_c=M_0$, as illustrated in Fig.~\ref{fig:phase_diagram_Z2}. The two magnetic ordered phases are proximate to two different $Z_2$ spin liquids corresponding to zero and $\pi$ flux phase of the Schwinger boson $b_b$ respectively at $x=0$\cite{wang2006spin}. We expect $\langle b_{b\sigma}\rangle=M_0-x$ and $\langle b_{t\sigma}\rangle =0$. When $x>x_c=M_0$, we should have a phase in the class IV of Table.~\ref{table:schwinger_boson_phases}. Given that both $\langle \epsilon_{\sigma \sigma'} b_{t\sigma} b_{t \sigma'} \rangle \neq 0$ and $\langle \epsilon_{\sigma \sigma'} b_{b\sigma} b_{b \sigma'} \rangle \neq 0$, we have the paired exciton condensation $\langle PP^\dagger \rangle \neq 0$, which coexists with the $Z_2$ spin liquids.  There are two such Z$_2$ spin liquids in the small $J^\prime_b$ and large $J^\prime_b$, which are connected by first order transition. With the above argument, we plot an illustrated phase diagram in Fig.~\ref{fig:phase_diagram_Z2}.

\begin{figure*}[ht]
\centering
\includegraphics[width=0.8\linewidth]{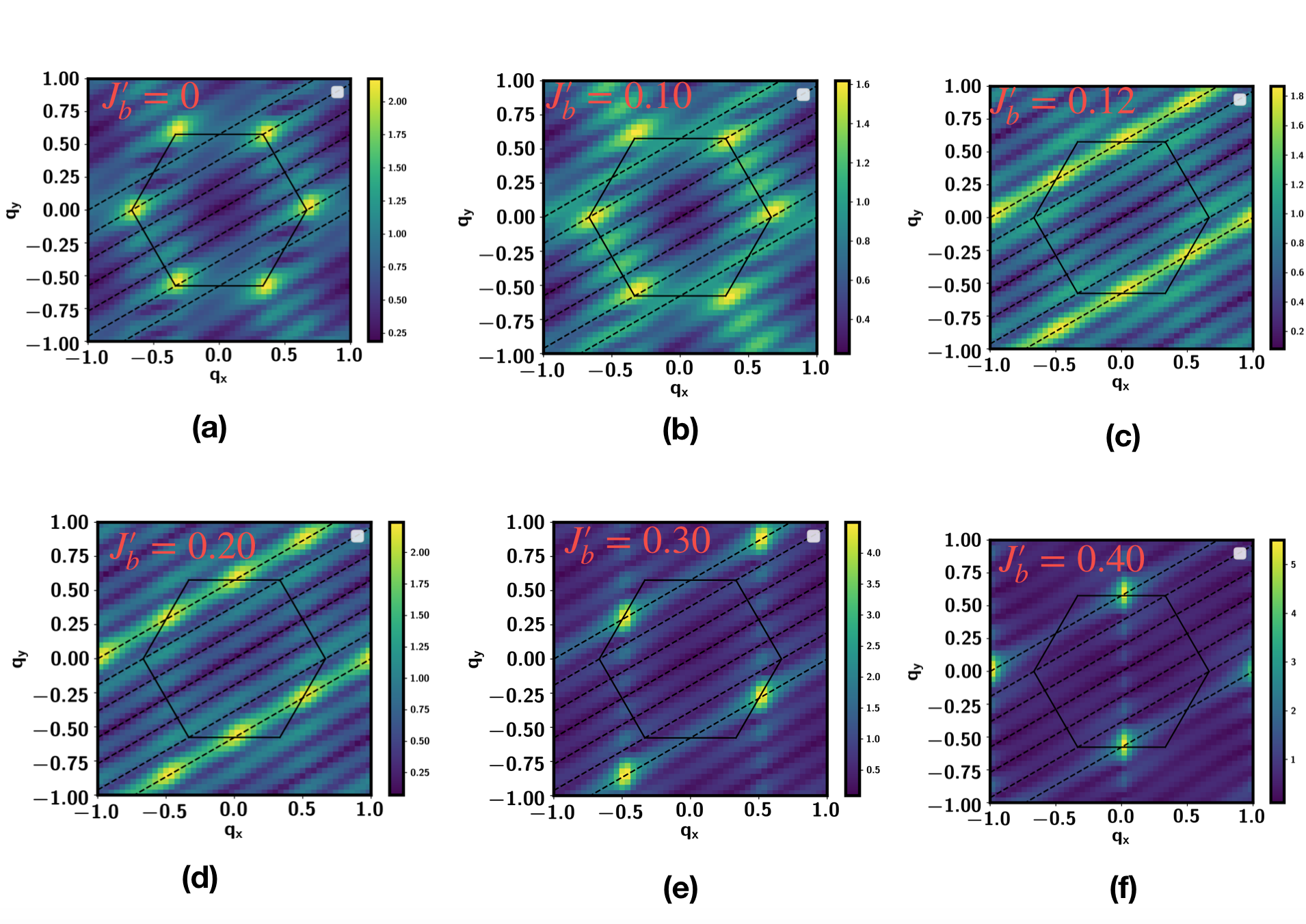}
\caption{Spin structure factor at the bottom layer $\langle \vec{S}_b(\mathbf q) \cdot \vec{S}_b(-\mathbf q)\rangle$ at different values of $J^\prime_b$ at fixed $J_t=4$. We use $J_p=J_{pz}=1$ and $x=\frac{1}{18}$. The results are from infinite DMRG with bond dimension $m=2000$. Around $J^\prime_b=0.12$, there is a clear change of the momentum of the peak from $\mathbf Q=K$ to $\mathbf Q=M$, consistent with a first order phase transition indicated by energy in Fig.~\ref{fig:energy_Z2}(b). }
\label{fig:SbSb_Z2}
\end{figure*}

\begin{figure*}[ht]
\centering
\includegraphics[width=0.8\linewidth]{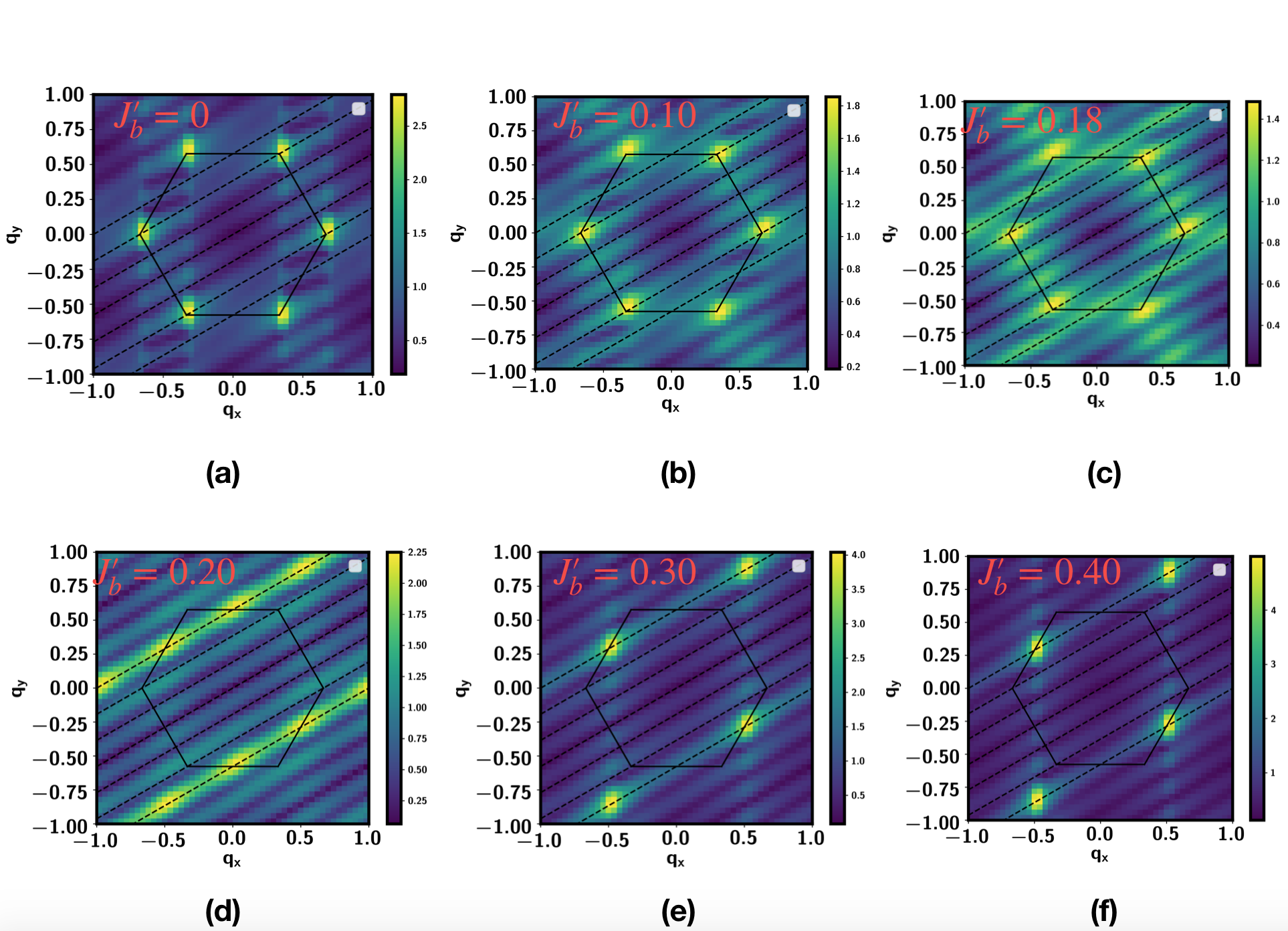}
\caption{Spin structure factor at the bottom layer $\langle \vec{S}_b(\mathbf q) \cdot \vec{S}_b(-\mathbf q)\rangle$ at different values of $J^\prime_b$ at fixed $J_t=1$. We use $J_p=J_{pz}=1$ and $x=\frac{1}{18}$. The results are from infinite DMRG with bond dimension $m=2000$. Around $J^\prime_b=0.18$, there is a clear change of the momentum of the peak from $\mathbf Q=K$ to $\mathbf Q=M$, consistent with a first order phase transition indicated by energy in Fig.~\ref{fig:energy_Z2}(a). }
\label{fig:SbSb_Z2_Jt1}
\end{figure*}

To support the phase diagram, we show DMRG results along the line of changing $J^\prime_b$ at fixed $x=\frac{1}{18}$. First, in Fig.~\ref{fig:energy_Z2} we find that the energy $E$ vs $J^\prime_b$ has a kink consistent with first order transition for both $J_t=1$ and $J_t=4$. The difference between the two sides across this first order transition line is most easily manifested in the spin structure factor of the bottom layer as shown in Fig.~\ref{fig:SbSb_Z2}. At $J_t=4$, when changing $J^\prime_b$ from $0.1$ to $0.12$, momentum of the peak of the structure factor $\langle \vec{S}_b(\mathbf q) \cdot \vec{S}_b(-\mathbf q)\rangle$ suddenly changes from $\mathbf Q=K$ to $\mathbf Q=M$. The same change also happens for $J_t=1$ shown in Fig.~\ref{fig:SbSb_Z2_Jt1}. Therefore it is quite obvious that there are two different phases separated by a first order transition. These two phases origin from the $120^\circ$ order and the stripe order phase at $x=0$, which are proximate to the zero flux and the $\pi$ flux ansatz of the Schwinger boson\cite{wang2006spin}. The main question now is whether the single boson $b_{a\sigma}$ is condensed or not, or equivalently whether the spin gap $\Delta_a$ is finite for both layers.

\begin{figure}[ht]
\centering
\includegraphics[width=0.95\linewidth]{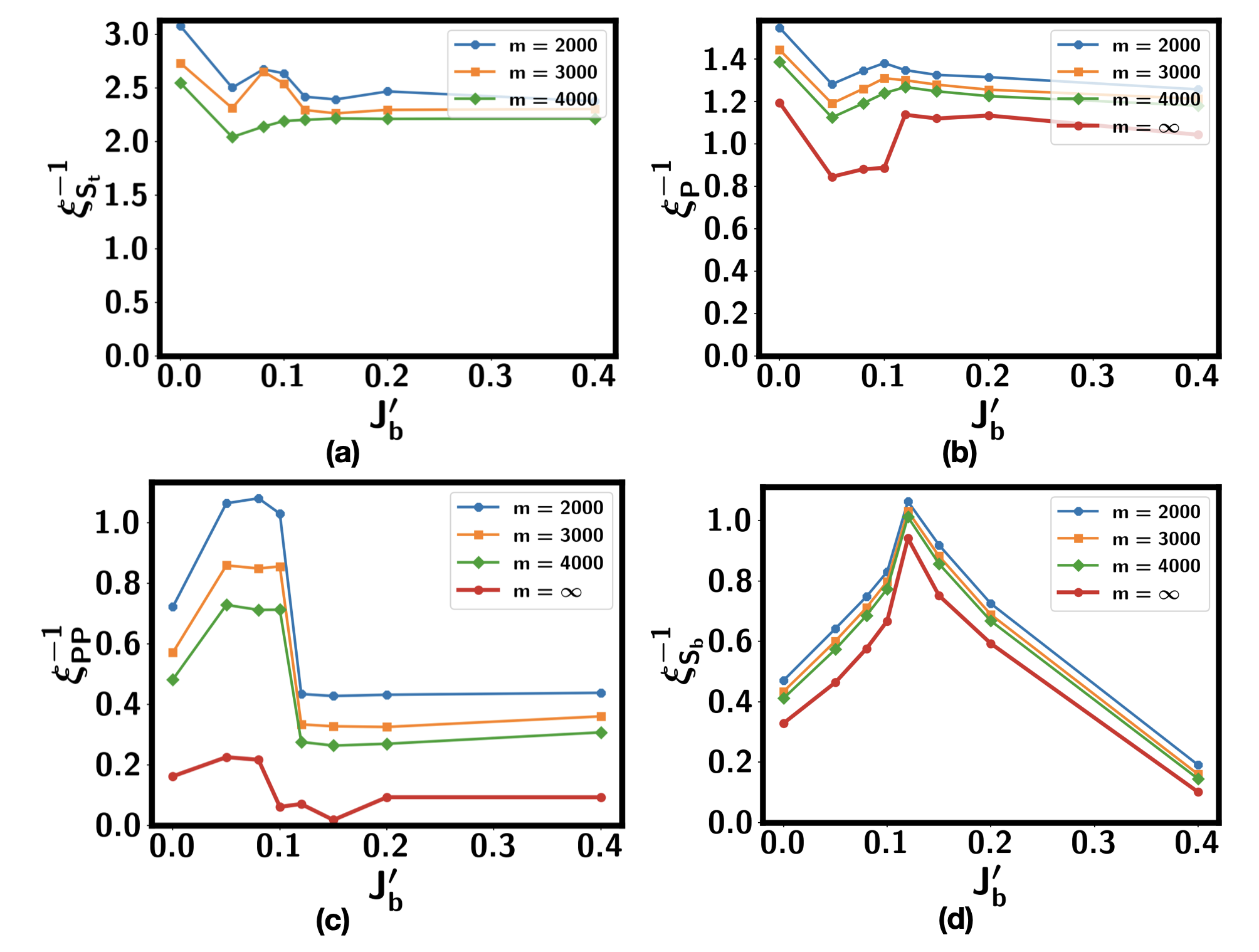}
\caption{Inverse of correlation lengths with the next-nearest neighbor spin-spin coupling $J^\prime_b$ from infinite DMRG with bond dimension $m=2000,3000,4000$. We fix $J_b=J_p=J_{pz}=1$ and $J_t=4$. We use system size $L_y=6$. The exciton density is fixed at $x=\frac{1}{18}$. The correlation length is obtained using the transfer matrix method (see Fig.~\ref{fig:corr_len} and texts around there). The value at $m=\infty$ is extrapolated using the formula $\xi^{-1}(m)=\xi^{-1}(m=\infty)+a (\frac{1}{m})^2+b\frac{1}{m}$. (a) $\xi_{S_t}$ is the correlation length of the spin in the top layer; (b) $\xi_{P}$ is the correlation length for the single exciton; (c) $\xi_{PP}$ is the correlation length of a pair of exciton; (d) $\xi_{S_b}$ is the correlation length for the spin in the bottom layer.}
\label{fig:corr_len_Z2_spin_liquids_m4000}
\end{figure}

We show the inverse of the correlation lengths from different operators in Fig.~\ref{fig:corr_len_Z2_spin_liquids_m4000}, using $J_b=J_p=J_{pz}=1$ and $J_t=4$. In DMRG the correlation length $\xi$ is always finite for a finite bond dimension $m$, but the scaling of the inverse of the correlation length with $\frac{1}{m}$ can provide the information on whether $\frac{1}{\xi}$ is finite for a given sector specified by the quantum numbers. We find that the correlation lengths for both $\vec S_t$ and the single exciton $P^\dagger$ are quite small with $\frac{1}{\xi}>1$ as shown in Fig.~\ref{fig:corr_len_Z2_spin_liquids_m4000}(a)(b). Meanwhile, in Fig.~\ref{fig:corr_len_Z2_spin_liquids_m4000}(c) we show that $\frac{1}{\xi}$ for the paired exciton decreases rapidly with increasing the bond dimension $m$ and the extrapolated value $\frac{1}{\xi}$ at $m=\infty$ is smaller than $0.2$ (almost $0$ at some value of $J^\prime_b$). We believe the finite value of $\frac{1}{\xi_{PP}}$ is a numerical artifact and $\frac{1}{\xi_{PP}}=0$ in the whole range if we really take the bond dimension to infinite. This means that the phase has a paired exciton condensation order, consistent with the expectation of the paired superfluid (PSF) order in Fig.~\ref{fig:phase_diagram_Z2}. The only remaining problem is whether it is in the AF+PSF phase or in the Z$_2$ SL+PSF phase illustrated in Fig.~\ref{fig:phase_diagram_Z2}.

\begin{figure*}[ht]
\centering
\includegraphics[width=0.8\linewidth]{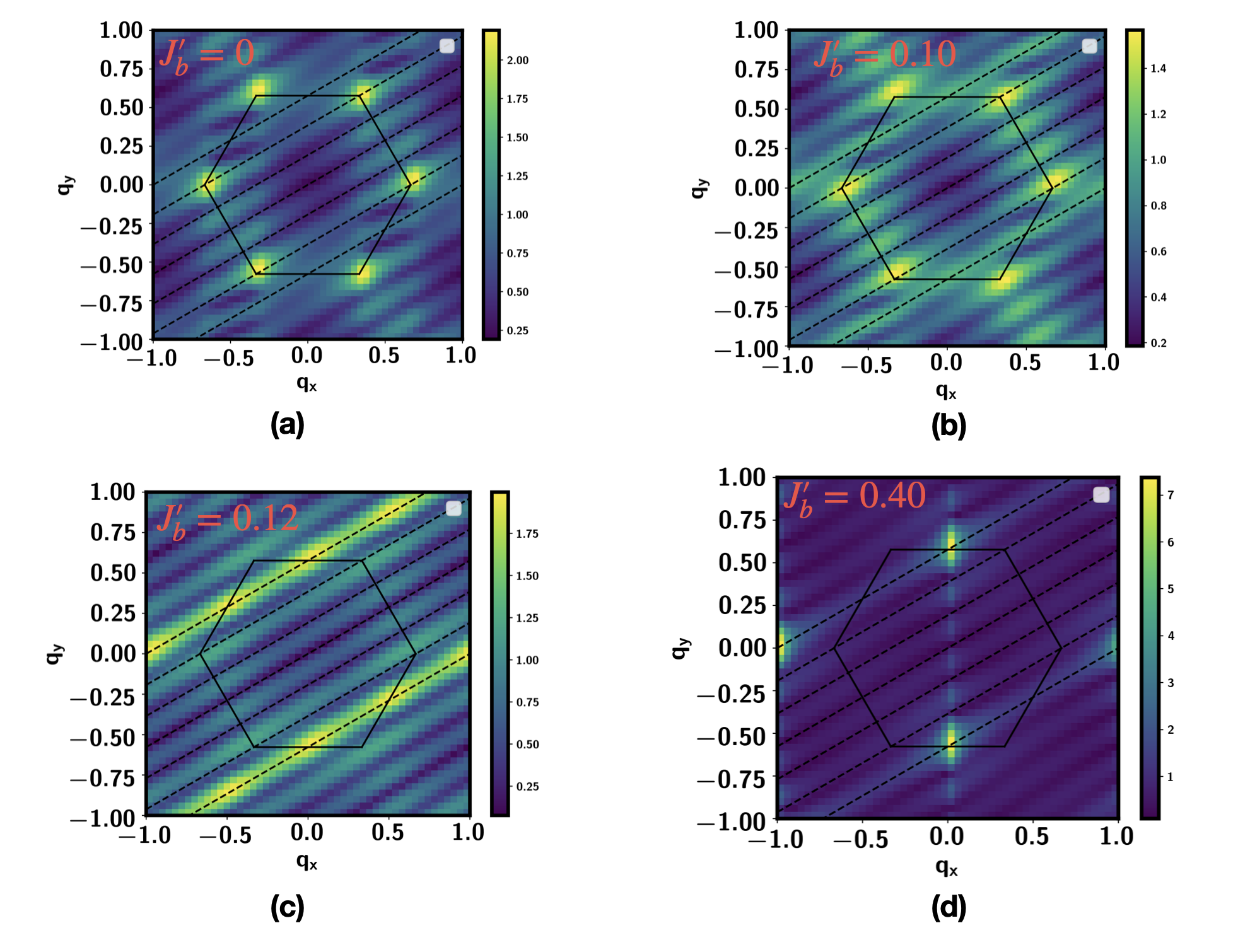}
\caption{$\langle \vec S_b(\mathbf q) \cdot \vec S_b(-\mathbf q)\rangle$ at various values of $J^\prime_b$. The parameters are the same as in Fig.~\ref{fig:corr_len_Z2_spin_liquids_m4000} and we use bond dimension $m=4000$.}
\label{fig:Sb_Sb_m=4000}
\end{figure*}

In Fig.~\ref{fig:corr_len_Z2_spin_liquids_m4000}(d) we show the inverse of the correlation length of the spin in the bottom layer.  We find that $\frac{1}{\xi_{S_b}}$ increases with $J^\prime_b$ and is maximized at the first order critical point at $J^\prime_b\approx 0.12$. $\frac{1}{\xi_{S_t}}$ extrapolated to $m=\infty$ at $J^\prime_b=0.10$ and $J^\prime_b=0.12$ are as large as $0.6$ and $0.9$ respectively, suggesting short correlation length ($\xi_{S_b}<2$). In Fig.~\ref{fig:Sb_Sb_m=4000} we show the spin-spin structure factor in the bottom layer for $J^\prime_b=0$, $0.1$, $0.12$, $0.4$. Obviously $J^\prime_b=0.1$ and $J^\prime_b=0.12$ are separated by the first order transition denoted as the dashed line in Fig.~\ref{fig:phase_diagram_Z2}. $\langle \vec S_b(\mathbf q) \cdot \vec S_b(-\mathbf q)\rangle$ shows peaks at $\mathbf Q=K$ and $\mathbf Q=M$ for $J^\prime_b=0.10$ and $J^\prime_b=0.12$ respectively, but the peaks are blurred compared to $J^\prime_b=0.0$ and $J^\prime_b=0.4$, consistent with the rapid increase of $\frac{1}{\xi_{S_b}}$ towards the intermediate regime. These results strongly suggest that there is a spin gap $\Delta_b>0$ at least at $J^\prime_b=0.10$ and $J^\prime_b=0.12$ and these two points have only short ranged antiferromagnetic correlations and belong to the Z$_2$ SL I + PSF phase and the Z$_2$ II + PSF phase in Fig.~\ref{fig:phase_diagram_Z2}.  This is also consistent with our theoretical expectation that the magnetic order moment $M=M_0-x$ must vanish when the doping density $x$ is finite given that $M_0$ should approach zero towards the intermediate regime. Note that for $J^\prime_b=0.12$ the peaks of the spin structure factor are at only two of the three M points, which breaks the $C_3$ rotation symmetry. However, this may be because our system in DMRG has a small $L_y=6$ and breaks the $C_3$ symmetry explicitly. A larger size calculation is needed to determine whether this is a nematic phase or a $C_3$ symmetric phase in the $L_y\rightarrow \infty$.

In our DMRG calculation it is not easy to decide the phase boundary between AF + PSF phase and Z$_2$ SL + PSF phase because it is always a crossover at finite bond dimension. For the particular parameter in Fig.~\ref{fig:corr_len_Z2_spin_liquids_m4000}, $\frac{1}{\xi_{S_b}}(m\rightarrow \infty)$ becomes as small as $0.1$ at $J^\prime_b=0.40$, but remains larger than $0.3$ at $J^\prime_b=0$. Thus we conjecture that $\Delta_b=0$ at $J^\prime_b=0.4$, but $\Delta_b>0$ even at $J^\prime_b=0$. This means that the AF I + PSF phase is absent for this parameter and we are in the Z$_2$ SL I + PSF phase even with $J^\prime_b=0$.  However, we note that it is tricky to distinguish $\Delta_b>0$ and $\Delta_b=0$ when $\Delta_b$ is small. Therefore we leave it to future to determine the precise boundary between Z$_2$ SL and AF phase using larger bond dimension and system size.  In Appendix.~\ref{appendix:fractional_SF} we show more results to demonstrate that the fractional superfluid phase (Z$_2$ SL + PSF) also exist with smaller value of $J_t=1$ or $J_t=2$ and with other parameters such as $J_p=2, J_{pz}=5$. Especially it survives a large $J_{pz}$ term, which is the repulsion between the excitons and does not matter much at the small $x$ regime.

Lastly we want to comment on the evolution from the spin liquid in the $x=0$ line to the Z$_2$ SL I + PSF phase in the Fig.~\ref{fig:phase_diagram_Z2}. Right now there are debates on whether this spin liquid in the intermediate $\frac{J^\prime_b}{J_b}$ at the $x=0$ limit is in a Z$_2$ spin liquid\cite{zhu2015spin,hu2015competing,saadatmand2016symmetry,jiang2022nature} or in a U(1) Dirac spin liquid\cite{iqbal2016spin,hu2019dirac}. If it is a Z$_2$ spin liquid, then it should be described by the zero flux ansatz of the Schwinger boson theory because it has short ranged $120^\circ$ correlations. In this case, doping the exciton induces a chemical potential tuned insulator to superfluid phase transition for the paired excitons on top of the Z2 spin liquid which leads to the  Z$_2$ SL I + PSF phase found in our numerical calculation. If the layer polarized spin liquid is a U(1) Dirac spin liquid, then we should use fermionic spinons such as in Eq.~\ref{eq:Schwinger_fermion} to describe the spin liquid phase. Then doping excitons will create spinon Fermi surfaces for both layers. But there is no signature of spinon Fermi surface in our numerical results at $x=\frac{1}{18}$. To reach the Z$_2$ SL I + PSF phase, we need the fermion spinons to further have pairing terms $\epsilon_{\sigma \sigma'} f_{a;\sigma}f_{a;\sigma'}$ for both layer $a=t,b$, which higgses the U(1) gauge field down to Z$_2$.  Because this is a $\pi$ flux ansatz of fermionic spinon, it is equivalent to the zero flux phase of the Schwinger boson\footnote{In this $Z_2$ spin liquid, the bosonic $e$ particle is a bound state of fermionic spinon and the vison. The vison feels a $\pi$ flux and hence the $\pi$ flux of fermionic spinon corresponds to the zero-flux of the bosonic spinon.} and is in the same Z2 SL I phase.  Thus we see that there need to be two phase transitions starting from the U(1) Dirac spin liquid phase to reach the Z$_2$ SL I +PSF phase.  Future calculations at infinitesimal $x$ regime may be able to resolve the debates between the Z$_2$ spin liquid and Dirac spin liquid in the $x=0$ limit.

Experimentally, in moir\'e + monolayer system, we expect $J_t$ to be larger than $J_b$ and $J_p$ because the top layer is less correlated. Then according to our analysis above it is easy to reach the paired superfluid phase with a spin gap $\Delta_t>0$ in the top layer. As long as there is a small $J^\prime_b\sim 0.1 J_b$ in the bottom layer, we can realize either the Z$_2$ SL I + PSF phase or the Z$_2$ SL II + PSF phases. These phases have spin gapes in both layers and have preformed cooper pairs, the same as the resonating-valence-bond (RVB) state first proposed for high Tc cuprates\cite{anderson1987resonating}. Therefore it is very interesting to search for possible superconductivity by doping the bilayer Mott insulator with the $Z_2$ spin liquids orders.

\section{Neutral Fermi surface formed by fermionic excitons\label{sec:neutral_fermi_surface} }

In Sec.~\ref{sec:boson_fermion_parton_construction} we argue that the doped exciton can be either bosonic or fermionic due to possible fractionalization. In Sec.~\ref{sec:bosonic_exciton_superfluids} we did a survey of various possible superfluid phases built with bosonic excitons, which are either local excitations on top of a magnetic order or fractional excitations coupled to a deconfined $Z_2$ gauge field. In this section, we turn to the more exotic possibility of fermionic excitons. In this case, the fermionic exciton obviously can not be local excitations and necessarily couples to a deconfined gauge field.

As shown in Sec.~\ref{sec:boson_fermion_parton_construction}, the spin operator $\vec{S}$ and exciton operator $\vec P$ can be expressed either with bosonic partons $b_{a\sigma}$ or fermionic parton $f_{a\sigma}$. When the layer polarized Mott insulator is in the magnetically ordered phase, the natural parton description should be the 
Schwinger boson theory shown in Sec.~\ref{sec:bosonic_exciton_superfluids}. When $t/U$ is in the intermediate regime,  the layer polarized Mott insulator may be in a spinon Fermi surface state\cite{motrunich2005variational,senthil2008theory}. The exact nature of the intermediate weak Mott insulator of the spin 1/2 Hubbard model is still under debate and may depend on details like whether further neighbor hoppings are included or not.  In the experimental study of the metal-insulator transition in the AA stacked MoTe$_2$/WSe$_2$ moir\'e system\cite{li2021continuous}, no magnetic order is found down to the lowest temperature.  The spin susceptibility in the weak Mott regime seems to be just like the metallic phase, suggesting a possible spinon Fermi surface ground state. However, such a neutral spinon Fermi surface is hard to detect even if it already exists in the MoTe$_2$/Wse$_2$ moir\'e system. In this section we propose to study a MoTe$_2$-hBN-MoTe$_2$/WSe$_2$ system similar to the moir\'e+monolayer setting up in Ref.~\onlinecite{gu2021dipolar,zhang2021correlated}. If the layer polarized Mott insulator indeed hosts a spinon Fermi surface, then at finite exciton density $x>0$, the most natural phase is a $U(1)$ spin liquid with spinon Fermi surfaces in both layers. Especially the neutral fermion in the top layer can be viewed as a fermionic exciton formed by electron $c_{t\sigma}^\dagger$ in the top layer bound to the holon $\varphi$ in the bottom layer, as illustrated in Fig.~\ref{fig:exciton_fractionalization}. Possibility of fermionic excitons has also been discussed previously in Kondo insulator SmB$_6$\cite{chowdhury2018mixed} or in bilayer Landau levels\cite{barkeshli2018topological,zaletel2018evidence}. In our realization the neutral fermion carries a charge under a $U(1)$ symmetry generated by $P_z$ (the same is true for the Fermionic exciton in the bilayer quantum Hall system\cite{barkeshli2018topological,zaletel2018evidence}). This makes it possible to detect the movement of the neutral fermionic excitons through counter-flow transport.

To favor a U(1) spin liquid with spinon Fermi surface state, it is necessary to include higher order ring exchange terms beyond our simple model in Eq.~\ref{eq:spin_layer_model}. The task to establish the spinon Fermi surface state as a ground state of a microscopic lattice model is very challenging and we leave it to future work. Here our focus is to explore the fate under doping excitons while assuming the $x=0$ limit indeed hosts a spinon Fermi surface.  As said before, this assumption is encouraged by the recent experiment\cite{li2021continuous}. But it is hard to detect a neutral Fermi surface even if it already exists.  The purpose of this section is to demonstrate that doping excitons can lead to smoking gun evidences of neutral Fermi surface from electric measurements.

Under the assumption that the $x=0$ limit has a spinon Fermi surface in the bottom layer, we should use the Abrikosov fermion partons $f_{t\sigma}$ and $f_{b\sigma}$ to represent the spin operators, as introduced in Eq.~\ref{eq:Schwinger_fermion}. A typical mean field theory in terms of these fermionic partons is in the form:

\begin{align}
	H_M&=- \sum_{ij}t_{b;ij}f^\dagger_{i;b\sigma}f_{j;b\sigma}-(\mu-D)\sum_i f^\dagger_{i;b\sigma}f_{i;b\sigma}\notag\\
	&~~~- \sum_{ij}t_{t;ij}f^\dagger_{i;t\sigma}f_{j;t\sigma}-(\mu+D)\sum_i f^\dagger_{i;t\sigma}f_{i;t\sigma}
	\label{eq:mean_field_fermionic_spinon}
\end{align}
where $\mu$ is the chemical potential to fix $n_{t}+n_b=1$ and $D$ is the displacement field.

\begin{figure}[ht]
\centering
\includegraphics[width=0.95\linewidth]{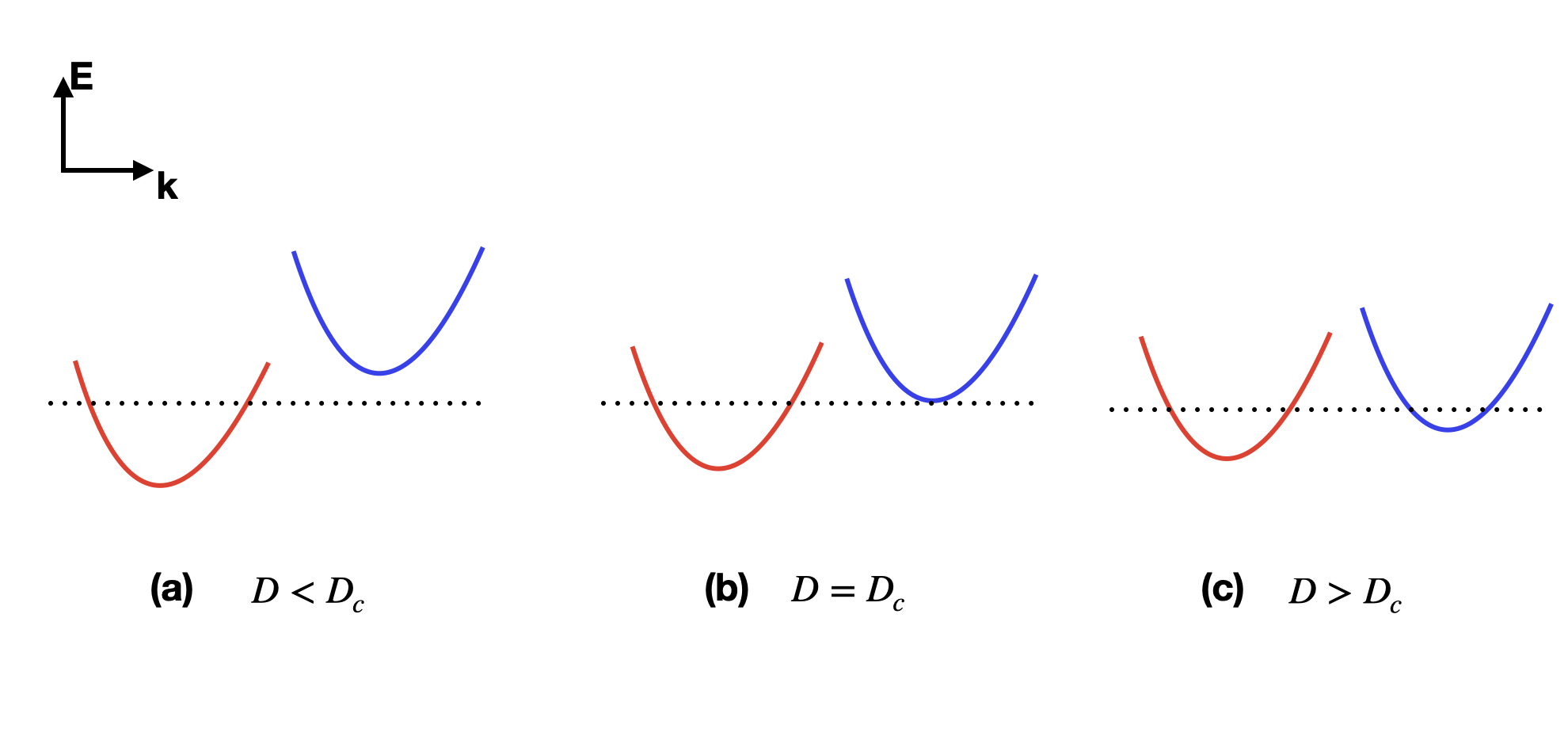}
\caption{Illustration for the displacement field $D$ tuned transition.  The red line is the dispersion of the spinon in the bottom layer and the blue is the dispersion of the spinon in the top layer. The dashed line is the chemical potential as defined in Eq.~\ref{eq:mean_field_fermionic_spinon}. (a) A layer polarized Mott insulator with spinon Fermi surface in the bottom layer. $\frac{\partial P_z}{\partial_D}=0$ and counterflow transport is insulating. (b) At critical $D_c$, the gap of the exciton is now zero.  (c) A $U(1)$ spin liquid phase with spinon Fermi surfaces in both layers. $\frac{\partial P_z}{\partial D}$ is finite and the counterflow transport is now metallic.}
\label{fig:spinon_FS_dispersion_transition}
\end{figure}

When $D\rightarrow-\infty$, the system is in a layer polarized Mott insulator with spinon Fermi surface only in the bottom layer, as illustrated in Fig.~\ref{fig:spinon_FS_dispersion_transition}(a). Then when increasing $D$, there is a Lifshitz transition at $D_c$ (see Fig.~\ref{fig:spinon_FS_dispersion_transition}(b)), after which we have spinon Fermi surfaces in both layers, illustrated in Fig.~\ref{fig:spinon_FS_dispersion_transition}(c).

A low energy theory can be written down to describe this spinon Fermi surface phase which needs to couple to an internal gauge field:

\onecolumngrid

\begin{align}
L&=\psi^\dagger_{t;\sigma}(\tau,x)\big(\partial_\tau-(\mu+D)-ia_0(\tau,x)-\frac{1}{2}iA^s_0(\tau,x)\big)\psi_{t;\sigma}(\tau,x)-\frac{\hbar^2}{2m_t}\psi^\dagger_{t;\sigma}(\tau,x)\big(-i\vec \partial-\vec a(\tau,x)-\frac{1}{2}\vec{A^s}(\tau,x)\big)^2\psi_{t;\sigma}(\tau,x) \notag\\
&~+\psi^\dagger_{b;\sigma}(\tau,x)\big(\partial_\tau-(\mu-D)-ia_0(\tau,x)+\frac{1}{2}iA^s_0(\tau_x)\big)\psi_{b;\sigma}(\tau,x)-\frac{\hbar^2}{2m_b}\psi^\dagger_{b;\sigma}(\tau,x)\big(-i\vec \partial-\vec a(\tau,x)+\frac{1}{2}\vec{A^s}(\tau,x)\big)^2\psi_{b;\sigma}(\tau,x)
\label{eq:spinon_fs_gauge_field_action}
\end{align}

\twocolumngrid
where $\psi_{a;\sigma}(\tau,x)$ is the fermionic field in the layer $a=t,b$. $a_{\mu}(\tau,x)$ is the internal U(1) gauge field associated with the gauge symmetry: $f_{i;a\sigma}\rightarrow f_{i;a\sigma}e^{i \alpha_i}$ for $a=t,b$. $A^s_\mu(\tau,x)$ is a probing field for the global U(1) symmetry associated with $P_z$. Physically its electric field $\vec{E}_s=-\vec{\partial}A^s_0+\frac{\partial \vec{A}^s}{\partial t}$ corresponds to a counter-flow electric field configuration: $\vec{E}_t=-\vec{E}_b$ in the two layers. The probing field $A^s_\mu$ minimally couples to the counter-flow current: $-J^\mu_s A^s_\mu$. Here $J^s_0=P_z$ is the layer polarization and $\vec{J}^s$ is its current. We do not include the coupling to $A^c_\mu$ because it can always be absorbed to the coupling to $a_\mu$ by redefinition $a_\mu \rightarrow a_\mu-A^c_\mu$. This is another way to understand that the above theory describes a Mott insulator without response to $A^c_\mu$.

In the following we provide two clear experimental signatures to detect this exotic phase with neutral Fermi surfaces in both layers.

\textbf{Counter-flow resistivity} We first show that the counter-flow transport is insulating when $D<D_c$ and metallic when $D>D_c$. Thus the displacement field tunes an counter-flow insulator to metal transition. The counter-flow resistivity is defined as $\rho_{s;xx}=\frac{E^s_x}{J^s_x}$, where $\vec E=\vec E^t-\vec E^b$ and $\vec J^s=\frac{1}{2}(\vec J^t-\vec J^b)$. As shown in Eq.~\ref{eq:ioffe_larkin_rule_resistivity}, the counter-flow resistivity $\rho_{s}=\rho_t+\rho_b$, where $\rho_t,\rho_b$ are the resistivities of the fermionic partons $f_{t\sigma}$ and $f_{b\sigma}$. When $D<D_c$, $\rho_t=\infty$ and hence $\rho_s=\infty$. When $D>D_c$, we expect that $\rho_t=A_t T^{\alpha}$ and $\rho_b=A_b T^{\alpha}$. Therefore the counter-flow resistivity is metallic $\rho_s(T)=T^{\alpha}$. The exponent $\alpha$ on the temperature dependence is likely deviating from the Fermi liquid result $\alpha=2$ due to the coupling to the gauge field $a_\mu$\cite{lee1992gauge,maslov2011resistivity,hartnoll2014transport,lee2018recent,lee2021low}, though the exact value of $\alpha$ is still not well established theoretically.  We will not try to solve this problem in this paper. Instead, we propose to measure the exponent $\alpha$ through counter-flow transport in experiments.

In addition to a metallic counter-flow transport, the spinon Fermi surface state when $D>D_c$ also has a metallic spin susceptibility and inter-layer polarizability.  As shown in Eq.~\ref{eq:ioffe_larkin_rule_compressibility}, the inter-layer polarizability $\kappa_s=(\frac{1}{\kappa_t}+\frac{1}{\kappa_b})^{-1}$, where $\kappa_t$ and $\kappa_b$ are the compressibilities of the fermionic partons in the two layers. When $D>D_c$, $\kappa_a$ is a constant for $a=t,b$. Therefore the inter-layer polarizability $\kappa_s$ is also a constant just like a Fermi liquid. Note that the fermionic parton $f_{a;\sigma}$ acquires a non-Fermi-liquid self energy $\Sigma(i\omega)\propto |\omega|^{\frac{2}{3}}$\cite{lee1992gauge} from coupling to the gauge field $a_\mu$, but it does not enter the spin susceptibility because of the cancellation with the vertex correction\cite{kim1994gauge}. Thus the spin susceptibility (or inter-layer polarizability) behaves as a Fermi liquid. We need to emphasize that an exciton superfluid as discussed in Sec.~\ref{sec:bosonic_exciton_superfluids} also has constant inter-layer polarizability. Then a constant inter-layer polarizability does not distinguish the superfluid and neutral Fermi surface phase. To better characterize these two different kind of phases, we really need counter-flow transport, which is either superfluid or metallic in the exciton condensation and neutral Fermi surface phase.

\textbf{Friedel oscillation in layer polarization} It is known that the existence of a Fermi surface can lead to Friedel oscillation because of the $2k_F$ scattering. In a Mott insulator with only spinon Fermi surface, the dominant Friedel oscillation is in the magnetization and hence is not easy to detect using the electric probe such as Scanning Tunneling Microscope (STM). An electric signal may be found in the weak Mott regime where the holon density is not exactly frozen to be one per site\cite{mross2011charge}, but the amplitude is suppressed if the charge gap is large.

In our case at $D>D_c$, there are active neutral fermions $f_{a;\sigma}$ in both layers and the fermionic parton carries not only spin $\vec S_a$ but also the layer polarization $P_z$. Because $P_z(\mathbf r)=\frac{1}{2} \big(n_t(\mathbf r)-n_b(\mathbf r)\big)$, its correlation function manifests $2k_F$ singularity of spinon Fermi surfaces in both layers. Therefore, there will also be Friedel oscillations in terms of the layer polarization:

\begin{equation}
    \langle P_z(\mathbf r) P_z(0) \rangle=\sum_{\mathbf Q=2\mathbf{k_F^t}} \frac{A_t}{|\mathbf r|^{\alpha_t(\mathbf Q)}} e^{i \mathbf Q \cdot \mathbf r}+\sum_{\mathbf Q=2\mathbf{k_F^b}} \frac{A_b}{|\mathbf r|^{\alpha_b(\mathbf Q)}} e^{i \mathbf Q \cdot \mathbf r}
\end{equation}
where $2\mathbf{k_F^a}$ is a vector connection two points in the Fermi surface in layer $a=t,b$.

We can calculate the Fourier transformation of $\langle P_z(\mathbf r)P_z(0)\rangle$ in the mean field level:

\begin{align}
&\chi_{zz}(\mathbf q)=\langle P_z(q_0=0,\mathbf q) P_z(-q_0=0,-\mathbf q) \rangle \notag \\  
&=-\frac{1}{4} \sum_{a\sigma}\sum_{\omega}G_{a\sigma}(i\omega,\mathbf k+\mathbf q) G_{a\sigma}(i\omega,\mathbf k) \notag \\
&=\frac{1}{2} \sum_{a=t,b}\sum_{\mathbf k} \frac{f(\xi_a(\mathbf k))-f(\xi_a(\mathbf k+\mathbf q))}{\xi_a(\mathbf k+\mathbf q)-\xi_a(\mathbf k)}
\end{align}
where $f(\xi(\mathbf k))=\theta(-\xi(\mathbf k))$ is the Fermi-Dirac distribution at $T=0$.

\begin{figure}[ht]
\centering
\includegraphics[width=0.95\linewidth]{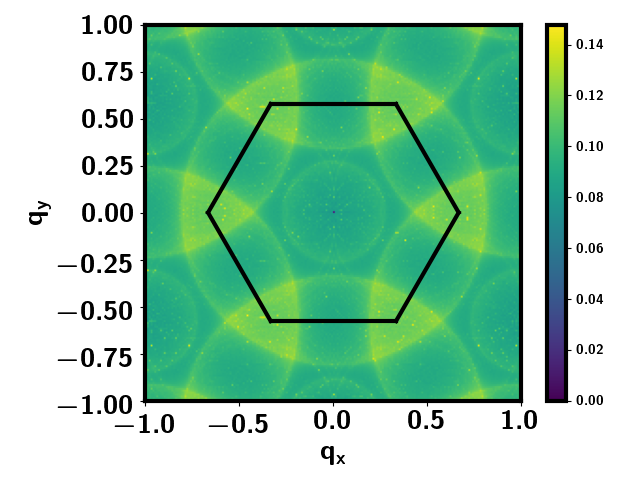}
\caption{ The Friedel oscillation in terms of the layer polarization. The color plot shows $\chi_{zz}(\mathbf q)=\langle P_z(q_0=0,\mathbf q)P_z(q_0=0,-\mathbf q)\rangle$ at the mean field level without gauge fluctuation. $q$ is in units of $2\pi$. The solid black line is the boundary of the first Brillouin zone (BZ).  We have used $t_t=t_b=1$ in Eq.~\ref{eq:mean_field_fermionic_spinon}. There are two circles corresponding to $2k_{F;b}$ and $2k_{F;t}$, which are fixed to be $\frac{\pi k_{F;t}^2}{4\pi^2}=\frac{x}{2}$ and $\frac{\pi k_{F;b}^{2}}{4\pi^2}=\frac{1-x}{2}$.}
\label{fig:Pz_Pz_friedel}
\end{figure}

We show the plot of $\chi_{zz}(\mathbf q)$ in Fig.~\ref{fig:Pz_Pz_friedel}. As expected, there are features along two circles centering at the $\Gamma$ point.  They correspond to $2k_{F;t}$ and $2k_{F;b}$. The one from $2k_{F;b}$ is outside the Brillouin zone (BZ) and needs to be folded back to the first BZ. We have ignored the gauge fluctuation in this calculation, which can further enhance the Friedel oscillations\cite{mross2010controlled}.

 The Friedel oscillation of the layer polarization $P_z$ may be detected in the following way: one can apply a local electric field in z direction $E_z(\mathbf r_0)$, then we expect that $\langle P_z(\mathbf r) \rangle\sim \sum_{\mathbf Q=2\mathbf{k_F^t}} \frac{A_t}{|\mathbf r-\mathbf r_0|^{\alpha_t(\mathbf Q)}} e^{i \mathbf Q \cdot (\mathbf r-\mathbf r_0)}+\sum_{\mathbf Q=2\mathbf{k_F^b}} \frac{A_b}{|\mathbf r-\mathbf r_0|^{\alpha_b(\mathbf Q)}} e^{i \mathbf Q \cdot (\mathbf r-\mathbf r_0)}$. The local profile of layer polarization $P_z(\mathbf r)$ will further induce electric field profile $E_z(\mathbf r)\sim P_z(\mathbf r)$, which may be detected by electric probes. The Fourier transformation of $E_z(\mathbf r-\mathbf r_0)$ should be similar to the plot in Fig.~\ref{fig:Pz_Pz_friedel}.

In addition to the counter-flow transport and the Friedel oscillation, the spinon Fermi surface state also has  $C/T \sim T^{-\frac{1}{3}}$\cite{lee1992gauge} where $C$ is the specific heat. Meanwhile the neutral Fermi surface may show quantum oscillations under external magnetic field due to coupling between the internal gauge flux $b=\nabla \times \vec a$ and the external magnetic field\cite{motrunich2006orbital,chowdhury2018mixed,sodemann2018quantum}. But there is a large g factor in TMD material and whether spinon Fermi surface state can survive under a finite field is not clear. In this sense the electric probes such as counter-flow or Friedel oscillation at zero magnetic field may be a better detection scheme.

In the above we start from the assumption that the layer polarized Mott insulator at the weak Mott regime is a spinon Fermi surface state and then argue that the small $x$ regime naturally hosts neutral Fermi surfaces in both layers. Whether the assumption is correct is a hard problem both theoretically and experimentally. Although there are suggestive signatures of constant spin susceptibility at the $x=0$ limit in the MoTe$_2$/WSe$_2$ system\cite{li2021continuous}, it is hard to prove or rule out the possibility of spinon Fermi surface in the $x=0$ sample because of the lack of the probe of neutral spin excitations.  Our proposal at finite $x$  in the MoTe$_2$-hBN-MoTe$_2$/WSe$_2$ system is thus also a good test of the nature of weak Mott insulator at the $x=0$ limit.

\subsection{Numerical results in one dimension}
In the above we discussed the experimental signatures of the spinon Fermi surface state, but a microscopic calculation of a 2D lattice model is avoided due to its technical challenge.  Here we provide numerical simulation for analog of the spinon Fermi surface state in the one dimension chain. We will show the 1D model has qualitatively the same physics as discussed in the above. Note that the existence of exotic spinon Fermi surface-like phase in one dimension is very special and can not be easily generalized to two dimension.  Indeed, we already showed in previous sections that the simple model in Eq.~\ref{eq:spin_layer_model} in two dimension hosts either single exciton or paired exciton superfluid and we never found a neutral Fermi surface. To favor a spinon Fermi surface in two dimension, more complicated higher order ring exchange terms must be included and it is still not clear when a spinon Fermi surface is a grond state.  In this paper we avoid this challenging energetical problem. The purpose of the 1D calculation is to demonstrate the behavior we expect when doping excitons into a spinon Fermi surface Mott insulator at the x=0 limit. We leave it to future work to  understand when and why  a spinon Fermi surface state can be stabilized in a two dimension model. As said in the introduction of this section, there is already encouraging experimental evidence for such a state at the $x=0$ limit in the MoTe$_2$/WSe$_2$ system\cite{li2021continuous} and we hope our calculation of the 1D model below can provide some insights on the fate of doping excitons into the weak Mott insulator in the MoTe$_2$/WSe$_2$ system. 

In one dimension, the $x=0$ limit of Eq.~\ref{eq:spin_layer_model} is a spin $1/2$ chain and is in a gapless phase without magnetic order. Such a phase described by SU(2)$_1$ conformal field theory (CFT) is known to be an analog of the spinon Fermi surface in higher dimension. Actually a good model wavefunction for the state is a Gutzwiller projection of Fermi sea\cite{haldane1988exact}, the same as the spinon Fermi surface state in higher dimension.  According to our analysis above, we should expect the phase at finite x to host spinon Fermi surfaces in both layers. Such a phase has a metallic counter-flow conductance and also show $2k_F$ singularities in spin-spin correlation and dipole-dipole correlation. We will demonstrate this by DMRG simulation of the model in Eq.~\ref{eq:spin_layer_model} in one dimension.

\begin{figure*}[ht]
\centering
\includegraphics[width=0.8\linewidth]{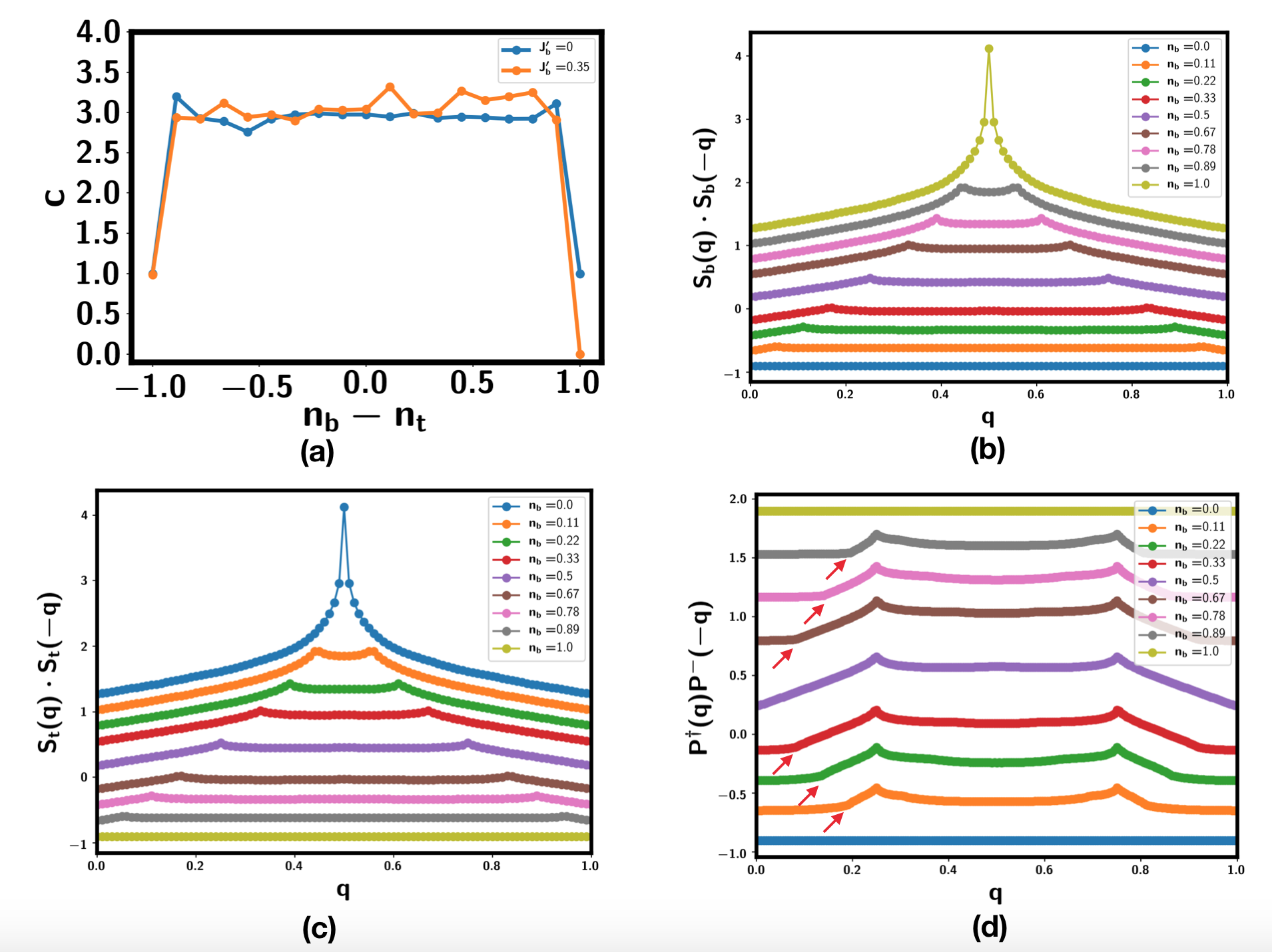}
\caption{DMRG results of the model in Eq.~\ref{eq:spin_layer_model} in one dimension. Here we use $J_t=J_b=J_p=J_{pz}=1$. (a) Central charge with the layer polarization $n_b-n_t$ for $J^\prime_b=0,0.35$. We can see $c=3$ when $0<x<1$.  $c$ is fit from the relation $S=\frac{c}{6}\log \xi$, where $S$ is the entanglement entropy and $\xi$ is the largest correlation length. (b) Spin structure factor in the bottom layer. (c) Spin structure factor in the top layer. (d) Exciton correlation function. The red arrow denotes the momentum $q_1=k_F^t-k_F^b=2\pi(\frac{n_t}{4}-\frac{n_b}{4})=2\pi(\frac{1}{2}n_t-\frac{1}{4})$. There is always a clear peak at $q_2=0.25\times 2\pi$. The momentum $q$ is in units of $\frac{2\pi}{a}$.}
\label{fig:1d_spinon_FS_1}
\end{figure*}

First we describe the 1D analog of the spinon Fermi surface state using bosonization. We will also argue that it can be understood as a one dimensional descendant of spinon Fermi surface introduced in the previous part for the two dimension limit.  Considering spinon Fermi surface formed by $f_{t\sigma}$ and $f_{b\sigma}$, in one dimension there are four modes coming from the combination of layer and spin.  We have $2k_F^t=\frac{x}{2}\times 2\pi$ and $2k_F^b=\frac{1-x}{2} \times 2\pi$. The left and right moving modes around $\pm k_F^a$ can be written using the standard bosonization language:

\begin{equation}
	\psi^{a}_{r\sigma}=\frac{1}{\sqrt{2\pi\alpha}} U^a_{r\sigma} e^{ir k_F^a x} e^{-\frac{i}{\sqrt{2}}(r \phi_{c;a}-\theta_{c;a}+\sigma(r \phi_{s;a}-\theta_{s;a}))}
\end{equation}
where $a=t,b$ labels the layer. $r=R,L$ labels the right and left moving mode. $\sigma=\uparrow,\downarrow$ labels the spin index. $U^a_{r\sigma}$ is the Klein factor to enforce the fermion statistics.

We also define new variables with a linear recombination:
\begin{equation}
	\theta_{c;\pm}=\frac{1}{\sqrt{2}}(\theta_{c;t}\pm \theta_{c;b})
\end{equation}
and
\begin{equation}
	\phi_{c;\pm}=\frac{1}{\sqrt{2}}(\phi_{c;t} \pm \phi_{c;b})
\end{equation}
Similarly one can define $\phi_{s;\pm}$ and $\theta_{s;\pm}$ for the two spin modes.

As analog of the spinon Fermi surface state, the fermion fields need to couple to a U(1) gauge field $a_\mu$ as in Eq.~\ref{eq:spinon_fs_gauge_field_action}.   In one dimensional version of spinon Fermi surface state, the U(1) gauge field and the charge mode $\theta_{c;+}$ higgs each other with a term like $L'=\big((\partial_\mu-ia_\mu)\theta_{c;+}\big)^2$. This makes sense because the total charge density $\frac{2}{\pi} \partial_x \phi_{c;+}$ is frozen in a Mott insulator. We will fix $\phi_{c;+}=0$.  In then end, we are left with three modes corresponding to $\phi_{c;-}$, $\phi_{s;t}$ and $\phi_{s;b}$.

The effective Hamiltonian for these three modes are:

\begin{align}
H&=\frac{\upsilon_-}{2\pi} \int dx K_{-}(\partial_x \theta_{c;-})^2+\frac{1}{K_-}(\partial_x \phi_{c;-})^2 \notag\\
&+\frac{\upsilon_t}{2\pi} \int dx (\partial_x \theta_{s;t})^2+(\partial_x \phi_{s;t})^2 \notag\\
&+\frac{\upsilon_b}{2\pi} \int dx (\partial_x \theta_{s;b})^2+(\partial_x \phi_{s;b})^2
\end{align}
where the SU(2)$\times$ SU(2) symmetry constrain the Luttinger parameter for the spin modes in the two layers to be $1$ and the spin modes in the two layers do not hybridize.  We expect $K_{-}<1$ because of the repulsive $J_{pz} P_z(i) P_z(j)$ term, as $P_z(r)\sim \frac{1}{\pi} \partial_x \phi_{c;-}$.

The spin-spin structure factor in the two layers will have peaks at $q=2k_F^t$ and $q=2k_F^b$ respectively. For example, we expect that the $2k_F^a$ part of the spin operator in the layer $a=t,b$ is

\begin{equation}
	S^\dagger_a(x)\sim \frac{1}{2\pi \alpha}i(e^{-i 2k_F^a x} e^{ \pm i \phi_{c;-}}e^{-i \sqrt{2}\theta_{s;a}}+e^{2ik_F^a x} e^{\mp i \phi_{c;-}}e^{-i \sqrt{2}\theta_s^a})
\end{equation}
where we have used $\phi_{c;+}=0$ and thus $\phi_{c;a}=\pm \frac{1}{\sqrt{2}}\phi_{c;-}$. Here $\pm$ is for $a=t,b$ respectively.

Then we obtain:

\begin{equation}
	\langle \vec{S}_a(x) \cdot \vec{S}_a(0)\rangle \sim \frac{1}{|x|^{\frac{K_-}{2}+1}}(\cos 2k_F^a x+\varphi_a)
\end{equation}
where we have ignored the zero momentum part.

In addition to gapless spin modes, there is also gapless mode corresponding to the exciton operator $P^\dagger$:

\begin{align}
P^\dagger(x)&=\frac{1}{\pi \alpha}\big(e^{-i (k_F^t-k_F^b)x} e^{i (\phi_{c;-}-\theta_{c;-})} \cos (\phi_{s;-}-\theta_{s;-})\notag\\
&+e^{i (k_F^t-k_F^b)x} e^{-i (\phi_{c;-}+\theta_{c;-})} \cos (\phi_{s;-}+\theta_{s;-})\notag\\
&+e^{-i(k_F^t+k_F^b)x}e^{-i\theta_{c;-}}\cos(\phi_{s;+}-\theta_{s;-})\notag\\
&+e^{i(k_F^t+k_F^b)x}e^{-i\theta_{c;-}}\cos(\phi_{s;+}+\theta_{s;-})\big)
\end{align}
where we again have used $\phi_{c;+}=0$.

Then its correlation function is in the form:

\begin{align}
\langle P^\dagger(x) P^{-}(0)\rangle&=A \frac{1}{|x|^{1+\frac{K_{-}+\frac{1}{K_-}}{2}}}\cos(q_1 x +\varphi_1)\notag\\
&+B \frac{1}{|x|^{1+\frac{1}{2K_-}}}\cos(q_2 x +\varphi_2)
\label{eq:PxPx}
\end{align}
where $q_1=k_F^t-k_F^b=2\pi(\frac{n_t}{4}-\frac{n_b}{4})=2\pi(\frac{1}{2}n_t-\frac{1}{4})$ and $q_2=k_F^t+k_F^b=2\pi (\frac{n_t}{4}+\frac{n_b}{4})=\frac{\pi}{2}$. The peak at $q_2$ should be more pronounced due to smaller decaying exponent.

Finally we discuss the Friedel oscillation of the layer polarization $P_z(x)$. We have
\begin{align}
P_z(x)&=\frac{1}{\pi}\partial_x \phi_{c;-}\notag\\
&+A_t(e^{-i\phi_{c;-}}\cos \sqrt{2}\phi_{s;t}e^{2ik_F^t x}+h..c)\notag\\
&-A_b(e^{-i\phi_{c;-}}\cos \sqrt{2}\phi_{s;b}e^{2ik_F^b x}+h..c)\notag\\
&+B(e^{-2i\phi_{c;-}}e^{4ik_F x}+h..c)\notag\\
\end{align}
where we have used the fact that $4k_F^t=2\pi-4k_F^b$ because $n_t+n_b=1$. So there is only one $4k_F=4k_F^t$ momentum.

From the above expression, we can quickly get the correlation function:

\begin{align}
\langle P_z(x) P_z(0) \rangle_c&=\frac{K_-}{2\pi^2}\frac{1}{x^2}\notag\\
&~~+|A_t|^2 \frac{1}{|x|^{1+\frac{1}{2}K_{-}}}\cos(2k_F^t x+\varphi_t)\notag\\
&~~+|A_b|^2 \frac{1}{|x|^{1+\frac{1}{2}K_{-}}}\cos(2k_F^b x +\varphi_b)\notag\\
&~~+|B|^2 \frac{1}{|x|^{\frac{3}{2}K_-}}\cos(4k_F x+\varphi)
\label{eq:PzPz}
\end{align}

This is the Friedel oscillation of the layer polarization also discussed in the spinon Fermi surface state in 2D (see Fig.~\ref{fig:Pz_Pz_friedel}). Here in 1D one special feature is that the $4k_F$ component is enhanced because $K_{-}<1$. Especially, the $4k_F$ part can dominate over $2k_F$ if $K_{-}<\frac{2}{3}$.

To test the above results, we simulated the model in Eq.~\ref{eq:spin_layer_model} in one dimension using DMRG. We show our results in Fig.~\ref{fig:1d_spinon_FS_1} at different values of $-2P_z=n_b-n_t$ while fixing the parameters $J_t=J_b=J_p=J_{pz}=1$.  We have tried both $J^\prime_b=0$ and $J^\prime_b=0.35$. When $J^\prime_b=0.35$, the $n_b=1,n_t=0$ is in a valence bond solid (VBS) phase and we have central charge $c=0$. If $J^\prime_b=0$, the $n_b=1,n_t=0$ point is in the gapless spin $1/2$ chain phase with $c=1$. The $n_t=1,n_b=0$ limit always has $c=1$ because we do not include next nearest neighbor coupling $J^\prime_t$ for the top layer.  Regardless of the fate in the layer polarized limit, as long as we dope excitons to reach $n_b=1-x,n_t=x$ with $x\in (0,1)$, we have a phase with $c=3$ as expected. In Fig.~\ref{fig:1d_spinon_FS_1}(b)(c), we show that the spin-spin structure factor indeed shows peak at $2k_F^t$ and $2k_F^b$ as we vary the layer polarization $n_b-n_t$.  In Fig.~\ref{fig:1d_spinon_FS_1}(d) we show the Fourier transformation of $\langle P^\dagger(x) P^{-}(y)\rangle$. Consistent with Eq.~\ref{eq:PxPx}, it shows peak at momentum $q_2=\frac{1}{4}\times 2\pi$ and an additional weaker singularity at momentum $q_1=k_F^t-k_F^b$. Therefore we conclude that the ground state of the model Eq.~\ref{eq:spin_layer_model} in 1D is an analog of the spinon Fermi surface state with both gapless spin mode and exciton mode.

\begin{figure*}[ht]
\centering
\includegraphics[width=0.8\linewidth]{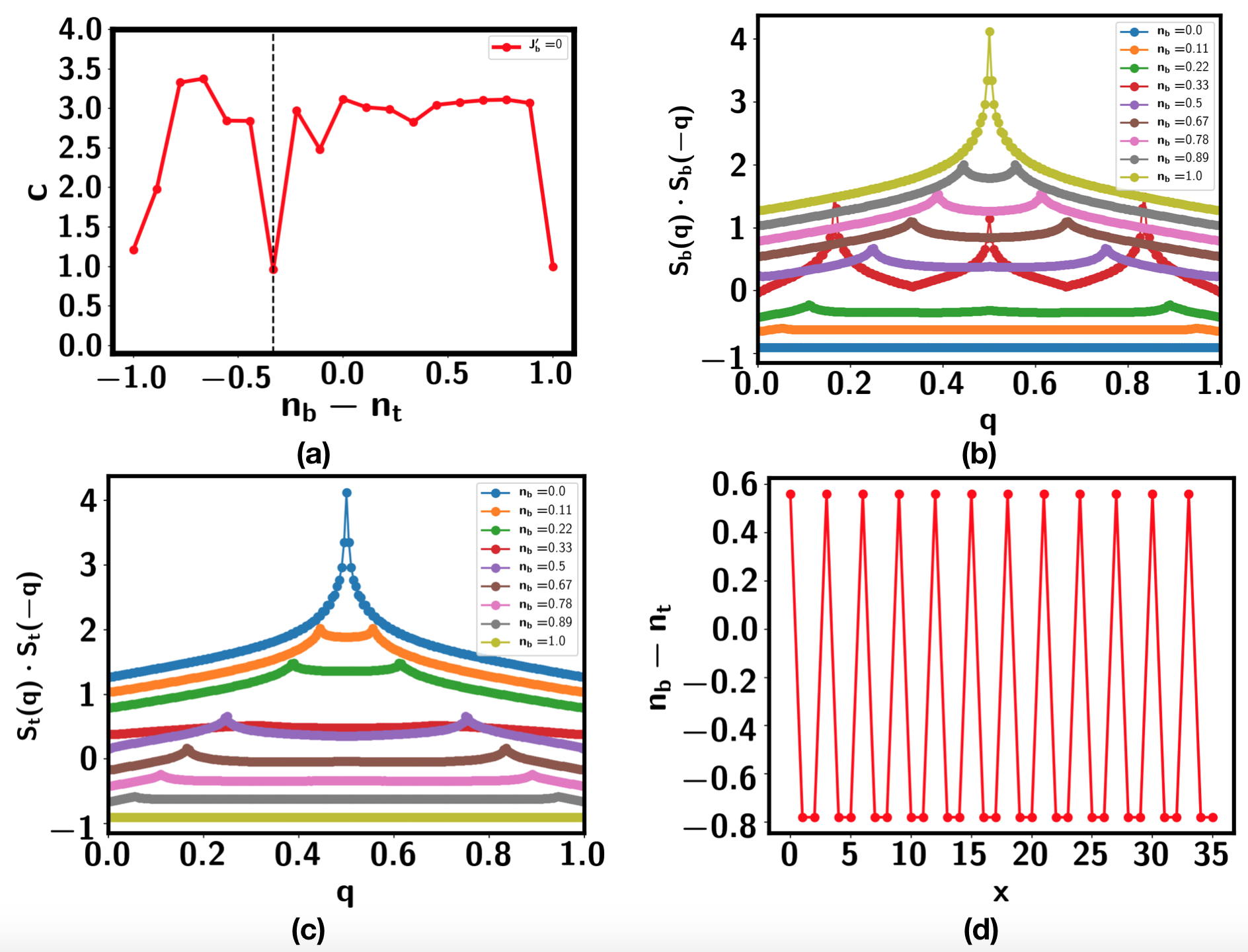}
\caption{DMRG results of the model in Eq.~\ref{eq:spin_layer_model} in one dimension. Here we use $J_b=1$, $J_t=5$, $J_p=2$, $J_{pz}=10$. (a) Central charge with the layer polarization $n_b-n_t$ for $J^\prime_b=0,0.35$. We can see $c=3$ when $0<x<1$ except at $n_t=\frac{2}{3},n_b=\frac{1}{3}$.  (b) Spin structure factor in the bottom layer. (c) Spin structure factor in the top layer. (d) Exciton density $n_b(x)-n_t(x)$ for the filling $n_t=\frac{2}{3},n_b=\frac{1}{3}$. One can see a density wave of the exciton (or layer polarization $P_z$) with period $3$.}
\label{fig:1d_spinon_FS_2}
\end{figure*}

To test whether the $c=3$ spinon Fermi surface phase is robust, we also tried a different set of parameter with $J_p=2$, $J_{pz}=10$, $J_t=5$, shown in Fig.~\ref{fig:1d_spinon_FS_2}. We still find the gapless phase with $c=3$ except at the commensurate filling $n_t=\frac{2}{3},n_b=\frac{1}{3}$.  At this special filling, the exciton is localized in a density wave state with period $3$ as shown in Fig.~\ref{fig:1d_spinon_FS_2}(d). There is a bottom, top, top pattern within the 3 site unit cell. Then the mode corresponding to $P^\dagger$ is gapped. The spin mode in the top layer is also gapped out because two nearby spins can just form a spin-singlet. The only gapless mode is from the spin in the bottom layer, which gives $c=1$.  Note that a large $J_t$ is crucial to favor the spin-singlet formation in the top layer and the localization of the exciton. Away from the commensurate filling, we always have the $c=3$ spinon Fermi surface phase at generic filling $x$.

\begin{figure*}[ht]
\centering
\includegraphics[width=0.8\linewidth]{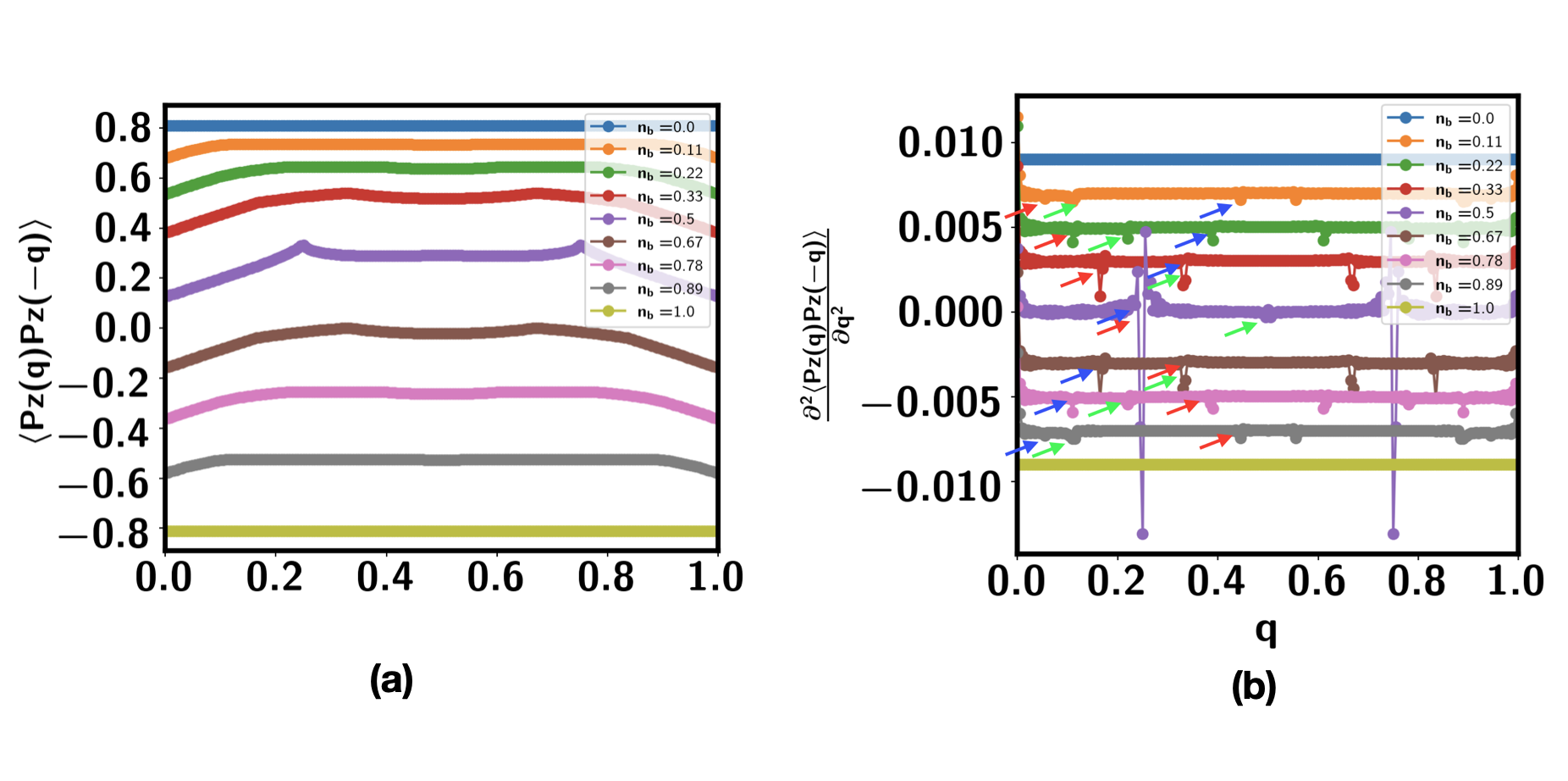}
\caption{(a) $\langle P_z(q) P_z(-q) \rangle$ for $J_b=1$, $J_t=1$, $J_p=1$, $J_{pz}=1$ and $J^\prime_b=0$. The momentum $q$ is in units of $2\pi$. (b) The second derivative  of  $\langle P_z(q) P_z(-q) \rangle$ with respect to $q$. The red and blue arrows denote $2k_F^b$ and $2k_F^t$ respectively. The green arrow denotes the $4k_F$. Due to the symmetry $q\leftrightarrow 2\pi -q$, we only label the peaks for $q\in [0,\pi]$.}
\label{fig:1d_Pz_Pz_first}
\end{figure*}

\begin{figure}[ht]
\centering
\includegraphics[width=0.8\linewidth]{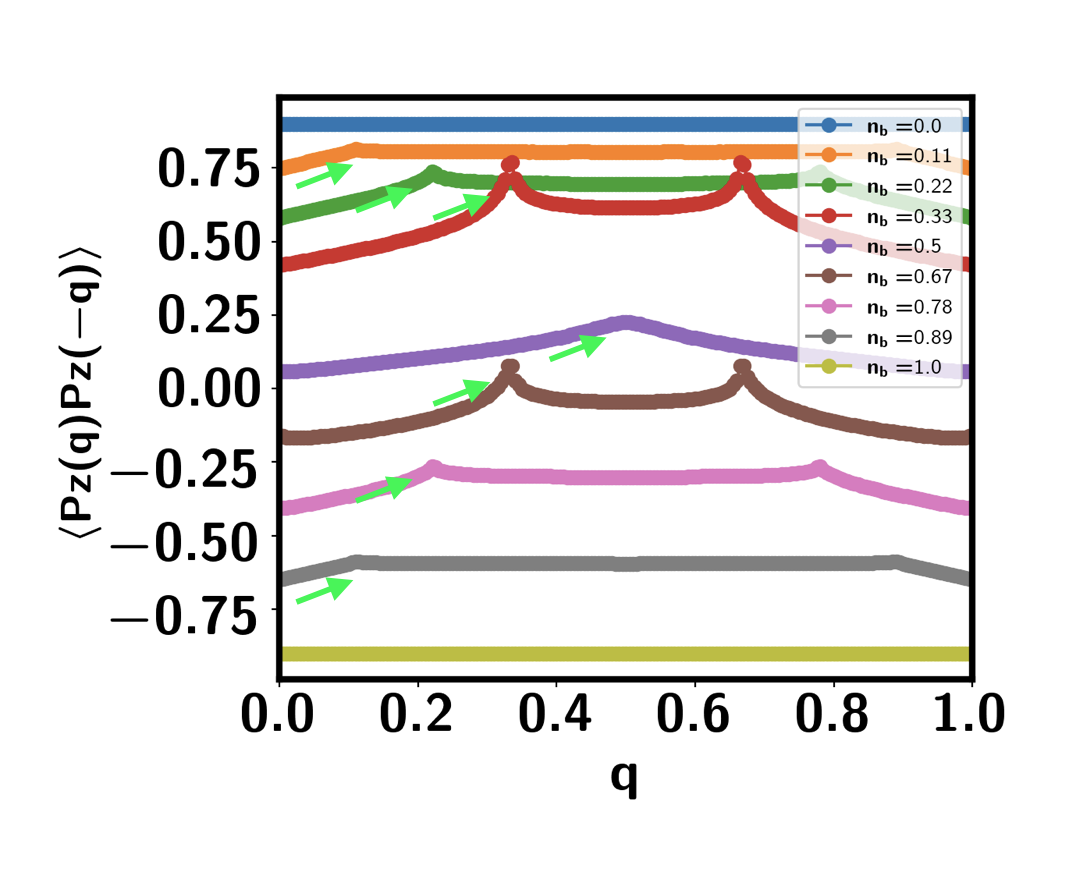}
\caption{$\langle P_z(q) P_z(-q) \rangle$ for $J_b=1$, $J_t=1$, $J_p=2$, $J_{pz}=10$ and $J^\prime_b=0$. The momentum $q$ is in units of $2\pi$. The green arrow denotes the $4k_F$ peak.}
\label{fig:1d_Pz_Pz_second}
\end{figure}

As discussed in the 2D case, the Friedel oscillation in the layer polarization may be experimentally easier to detect than spin-spin structure factor. We show $\chi_{zz}(q)=\langle P_z(q) P_z(-q)\rangle$ in Fig.~\ref{fig:1d_Pz_Pz_first} for the parameter $J_b=J_t=J_p=J_{pz}=1$ and $J^\prime_b=0$. In this case, the singularity in $\langle P_z(q) P_z(-q) \rangle$ is not very pronounced and we need its second derivative to $q$ to reveal the peaks. In Fig.~\ref{fig:1d_Pz_Pz_first}(b) we label the peaks corresponding to $2k_F^t$, $2k_F^b$ and $4k_F$. We also tried a different parameter $J_b=1$, $J_t=1$, $J_P=2$, $J_{pz}=10$ and $J^\prime_b=0$. This time the peak from $4k_F$ is enhanced, as shown in Fig.~\ref{fig:1d_Pz_Pz_second}.  A large $J_{pz}$ reduced $K_{-}$ and we indeed expect that the  $4k_F$ peak dominates according to Eq.~\ref{eq:PzPz}. 

In summary, our numerical simulation clearly demonstrates that the ground state of Eq.~\ref{eq:spin_layer_model} in one dimension is in an analog of spinon Fermi surface state. Similar to the higher dimension spinon Fermi surface phase, there is Friedel oscillation in the layer polarization $P_z$. There should also be a metallic conductance in the counter-flow transport. These results confirm our argument that doping excitons into a layer polarized Mott insulator with spinon Fermi surface naturally results in a phase with neutral Fermi surfaces in both layers. Due to the strong fluctuation in one dimension, magnetic order is not present and a spinon Fermi surface state can be found already in the simple spin model as in Eq.~\ref{eq:spin_layer_model}. In two dimension, the ground state of Eq.~\ref{eq:spin_layer_model} is more likely to be in an exciton superfluid phase described by bosonic parton theory as shown in Sec.~\ref{sec:bosonic_exciton_superfluids}. However, in the weak Mott regime, higher order ring exchange couplings needed to be included in Eq.~\ref{eq:spin_less_model}. In this case a spinon Fermi surface described by fermionic parton theory may be stabilized\cite{motrunich2005variational,sheng2009spin} and then we expect a phase with neutral Fermi surfaces in both layers at finite $x$. We leave the detailed numerical study of the more complicated problem with additional ring exchange couplings to future.  Experimentally we note that there is already encouraging signature of spinon Fermi surface at $x=0$ limit in MoTe$_2$/WSe$_2$ system\cite{li2021continuous} and we propose to search for smoking gun evidence of the neutral Fermi surface at finite $x$ in MoTe$_2$-hBN-MoTe$_2$/Wse$_2$ system from counter-flow transport and Friedel oscillation measurements. Although our numerical simulation in one dimension is mainly used as a guidance for two dimension, the recent experimental progress on one dimensional moir\'e superlattice\cite{wang2021one} may make the experimental realization of the 1D model also possible in near future.

\section{Metal insulator transition: a universal drag resistivity\label{sec:MIT}}

\begin{figure}[ht]
\centering
\includegraphics[width=0.95\linewidth]{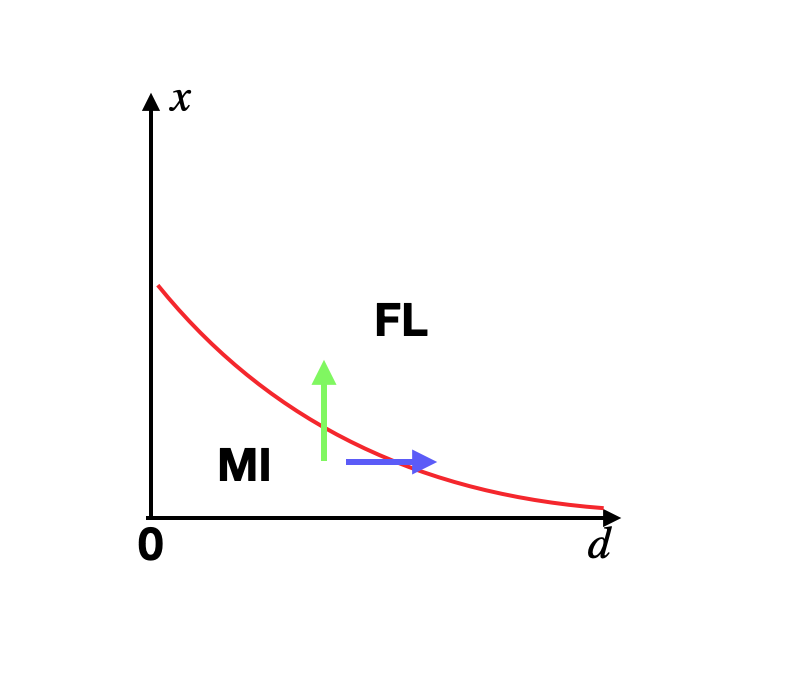}
\caption{Illustration of the phase diagram tuned by exciton density $x$ and the inter-layer distance $d$. MI and FL means Mott insulator and Fermi liquid. We assume that $U_b/t_b$ is large, so the $x=0$ limit is a Mott insulator regardless of $d$. For finite $x$, a metal-insulator transition (MIT) can be driven by tuning the inter-layer repulsion $U'$ through $d$, as denoted by the blue arrow. Because we assume the top layer is less correlated with a higher $t_t$ and smaller $U_t$, we expect $d_c$ to decrease with larger density $x$ in the top layer. Experimentally the same MIT transition can also be realized by tuning exciton density $x$ at a fixed $d$, denoted by the green arrow.}
\label{fig:MIT}
\end{figure}

In previous sections we focus on possible states when there is a finite charge gap. In this section we discuss possible experimental signature of a metal-insulator transition (MIT) tuned by the exciton density $x$. Let us now consider the bilayer Hubbard model defined in Eq.~\ref{eq:Hubbard_model_moire_bilayer}, a metal insulator transition can be driven by tuning $U'$ from $0$ to large when fixing $t_a$ and $U_a$. Experimentally $U'$ is the inter-layer interaction and is controlled by the inter-layer distance  $d$ (the thickness of the hBN barrier). When $d$ is large, the two layers are decoupled and each of them is in a metallic (if not superconducting) state because the density $n_a$ at either layer is less then $1$. When $d$ is reduced and $U'$ is enhanced, the simultaneous occupancy of both layers at each lattice site $i$
 is suppressed and the system enters a Mott insulator. A phase diagram in the parameter space $(x,d)$ is illustrated in Fig.~\ref{fig:MIT}. Because the top layer is less correlated, we expect the critical value $d_c$ to decrease when increasing the density $x$ at the top layer. This makes it possible to tune the same MIT transition through tuning the exciton density $x$ at a fixed inter-layer distance $d$. The closing of the charge gap when $x$ is larger than a critical value $x_c$ has already been observed in the WSe$_2$-hBN-WSe$_2$/WS$_2$ system\cite{gu2021dipolar,zhang2021correlated}. However, the nature of this transition is not clear now. We will provide one possible theory of this transition and propose experimental signatures in transport measurements.

First, from Fig.~\ref{fig:MIT} it is clear that the metal-insulator transition belongs the class of bandwidth controlled MIT because the total density is always fixed to be $n_t+n_b=1$. This transition could be continuous if the Mott insulator just across $x_c$ hosts neutral spinon Fermi surfaces, as discussed in Sec.~\ref{sec:neutral_fermi_surface}. In this case, the MIT is driven by condensation of the bosonic holon.  The theory of the MIT in our case is the same as in Ref.~\onlinecite{senthil2008theory}. But we will show that the bilayer structure enables a better probe of the criticality through measurement of the drag resistivity.

The transition can be described using the slave boson theory: $c_{i;a\sigma}=\varphi_i f_{i;a\sigma}$, where $\varphi_i$ is a bosonic holon or slave boson. $f_{i;a\sigma}$ is the neutral fermion.  There is a constraint that 

\begin{equation}
    \varphi_i^\dagger\varphi_i=\sum_{a\sigma} f^\dagger_{i;a\sigma} f_{i;a\sigma}
\end{equation}

Thus on average we have $\langle \varphi^\dagger_i \varphi_i\rangle=1$. There is also a U(1) gauge field arising from the gauge symmetry: $\varphi_i \rightarrow e^{i \alpha_i}$ and $f_{i;a\sigma}\rightarrow f_{i;a\sigma}e^{-i \alpha_i}$. There are two global U(1) symmetries generated by the total charge $Q$ and layer polarization $P_z$. The one generated by $Q$ acts as: $\varphi_i \rightarrow \varphi_i e^{i \theta_c}, f_{i;a\sigma}\rightarrow f_{i;a\sigma}$. The one corresponding to $P_z$ acts as $\varphi_i \rightarrow \varphi_i$, $f_{i;t\sigma}\rightarrow f_{i;t\sigma}e^{i\frac{1}{2} \theta_s}$ and $f_{i;b\sigma}\rightarrow f_{i;b\sigma}e^{-i\frac{1}{2} \theta_s}$.

We assume that the fermion $f_{i;a\sigma}$ is always in a mean field ansatz with Fermi surfaces, as in Eq.~\ref{eq:mean_field_fermionic_spinon}. When $x<x_c$, $\varphi_i$ is in a Mott insulator. When $x>x_c$, $\varphi_i$ is in a superfluid phase with $\langle \varphi_i \rangle \neq 0$. Therefore the MIT is driven by the superfluid-insulator transition fo the bosonic holon $\varphi_i$. The action of the critical theory is:

\onecolumngrid

\begin{align}
S&=\int d\tau d^2 x \psi^\dagger_{t;\sigma}(\tau,x)\big(\partial_\tau-\mu_t-ia_0(\tau,x)-\frac{1}{2}iA^s_0(\tau,x)\big)\psi_{t;\sigma}(\tau,x)-\frac{\hbar^2}{2m_t}\psi^\dagger_{t;\sigma}(\tau,x)\big(-i\vec \partial-\vec a(\tau,x)-\frac{1}{2}\vec{A^s}(\tau,x)\big)^2\psi_{t;\sigma}(\tau,x) \notag\\
&~+\int d\tau d^2 x \psi^\dagger_{b;\sigma}(\tau,x)\big(\partial_\tau-\mu_b-ia_0(\tau,x)+\frac{1}{2}iA^s_0(\tau_x)\big)\psi_{b;\sigma}(\tau,x)-\frac{\hbar^2}{2m_b}\psi^\dagger_{b;\sigma}(\tau,x)\big(-i\vec \partial-\vec a(\tau,x)+\frac{1}{2}\vec{A^s}(\tau,x)\big)^2\psi_{b;\sigma}(\tau,x)\notag \\ 
&+\int d\tau d^2 x|(\partial_\mu+i a_\mu-i A_\mu^c)\varphi|^2+s |\varphi|^2+g|\varphi|^4\notag \\ 
&+\int d\omega d^2 q \big(\frac{k_0|\omega|}{|q|}+\chi_d |q|^2+\sigma_b \sqrt{|\omega|^2+c^2|q|^2}\big)|a(\omega,q)|^2
\label{eq:critical_MIT}
\end{align}

\twocolumngrid
The first two lines describe the neutral Fermi surface coupled to the U(1) gauge field $a_\mu$. The third line describes the critical boson and the last line encodes the effective action for the transverse gauge field. Note that $a_0$ is screened by finite density of spinons and can be ignored at low energy. $\chi_d$ is from the diamagnetic susceptibility of the spinon Fermi surfaces.  $\sigma_b$ is the universal conductivity of the critical boson at $s=0$. This term is absent away from $s_c=0$. When $s<0$, the slave boson condenses: $\langle \varphi \rangle \neq 0$, which higgses $a_\mu=A^c_\mu$. After that, $\psi_t$ couples to $A^t_\mu=A^c_\mu+\frac{1}{2} A^s_\mu$ and $\psi_b$ couples to $A^b_\mu=A^c_\mu-\frac{1}{2}A^s_\mu$ and they can be identified as the electron operator: $c_{a\sigma}\sim \psi_{a\sigma}$. This is just the Fermi liquid phase. When $s>0$, we have $\langle \varphi \rangle=0$ and the slave boson is in a Mott insulator. Then $\varphi$ is gapped and drops out from the low energy. Meanwhile $\sigma_b=0$. The final action reduces to Eq.~\ref{eq:spinon_fs_gauge_field_action}, which describes a Mott insulator with neutral Fermi surfaces.

The critical property of Eq.~\ref{eq:critical_MIT} has been analyzed in Ref.~\onlinecite{senthil2008theory}. It can be shown that the coupling to the gauge field $a_\mu$ for $\varphi$ is irrelevant because the $\frac{|\omega|}{|q|}$ term quenches the gauge field. After that, the critical theory for the boson $\varphi$ is the same as the usual XY transition with exponent $\nu \approx 0.67$ and $z=1$.  The coupling to gauge field for the fermions $\psi$ leads to a singular self-energy: $\Sigma(i\omega)\sim i\omega \log \frac{1}{|\omega|}$. 

 \begin{figure}[ht]
\centering
\includegraphics[width=0.95\linewidth]{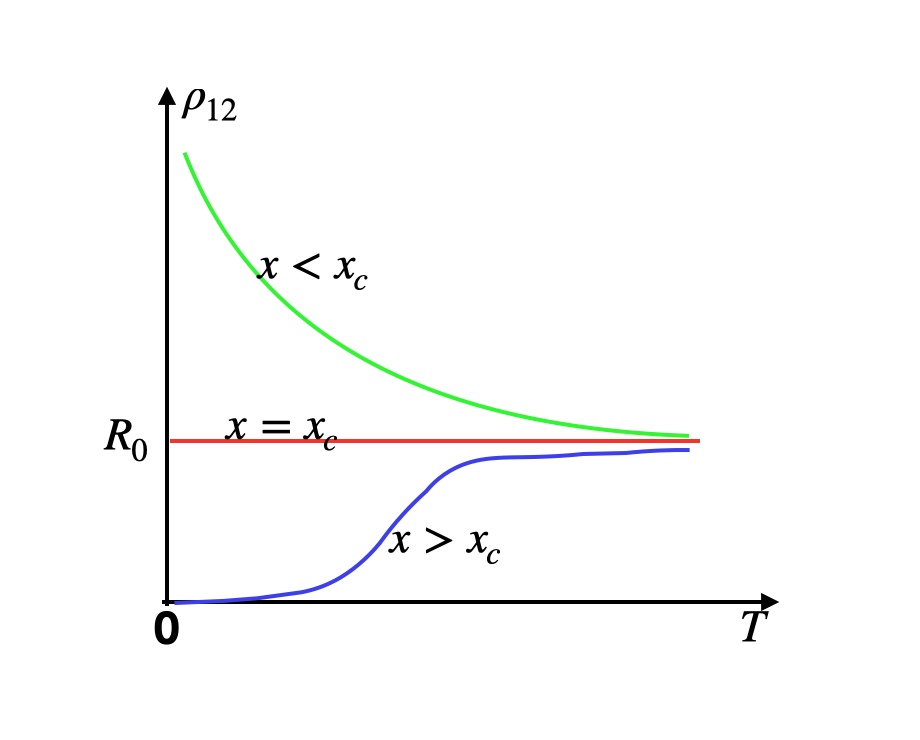}
\caption{Illustration of the drag resistivity $\rho_{12}$ with temperature $T$ in Fermi liquid regime $x>x_c$, Mott insulator regime $x<x_c$ and critical regime $x=x_c$. $R_0$ is a universal number at order $\frac{h}{e^2}$.}
\label{fig:drag_resistivity}
\end{figure}

Experimentally the easiest probe of the criticality may be from electric transport. It has already been pointed out that there is a universal jump of residual resistivity across the MIT\cite{senthil2008theory} coming from the universal conductivity $\sigma_b$ of the critical boson $\varphi$. Here we show that in our bilayer system the universal conductivity of the critical boson can be directly measured from the drag resistivity. 

As shown before, we have two currents $J^t_\mu, J^b_\mu$ coupled to the probing gauge field $A^t_\mu,A^b_\mu$. Therefore, we can define a $2\times 2$ resistivity tensor:

\begin{equation}
    \begin{pmatrix} E^t_x \\ E^b_x \end{pmatrix}=\begin{pmatrix} \rho_{11} & \rho_{12} \\ \rho_{21} & \rho_{22} \end{pmatrix}  \begin{pmatrix} J^t_x \\ J^b_x\end{pmatrix}
\end{equation}

We can easily derive an Ioffe-Larkin rule (see Appendix.~\ref{appendix:ioffe_larkin_rule_slave_boson}) for the resistivity tensor $\rho=\begin{pmatrix} \rho_{11} & \rho_{12} \\ \rho_{21} & \rho_{22} \end{pmatrix}$:

\begin{equation}
    \rho=\rho_b+\rho_f
\end{equation}
with
\begin{equation}
    \rho_f=\begin{pmatrix} \rho_{f;t} & 0 \\ 0 & \rho_{f;b} \end{pmatrix}
\end{equation}
and 
\begin{equation}
    \rho_b=\frac{1}{\sigma_b}\begin{pmatrix} 1 & 1 \\ 1 & 1 \end{pmatrix}
\end{equation}
Here $\rho_{f;a}$ is the resistivity of $f_{a}$ in the layer $a$. $\sigma_b$ is the conductivity for the slave boson $\varphi$ at the critical point.   Then we can see that $\rho_{11}=\rho_{f;t}+\rho_b$ has contributions from both fermion and critical boson as in single layer case\cite{senthil2008theory}. However, $\rho_{12}=\rho_b$ is dominated by the resistivity of the slave boson now. Because the boson $\varphi$ goes through a superfluid-insulator transition, we expect that the drag resistivity $\rho_{12}$ shows similar behavior as the resistivity across a superconductor-insulator transition, as illustrated in Fig.~\ref{fig:drag_resistivity}. Especially there is a large drag resistivity $R_0=\frac{1}{\sigma_b^c}$ at order $\frac{h}{e^2}$ in the critical regime. Around the critical point, we also anticipate a scaling in the form $\rho(T,x-x_c)= a F(\frac{T}{|x-x_c|^{\nu z}})$ with $\nu z \approx 0.67$. Such a scaling has been seen in experiment for the MoTe$_2$/Wse$_2$ system\cite{li2021continuous}. It is interesting to test the scaling also in the moire+monolayer system tuned by exciton density.

\section{Conclusion}

In summary, we modeled and analyzed a new problem in moir\'e bilayer or moir\'e+monolayer based on TMD moir\'e superlattice: we dope inter-layer excitons into a layer polarized Mott insulator. When the densities of the two layers are $n_t=x, n_b=1-x$, there is a finite charge gap when $x<x_c$. Below the charge gap, we argue that the exciton and spin intertwine with each other and are described by a four-flavor spin model. Using either Schwinger boson or Abrikosov fermion parton theory, we identify several possible interesting phases: (I) Exciton superfluid phase with or without spin gap. Especially there could be a fractional superfluid with a paired exciton condensation coexisting with $Z_2$ spin liquid; (II) A spin liquid phase with neutral Fermi surfaces in both layers. In this case the neutral fermion carries both spin and also layer polarization $P_z$. The existence of neutral Fermi surface can be tested  by the counter-flow transport and the Friedel oscillations in terms of layer polarization.  We also provide a theory of continuous metal insulator transition tuned by exciton density $x$ and predicted a universal drag resistivity in the critical regime. Our work opens a new direction to search for spin liquids and fractionalization using electrical probes such as counter-flow measurements. In future it is interesting to dope the bilayer Mott insulator and search for superconducting or exotic metallic phases.

\section{Acknowledgement}
We thank Ashvin Vishwanath for discussions. This work is supported by a startup funding from Johns Hopkins University.

\bibliographystyle{apsrev4-1}
\bibliography{csl}

\appendix

\onecolumngrid

\section{$t/U$ expansion \label{append:t_U_expansion}}

In this section we perform the $t/U$ expansion to derive the four-flavor spin model from Eq.~\ref{eq:Hubbard_model_moire_bilayer}. First, the low energy Hilbert space has four states per site, they are labeled as $\ket{1}_i=c^\dagger_{i;t \uparrow}\ket{0}$, $\ket{2}_i=c^\dagger_{i;t\downarrow}\ket{0}$, $\ket{3}_i=c^\dagger_{i;b\uparrow}\ket{0}$ and $\ket{4}_i=c^\dagger_{i;b\downarrow}\ket{0}$. The $t/U$ expansion is generated by virtual double occupied states, which are not included in the low energy Hilbert space.  

The Hamiltonian can be expressed purely in terms of generalized spin operators. We can view the four dimension Hilbert space as a tensor product of two spin $1/2$ Hilbert space. Each of the four state can be labeled as $\ket{a \sigma}=\ket{a} \otimes \ket{\sigma}$, where $a=t,b$ and $\sigma=\uparrow,\downarrow$. The first spin $1/2$ (spanned by $\ket{a=t}$ and $\ket{a=b}$) corresponds to a layer pseudospin $\vec P$ and the second spin $1/2$ (spanned by $\ket{\sigma=\uparrow}$ and $\ket{\sigma=\downarrow}$) corresponds to the real spin $\vec S$.  $\vec P$ is defined as $1/2 \vec{\sigma}$, where $\vec \sigma$ is the $2\times 2$ Pauli matrix within the corresponding spin $1/2$ Hilbert space spanned by $\ket{a=t}$ and $\ket{a=b}$. $\vec S$ is similarly defined.  We can also define $P_0$ and $S_0$ as an identity matrix.   Together $P_\mu S_\nu=P_\mu \otimes S_\nu$ with $\mu,\nu=0,x,y,z$ form $16$ generators of a $U(4)$ group. The effective spin Hamiltonian can be written in the similar form as a SU(4) spin model, though now we have large anisotropy terms.

\subsection{Zeroth order}
We note that the interaction term can enter the low energy Hamiltonian even at the zeroth order of the hopping $t$.  The on-site interactions $U_t,U_b,U'$ vanishes at the zeroth order because we always have the constraint $n_{i;t}+n_{i;b}=1$ at each site.  The nearest neighbor interaction, on the other hand, gives a term:

\begin{equation}
	H_S^{(0)}=\delta V \sum_{\langle ij \rangle}P_z(i) P_z(j)
\end{equation}
where $\delta V=V_t+V_b-2 V'$. We have used the identity $n_{i;t}=\frac{1}{2}+P_z(i)$ and $n_{i;b}=\frac{1}{2}-P_z(i)$ for the restricted Hilbert space.  We also ignored the constant term and the term linear to $P_z$, which just renormalizes the displacement field $D$. In the canonical ensemble with constant $P_z$, such a term linear to $P_z$ is meaningless as it is just a chemical potential term.

\subsection{Second order}
There is no terms generated at the linear order to $t$, and we focus on the second order here. For simplicity let us ignore the nearest neighbor interaction.  It is easy to show

\begin{align}
  	H_S^{(2)}&=-\frac{t_t^2}{U_t} \sum_{\langle ij \rangle} (c^\dagger_{i;t \sigma}c_{j;t \sigma} c^\dagger_{j;t \sigma'}c_{i;t \sigma'}+h.c.)n_t(i)n_t(j)\notag\\
  	&~~~-\frac{t_b^2}{U_b} \sum_{\langle ij \rangle} (c^\dagger_{i;b \sigma}c_{j;b \sigma} c^\dagger_{j;b \sigma'}c_{i;b \sigma'}+h.c.) n_b(i) n_b(j)\notag\\
  	&~~~-\frac{t_t^2}{U'} \sum_{\langle ij \rangle} (c^\dagger_{i;t \sigma}c_{j;t \sigma} c^\dagger_{j;t \sigma'}c_{i;t \sigma'}+h.c.)n_t(i)n_b(j)\notag\\
	&~~~-\frac{t_b^2}{U'} \sum_{\langle ij \rangle} (c^\dagger_{i;b \sigma}c_{j;b \sigma} c^\dagger_{j;b\sigma'}c_{i;b \sigma'}+h.c.)n_b(i)n_t(j)\notag\\
  	&~~~-\frac{t_t t_b}{U'} \sum_{\langle ij \rangle} (c^\dagger_{i;t \sigma}c_{j;t \sigma} c^\dagger_{j;b \sigma'}c_{i;b \sigma'}+c^\dagger_{j;b \sigma'}c_{i;b \sigma'}c^\dagger_{i;t \sigma}c_{j;t \sigma}) \notag\\
  	&~~~-\frac{t_t t_b}{U'} \sum_{\langle ij \rangle} (c^\dagger_{i;b \sigma}c_{j;b \sigma} c^\dagger_{j;t \sigma'}c_{i;t \sigma'}+c^\dagger_{j;t \sigma'}c_{i;t \sigma'}c^\dagger_{i;b \sigma}c_{j;b \sigma}) \notag\\
  \end{align}

Let us label $A=-(c^\dagger_{i;t \sigma}c_{j;t \sigma} c^\dagger_{j;t \sigma'}c_{i;t \sigma'}+h.c.)n_t(i)n_t(j)$. Within the restricted Hilbert space, it can be shown that $A \ket{1}_i \otimes \ket{2}_j=2(\ket{2}_i \otimes \ket{1}_j-\ket{1}_i \otimes \ket{2}_j)$, $A \ket{2}_i \otimes \ket{1}_j=2(\ket{1}_i \otimes \ket{2}_j-\ket{2}_i\otimes \ket{1}_j)$ and $A \ket{\alpha}_i \otimes \ket{\beta}_j=0$, if $(\alpha,\beta)\neq (1,2)\  \text{and}\  (2,1)$. It can be shown that $A=4\vec{S}_t(i)\otimes \vec{S}_t(j)-n_t(i)\otimes n_t(j)$. Here $\vec{S}_t(i)=n_t(i)\vec{S}(i)$ is the spin operator projected to the top layer. Note $n_t(i)=\frac{1}{2}+P_z(i)$ and $n_b(i)=\frac{1}{2}-P_z(i)$ can be viewed as projection operator within the restricted Hilbert space. Similarly, $-(c^\dagger_{i;b \sigma}c_{j;b \sigma} c^\dagger_{j;b \sigma'}c_{i;b \sigma'}+h.c.)n_b(i)n_b(j)=4 \vec{S}_b(i)\otimes \vec{S}_b(j)-n_b(i)\otimes n_b(j)$, where $\vec{S}_b(i)=n_b(i)\vec{S}(i)$ is the spin operator projected to the bottom layer.   $-\frac{t_t^2}{U'} \sum_{\langle ij \rangle} (c^\dagger_{i;t \sigma}c_{j;t \sigma} c^\dagger_{j;t \sigma'}c_{i;t \sigma'}+h.c.)n_t(i)n_b(j)=-\frac{2t_t^2}{U'}n_t(i)n_b(j)$ within the restricted Hilbert space.

Let us label $B=-(c^\dagger_{i;t \sigma}c_{j;t \sigma} c^\dagger_{j;b \sigma'}c_{i;b \sigma'}+c^\dagger_{j;b \sigma'}c_{i;b \sigma'}c^\dagger_{i;t \sigma}c_{j;t \sigma})$.  Within the restricted Hilbert space, it is easy to see that $B \ket{\alpha}_i \ket{\beta}_j=2 \ket{\beta}_i \ket{\alpha}_j$ if $\alpha=3,4$ and $\beta=1,2$. Otherwise $B \ket{\alpha}_i \ket{\beta}_j=0$.  We find that $B=P^\dagger(i) P^{-}(j) (4 \vec{S}(i) \cdot \vec{S}(j)+\vec{S}_0(i) \vec{S}_0(j))$. Where $P^{\pm}(i)=P_x(i)\pm P_y(i)$.

\begin{equation}
	H_S^{(2)}=4 \frac{t_t^2}{U_t} \sum_{\langle ij \rangle}\vec{S}_t(i) \vec{S}_t(j)+4 \frac{t_b^2}{U_b} \sum_{\langle ij \rangle}\vec{S}_b(i) \vec{S}_b(j)
\end{equation}

Finally, we have

\begin{align}
	H_S=&J_t \sum_{\langle ij \rangle}\vec{S}_t(i) \cdot \vec{S}_t(j)+J_b \sum_{\langle ij \rangle}\vec{S}_b(i) \cdot \vec{S}_b(j)+\frac{1}{2}J_{pz}\sum_{\langle ij \rangle} P_z(i)P_z(j) \notag\\
	&+J_p \sum_{\langle ij \rangle}\frac{1}{2}\big(P_x(i)P_x(j)+P_y(i)P_y(j)\big) \left(4\vec{S}(i) \cdot \vec{S}(j)+S_0(i)S_0(j)\right)
\end{align}
where $J_t=\frac{4 t_t^2}{U_t}$, $J_b=\frac{4 t_b^2}{U_b}$, $J_p=\frac{4 t_t t_b}{U'}$ and $J_{pz}=2\delta V-\frac{1}{2}(J_t+J_b)+(\frac{4 t_t^2}{U'}+\frac{4 t_b^2}{U'})$.  In the above we ignored the constant term and the term linear to $P_z(i)$.

In the $SU(4)$ symmetric limit with $J_t=J_b=J_p=J_{pz}=J$, we recover the SU(4) spin model
\begin{equation}
	H_S=\frac{J}{8}\left(4 \vec P(i)\cdot \vec P(j)+P_0(i)P_0(j)\right)\left(4 \vec S(i)\cdot \vec S(j)+S_0(i)S_0(j)\right)
\end{equation}
up to constant terms.

\section{DMRG implementation\label{append:dmrg}}

In DMRG, we use the spin operator $S_{ab}=\ket{a}\bra{b}$ with $a,b=1,2,3,4$.  We rewrite the Hamiltonian using the following formulas:

\begin{equation}
	\vec{S}_t(i)\cdot \vec{S}_t(j)=\frac{1}{2}S_{12}(i) S_{21}(j)+\frac{1}{2} S_{21}(i)S_{12}(j)+\frac{1}{4} S_{11}(i)S_{11}(j)+\frac{1}{4}S_{22}(i)S_{22}(j)-\frac{1}{4} S_{11}(i)S_{22}(j)-\frac{1}{4}S_{22}(i)S_{11}(j)
\end{equation}

\begin{equation}
	\vec{S}_b(i)\cdot \vec{S}_b(j)=\frac{1}{2}S_{34}(i) S_{43}(j)+\frac{1}{2} S_{43}(i)S_{34}(j)+\frac{1}{4} S_{33}(i)S_{33}(j)+\frac{1}{4}S_{44}(i)S_{44}(j)-\frac{1}{4} S_{33}(i)S_{44}(j)-\frac{1}{4}S_{44}(i)S_{33}(j)
\end{equation}

\begin{equation}
	P_z(i) P_z(j)=\frac{1}{4}(S_{11}(i)+S_{22}(i)-S_{33}(i)-S_{44}(i))(S_{11}(j)+S_{22}(j)-S_{33}(j)-S_{44}(j))
\end{equation}

and
\begin{align}
	&\frac{1}{2}\big(P_x(i)P_x(j)+P_y(i)P_y(j)\big) \left(4\vec{S}(i) \cdot \vec{S}(j)+S_0(i)S_0(j)\right)\notag\\
	&=\frac{1}{2}\left(S_{13}(i)S_{31}(j)+S_{14}(i)S_{41}(j)+S_{23}(i)S_{32}(j)+S_{24}(i)S_{42}(j)+(i \leftrightarrow j)\right)
\end{align}

In the limit $J_t=J_b=J_p=J_{pz}=J$, we find that 

\begin{equation}
	H_S=\frac{1}{2} J \sum_{\langle ij \rangle}\sum_{a,b=1,2,3,4}S_{ab}(i)S_{ba}(j)-\frac{1}{8}J_{pz}\sum_{\langle ij \rangle}n(i) n(j)
\end{equation}

Note that $-\frac{1}{2}J_{pz}\sum_{\langle ij \rangle}n(i) n(j)=-\frac{3}{8} J_{pz} N_s$, where $N_s$ is the number of sites.

For simplicity, we will define a new spin model:

\begin{equation}
	\tilde H_S=H_S+\frac{1}{8}J_{pz}\sum_{\langle ij \rangle}n(i) n(j)
\end{equation}

Note that the term proportional to $J_{pz}$ now becomes:

\begin{equation}
	\frac{1}{4}J_{pz}\left(\sum_{a=1,2,3,4}S_{aa}(i)S_{aa}(j)+S_{11}(i)S_{22}(j)+S_{22}(i)S_{11}(j)+S_{33}(i)S_{44}(j)+S_{44}(i)S_{33}(j) \right)
\end{equation}

We simulate $\tilde H_S$ in DMRG calculation, its full form is:

\begin{align}
\tilde H_S&= \sum_{\langle ij \rangle}\{ \frac{1}{4}(J_t+J_{pz})\left(S_{11}(i)S_{11}(j)+S_{22}(i)S_{22}(j)\right)+\frac{1}{4}(J_b+J_{pz})\left(S_{33}(i)S_{33}(j)+S_{44}(i)S_{44}(j)\right) \notag\\
&~~~+\frac{1}{4}(J_{pz}-J_t)\big(S_{11}(i)S_{22}(j)+S_{22}(i)S_{11}(j)\big)+\frac{1}{4}(J_{pz}-J_b)\big(S_{33}(i)S_{44}(j)+S_{44}(i)S_{33}(j)\big)\notag\\
&~~~+\frac{1}{2}J_t \big(S_{12}(i)S_{21}(j)+S_{21}(i)S_{12}(j)\big)+\frac{1}{2}J_b \big(S_{34}(i)S_{43}(j)+S_{43}(i)S_{34}(j)\big)\notag\\
&~~~+\frac{1}{2}J_p \big(S_{13}(i)S_{31}(j)+S_{14}(i)S_{41}(j)+S_{23}(i)S_{32}(j)+S_{24}(i)S_{42}(J)\notag\\
&~~~~+S_{31}(i)S_{13}(j)+S_{32}(i)S_{23}(j)+S_{41}(i)S_{14}(j)+S_{42}(i)S_{24}(j)\big) \}
\end{align}

We define the paired exciton operator as exciton of cooper pairs in the two layers. So $ PP^\dagger(i) = (\epsilon_{\sigma \sigma'} c^\dagger_{t;\sigma}(i)c^\dagger_{t;\sigma'}(j)) (\epsilon_{\alpha \beta} c_{b;\alpha}(i)c_{b;\beta}(j))$, where $j$ is a nearest neighbor site of $i$.  It can be expressed using our spin operators as 

\begin{equation}
	PP^\dagger(i)=S_{13}(i) S_{24}(j)+S_{24}(i)S_{13}(j)-S_{14}(i)S_{23}(j)-S_{23}(i)S_{14}(j)
    \label{eq:PP_definition_dmrg}
\end{equation}

\section{Ioffe-Larkin rule in slave boson theory\label{appendix:ioffe_larkin_rule_slave_boson}}

We use the slave boson theory $c_{i;a\sigma}=\varphi_i f_{i;a\sigma}$. We assign the charge that $\varphi_i$ couples to $a_\mu$, $f_t$ couples to $A^t_\mu-a_\mu$ and $f_b$ couples to $A^b_\mu-a_\mu$.  Note here we assume that the slave boson is neutral. One can also let $\varphi$ couples to $A^c_\mu$, but the conclusion will not change.  We need to emphasize that there is only one common U(1) gauge field $a_{\mu}$ for the two layers. This is because we are at total filling $n_t+n_b=1$ and the gauge field is introduced to fix the total filling of the fermions to be one in the Mott insulator, so they must share the same gauge field.  Alternatively, one may use slave boson theory $c_{i;a\sigma}=\varphi_{i;a} f_{i;a\sigma}$ with independent slave boson $\varphi_t, \varphi_b$ in the two layers. In this approach the two layers will have independent U(1) gauge field $a_t, a_b$.  However, only the total filling of $\varphi_t,\varphi_b$ is an integer one. So slave bosons can be put into a Mott insulator with $\varphi^\dagger_t \varphi_t+\varphi^\dagger_b \varphi_b=1$ at each site. But it is impossible to freeze the density at each layer and generically we will have $\langle \varphi^\dagger_t \varphi_b\rangle \neq 0$, which locks $a_t=a_b=a$. For simplicity we will use the formalism with a common slave boson in the two layers.

Let us label $J^a_\mu$ as the physical current for the layer $a$. $J^a_{f;\mu}$ labels the current for $f_{a\sigma}$. $J_{\varphi;\mu}$ labels the current of the slave boson $\varphi$. We have constraint:

\begin{equation}
    J^a_\mu=J^a_{f;\mu}
\end{equation}
and
\begin{equation}
    J_{\varphi;\mu}=J^t_{f;\mu}+J^b_{f;\mu}
\end{equation}

Boson couples to $a$, so we get:
\begin{equation}
    J_{\varphi;x}= \sigma_b e_x
\end{equation}
where $e_\mu$ is the electric field of the internal gauge field $a_\mu$.

Fermion $f_a$ couples to $A^a_\mu-a_\mu$, so we get:

\begin{align}
J^t_{f;x}&=\sigma_{f;t} (E^t_x-e_x)\notag\\
J^b_{f;x}&=\sigma_{f;b} (E^b_x-e_x)
\end{align}

With these equations together, we can get $\begin{pmatrix} J_{t} \\ J_{b} \end{pmatrix}=\sigma \begin{pmatrix} E_t \\ E_b \end{pmatrix}$ with the $2\times 2$ conductivity tensor as:

\begin{equation}
    \sigma=\left(
\begin{array}{cc}
 \frac{\left(\sigma _b+\sigma _{\text{f;b}}\right) \sigma _{\text{f;t}}}{\sigma _b+\sigma _{\text{f;b}}+\sigma _{\text{f;t}}} & -\frac{\sigma _{\text{f;b}} \sigma _{\text{f;t}}}{\sigma _b+\sigma _{\text{f;b}}+\sigma _{\text{f;t}}} \\
 -\frac{\sigma _{\text{f;b}} \sigma _{\text{f;t}}}{\sigma _b+\sigma _{\text{f;b}}+\sigma _{\text{f;t}}} & \frac{\sigma _{\text{f;b}} \left(\sigma _b+\sigma _{\text{f;t}}\right)}{\sigma _b+\sigma _{\text{f;b}}+\sigma _{\text{f;t}}} \\
\end{array}
\right)
\end{equation}

and its corresponding resistivity tensor

\begin{equation}
    \rho_c=\begin{pmatrix} \rho_b+\rho_{f;t} & \rho_b \\ \rho_b & \rho_b+\rho_{f;b} \end{pmatrix}
    \label{resistivity_Ioffe_larkin_rule}
\end{equation}
where $\rho_b=\frac{1}{\sigma_b}$ and $\rho_{f;a}=\frac{1}{\sigma_{f;a}}$.

\subsection{A path integral derivation}

We can also derive the same Ioffe-Larkin rule using the path integral. Let us start from the critical theory:

\begin{align}
S&=\int d\tau d^2 x \psi^\dagger_{t;\sigma}(\tau,x)\big(\partial_\tau-\mu_t-ia_0(\tau,x)-\frac{1}{2}iA^s_0(\tau,x)\big)\psi_{t;\sigma}(\tau,x)-\frac{\hbar^2}{2m_t}\psi^\dagger_{t;\sigma}(\tau,x)\big(-i\vec \partial-\vec a(\tau,x)-\frac{1}{2}\vec{A^s}(\tau,x)\big)^2\psi_{t;\sigma}(\tau,x) \notag\\
&~+\int d\tau d^2 x \psi^\dagger_{b;\sigma}(\tau,x)\big(\partial_\tau-\mu_b-ia_0(\tau,x)+\frac{1}{2}iA^s_0(\tau_x)\big)\psi_{b;\sigma}(\tau,x)-\frac{\hbar^2}{2m_b}\psi^\dagger_{b;\sigma}(\tau,x)\big(-i\vec \partial-\vec a(\tau,x)+\frac{1}{2}\vec{A^s}(\tau,x)\big)^2\psi_{b;\sigma}(\tau,x)\notag \\ 
&+\int d\tau d^2 x|(\partial_\mu+i a_\mu-i A_\mu^c)\varphi|^2+s |\varphi|^2+g|\varphi|^4\notag \\ 
&+\int d\omega d^2 q \big(\frac{k_0|\omega|}{|q|}+\chi_d |q|^2+\sigma_b \sqrt{|\omega|^2+c^2|q|^2}\big)|a(\omega,q)|^2
\end{align}

Here $A^c_\mu$ and $A^s_\mu$ are probing fields and their responses are captured by the partition function $Z[A^c_\mu, A^s_\mu]$ after doing the path integral:
\begin{equation}
    Z[A^c_\mu, A^s_\mu]=\int d[a] d [\psi_t] d[\psi_b] e^{-S}
\end{equation}

We can do a simple redefinition $a_\mu \rightarrow a_\mu+A^c_\mu$, then we have

\begin{align}
S&=\int d\tau d^2 x \psi^\dagger_{t;\sigma}(\tau,x)\big(\partial_\tau-\mu_t-ia_0(\tau,x)-iA^t_0(\tau,x)\big)\psi_{t;\sigma}(\tau,x)-\frac{\hbar^2}{2m_t}\psi^\dagger_{t;\sigma}(\tau,x)\big(-i\vec \partial-\vec a(\tau,x)-\vec{A^t}(\tau,x)\big)^2\psi_{t;\sigma}(\tau,x) \notag\\
&~+\int d\tau d^2 x \psi^\dagger_{b;\sigma}(\tau,x)\big(\partial_\tau-\mu_b-ia_0(\tau,x)-iA^b_0(\tau_x)\big)\psi_{b;\sigma}(\tau,x)-\frac{\hbar^2}{2m_b}\psi^\dagger_{b;\sigma}(\tau,x)\big(-i\vec \partial-\vec a(\tau,x)-\vec{A}^b(\tau,x)\big)^2\psi_{b;\sigma}(\tau,x)\notag \\ 
&+\int d\tau d^2 x|(\partial_\mu+i a_\mu)\varphi|^2+s |\varphi|^2+g|\varphi|^4\notag \\ 
&+\int d\omega d^2 q \big(\frac{k_0|\omega|}{|q|}+\chi_d |q|^2+\sigma_b \sqrt{|\omega|^2+c^2|q|^2}\big)|a(\omega,q)|^2
\end{align}

Integration of $\psi_a$ leads to an effective action:

\begin{equation}
    S_{eff}=\int d\omega d^2 q \Pi_b(i\omega,q)|a(\omega,q)|^2+ \Pi_{f;t}(i\omega,q)|a(\omega,q)+A^t(\omega,q)|^2+\Pi_{f;b}(i\omega,q)|a(\omega,q)+A^b(\omega,q)|^2
\end{equation}
where we use the gauge $\vec{q}\cdot \vec A=0$ and $\vec{q}\cdot \vec a=0$.

From $Z[A^t,A^b]=\int d[a] e^{-S_{eff}}$ we can integrate $a_\mu$ and get $Z[A^t,A^b]=e^{- \int d^2 q d\omega  A^a(\omega,\mathbf q)\Pi_{ab} A^b(-\omega,-\mathbf q)}$, with 

\begin{equation}
    \Pi(i\omega,\mathbf q)=\begin{pmatrix} \frac{\Pi_{f;t}(\Pi_b+\Pi_{f;b})}{\Pi_b+\Pi_{f;t}+\Pi_{f;b}} & -\frac{\Pi_{f;t}\Pi_{f;b}}{\Pi_b+\Pi_{f;t}+\Pi_{f;b}} \\  -\frac{\Pi_{f;t}\Pi_{f;b}}{\Pi_b+\Pi_{f;t}+\Pi_{f;b}} & \frac{\Pi_{f;b}(\Pi_b+\Pi_{f;t})}{\Pi_b+\Pi_{f;t}+\Pi_{f;b}} \end{pmatrix}
\end{equation}

Its inverse is

\begin{equation}
    \Pi^{-1}=\frac{1}{\Pi_b} \begin{pmatrix} 1 & 1 \\ 1 &1 \end{pmatrix}+\begin{pmatrix} \frac{1}{\Pi_{f;t}} & 0 \\ 0 & \frac{1}{\Pi_{f;b}}\end{pmatrix}
\end{equation}

Conductivity tensor is related to $\Pi(\omega,\mathbf q)=\Pi(i \omega \rightarrow \omega +i \delta,\mathbf q)$ through $\sigma_{ab}(\omega,q)=\frac{1}{i\omega} \Pi_{ab}(\omega,\mathbf q)$. Then we immediately derive the Ioffe-Larkin rule as in Eq.~\ref{resistivity_Ioffe_larkin_rule}.

\section{More DMRG results for the fractional superfluid\label{appendix:fractional_SF}}

\begin{figure}[H]
\centering
\includegraphics[width=0.95\linewidth]{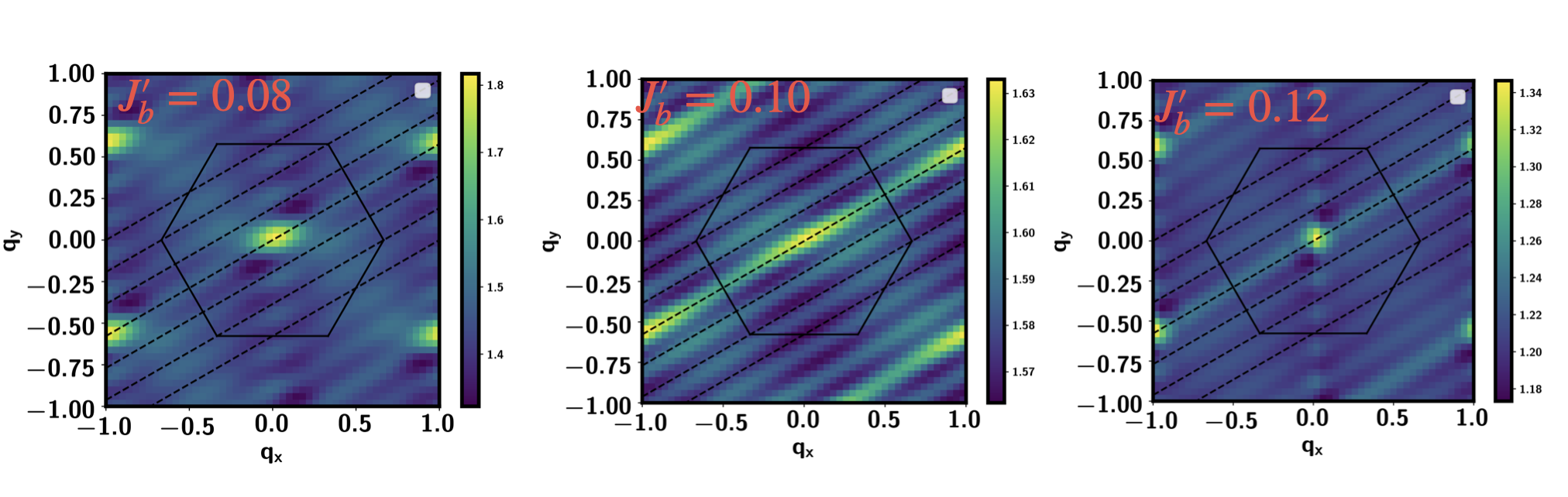}
\caption{$\langle PP^\dagger(\mathbf q) PP^{-}(-\mathbf q)\rangle$ at $J_b=J_p=J_{pz}=1$ and $J_t=4$ for system size $L_y=6$, $x=\frac{1}{18}$ and bond dimension $m=4000$. $PP^\dagger(\mathbf q)$ is the Fourier transformation of the paired exciton operator ($PP^\dagger(i)$ in Eq.~\ref{eq:PP_definition_dmrg}) living on a bond $(i,i+\hat{x})$. (a)$J^\prime_b=0.08$. (b)$J^\prime_b=0.10$. (c)$J^\prime_b=0.12$.  }
\label{fig:PP_appendix}
\end{figure}

Here we provide more DMRG results to support the existence of the fractional superfluid phase with paired exciton condensation coexists with $Z_2$ spin liquid discussed in Sec.~\ref{subsection:fractional_SF}.

First, we still use the parameter $J_b=J_p=J_{pz}=1$ and $J_t=4$ as in the main text in Sec.~\ref{subsection:fractional_SF}. In the main text we already show that there is a large gap for single exciton and the spin $\vec S_t$ in the top layer at $J^\prime_b=0.10, 0.12$. The spin in the bottom layer has short ranged antiferromagnetic correlations with momentum $\mathbf Q=K$ and $\mathbf Q=M$ respectively. The correlation length of the paired exciton is quite large with almost zero $\frac{1}{\xi_{pp}}$ extrapolated to infinite bond dimension. Here in Fig.~\ref{fig:PP_appendix} we show that $\langle PP^\dagger(\mathbf q) PP^{-}(-\mathbf q)\rangle $ has peak at $\mathbf Q=0$ for both phases separated by the first order critical point around $J^\prime_b \approx 0.11$. This establishes a uniform paired exciton condensation order as we expected theoretically for both the Z$_2$ SL I + PSF phase and the Z$_2$ SL II + PSF phase.

\begin{figure}[ht]
\centering
\includegraphics[width=0.8\linewidth]{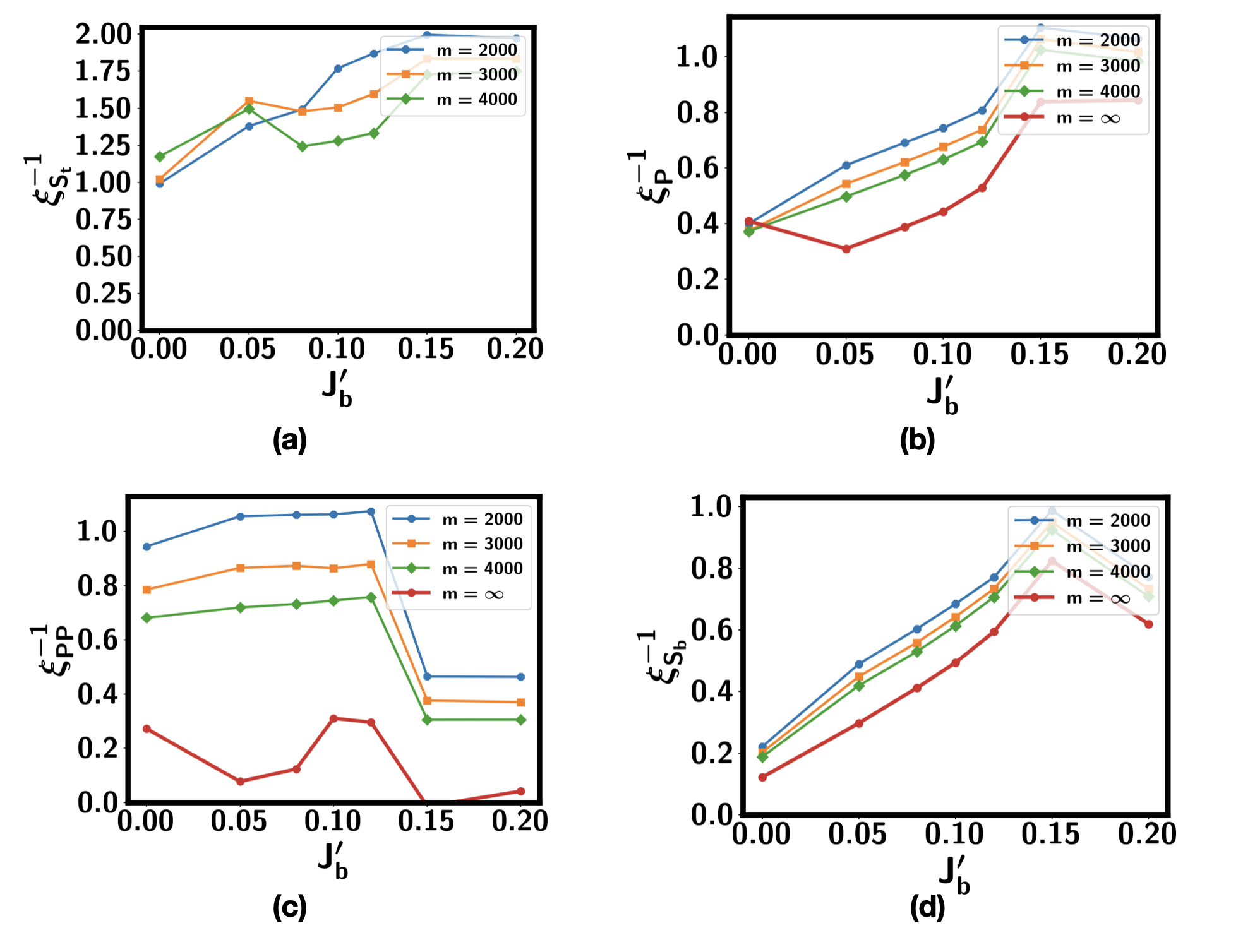}
\caption{Inverse correlation lengths at $J_b=J_p=J_{pz}=1$ and $J_t=2$ for system size $L_y=6$, $x=\frac{1}{18}$ and $m=2000,3000,4000$.  }
\label{fig:corr_len_Jt=2_appendix}
\end{figure}

\begin{figure}[ht]
\centering
\includegraphics[width=0.8\linewidth]{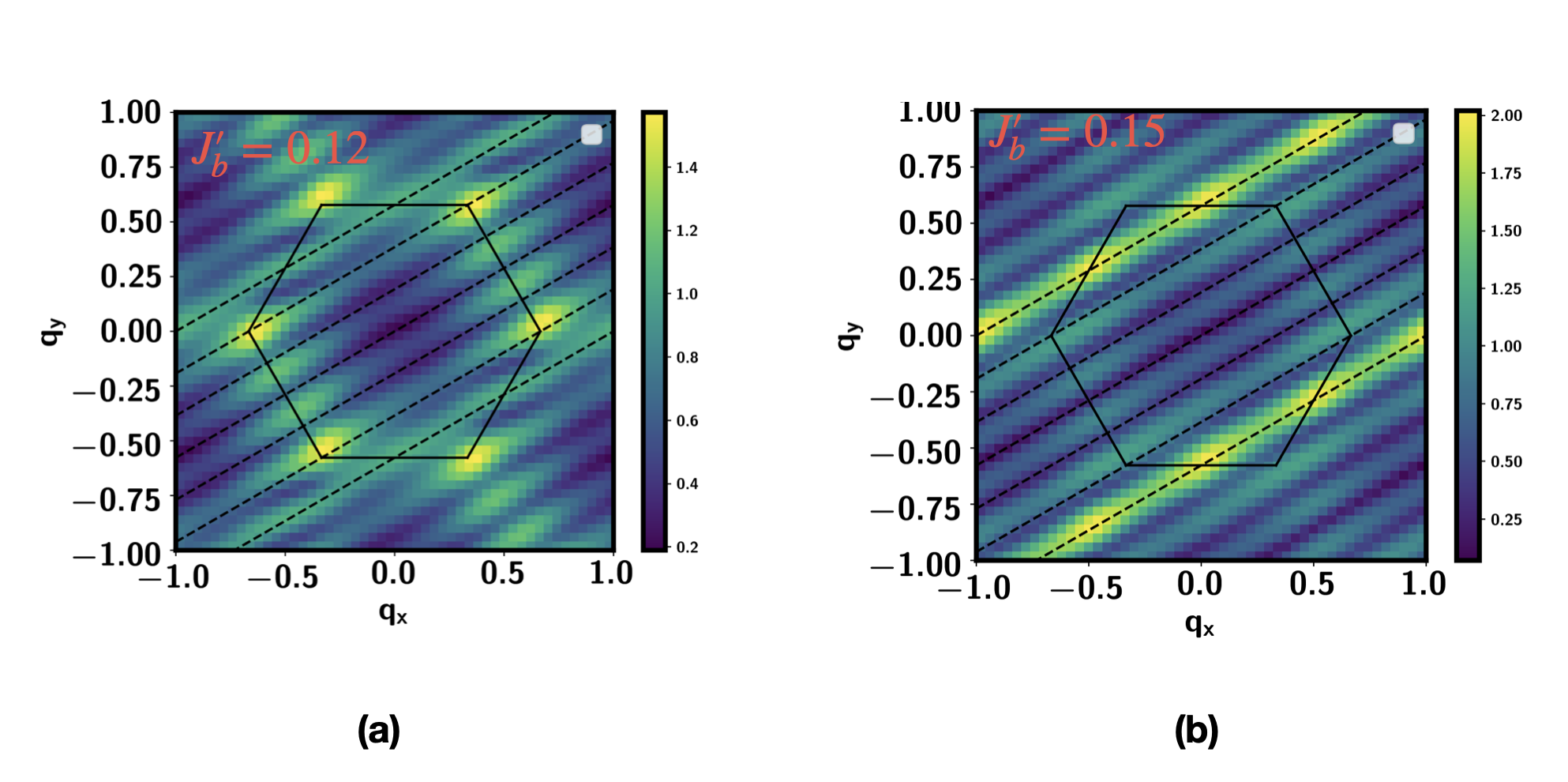}
\caption{$\langle \vec{S}_b(\mathbf q)\cdot \vec{S_b}(-\mathbf q)\rangle$ at $J_b=J_p=J_{pz}=1$ and $J_t=2$ for system size $L_y=6$, $x=\frac{1}{18}$ and $m=2000,3000,4000$.  }
\label{fig:spin_Jt=2_appendix}
\end{figure}

In the above we use a large value of $J_t$ so the spin in the top layer is strongly gapped. Next we use smaller values of $J_t$ and show that the Z$_2$ spin liquid + PSF phases can still exist. In Fig.~\ref{fig:corr_len_Jt=2_appendix} we show the inverse correlation lengths $\frac{1}{\xi}$ for various operators at $J_t=2$. Again $\frac{1}{\xi_{S_t}}$ and $\frac{1}{\xi_{P}}$ are large and increase with $J^\prime_b$. Actually we believe that the spin gap $\Delta_t$ is still finite even at $J^\prime_b=0$ for $J_t=2$, as discussed in Sec.~\ref{subsection:with_magnetic_order}. $\xi^{-1}_{S_b}$ still gets maximized around  $J^\prime_b \approx 0.15$. In Fig.~\ref{fig:spin_Jt=2_appendix} we show later that there is still a transition  separating $Z_2$ SL I phase at $J^\prime_b=0.12$ and $Z_2$ SL II phase at $J^\prime_b=0.15$.

\begin{figure}[ht]
\centering
\includegraphics[width=0.8\linewidth]{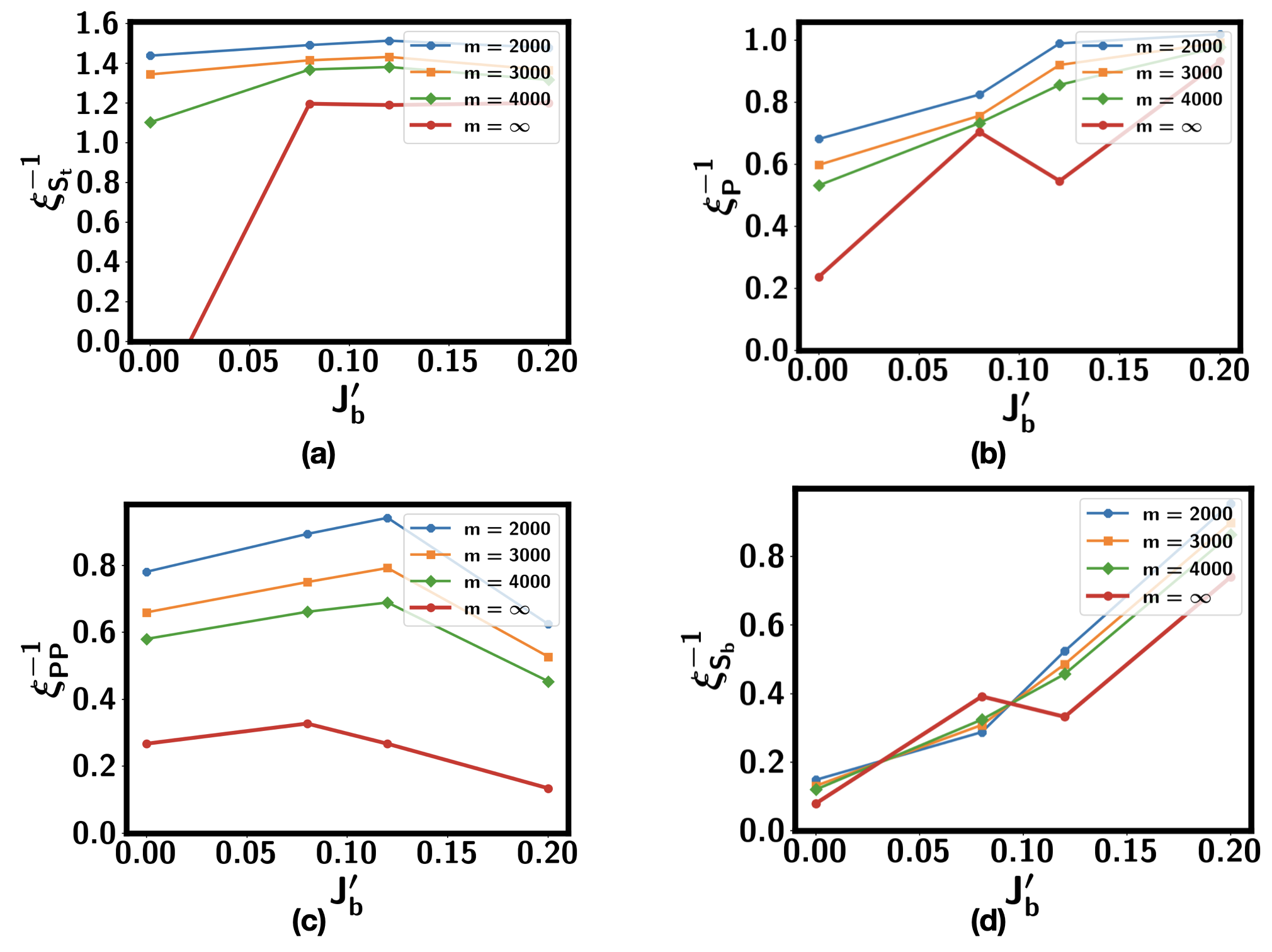}
\caption{Inverse correlation lengths at $J_b=1$, $J_p=2$, $J_{pz}=5$ and $J_t=4$ for system size $L_y=6$, $x=\frac{1}{18}$ and $m=2000,3000,4000$.  }
\label{fig:corr_len_Jt=2_Jp=2_appendix}
\end{figure}

\begin{figure}[ht]
\centering
\includegraphics[width=0.8\linewidth]{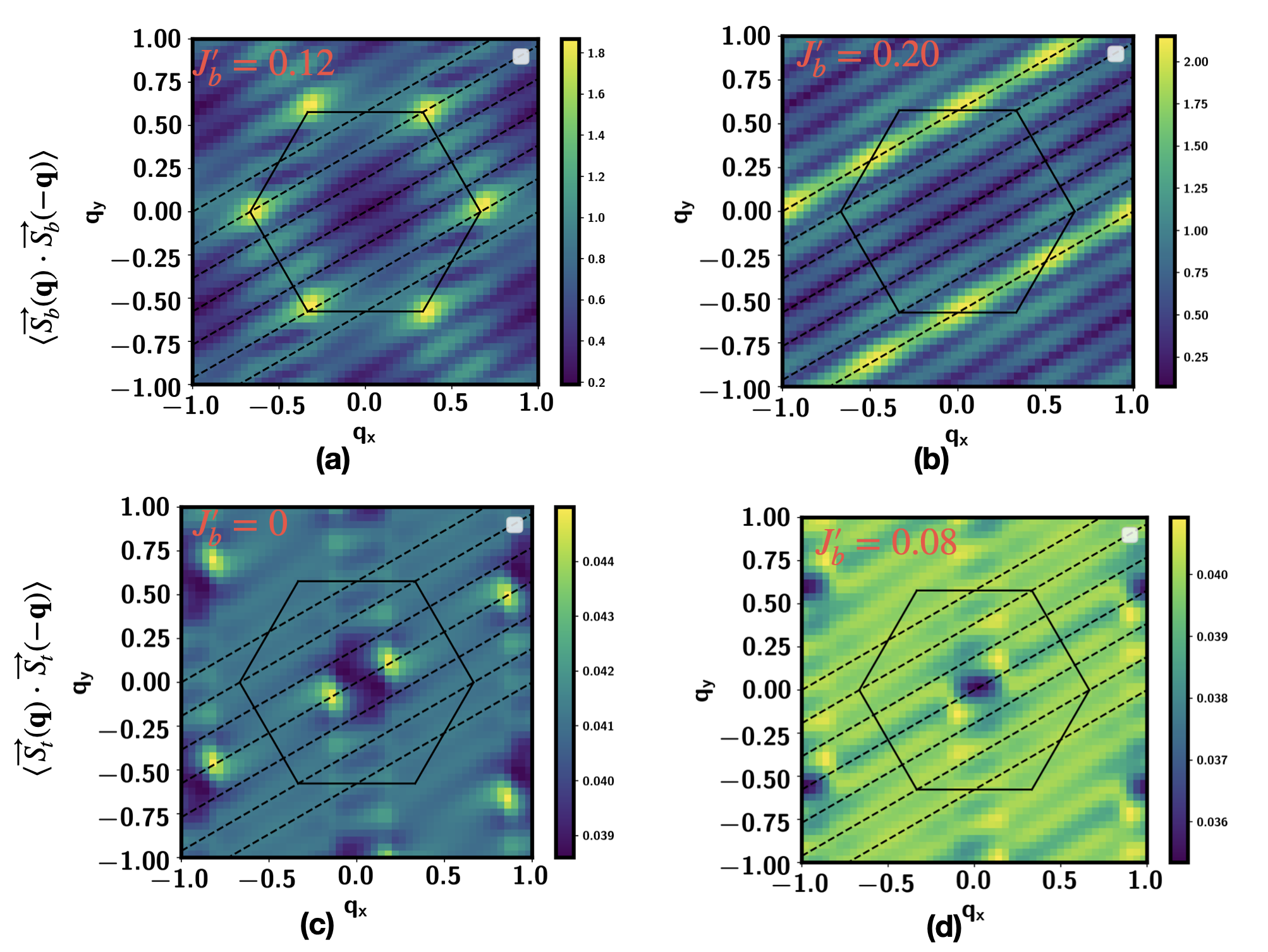}
\caption{Spin structure factor at $J_b=1$, $J_p=2$, $J_{pz}=5$ and $J_t=4$ for system size $L_y=6$, $x=\frac{1}{18}$ and $m=2000,3000,4000$.  (a)(b) $\langle \vec{S}_b(\mathbf q)\cdot \vec{S_b}(-\mathbf q)\rangle$; (c)(d)$\langle \vec{S}_t(\mathbf q)\cdot \vec{S_t}(-\mathbf q)\rangle$ }
\label{fig:spin_Jt=4_Jp=2_appendix}
\end{figure}

 We also show the inverse correlation lengths $\frac{1}{\xi}$ for $J_b=1$, $J_p=2$, $J_{pz}=5$ and $J_t=4$ in Fig.~\ref{fig:corr_len_Jt=2_Jp=2_appendix}. This time at $J^\prime_b=0$ there seems to have a very small $\frac{1}{\xi_{S_t}}$ and $\frac{1}{\xi_{S_b}}$ and is more consistent with a single exciton condensation phase with magnetic orders in both layers. Actually $\langle \vec{S_t}(\mathbf q)\cdot \vec{S_t}(-\mathbf q)\rangle$ is consistent with the spiral phase discussed in Sec.~\ref{subsubsection:intermediate_spiral} (see Fig.~\ref{fig:spin_Jt=4_Jp=2_appendix}(c)).  However, when $J^\prime_b\geq 0.08$, we again see a large $\frac{1}{\xi_{S_t}}$, $\frac{1}{\xi_P}$ and a finite value of $\frac{1}{\xi_{S_b}}$ which is maximized at $J^\prime_b=0.20$. At $J^\prime_b=0.12$ and $J^\prime_b=0.20$, the only small $\frac{1}{\xi}$ is from the paired exciton operator. From $\langle \vec{S}_b(\mathbf q)\cdot \vec{S}_b(-\mathbf q)\rangle$ (see Fig.~\ref{fig:spin_Jt=4_Jp=2_appendix}(a)(b)), we can see they are consistent with Z$_2$ SL I + PSF and Z$_2$ SL II + PSF phase respectively.  Therefore we conclude that a large value of $J_{pz}$ does not destroy the fractional superfluid phases.

\end{document}